\documentclass[a4paper,10pt]{article}
\pdfoutput=1
\usepackage{jheppub}
\usepackage{dsfont}
\usepackage{graphicx}
\usepackage{amsmath}
\usepackage{empheq}
\usepackage{subcaption}
\usepackage{mathtools}
\usepackage[shortlabels]{enumitem}
\usepackage{tikz}
\usetikzlibrary{arrows,shapes}
\usetikzlibrary{trees}
\usetikzlibrary{patterns}
\usetikzlibrary{matrix,arrows} 				
\usetikzlibrary{positioning}				
\usetikzlibrary{calc,through}				
\usetikzlibrary{decorations.pathreplacing}  
\usepackage{pgffor}							

\usetikzlibrary{decorations.pathmorphing}	
\usetikzlibrary{decorations.markings}
\usepackage{todonotes}
\tikzset{
    vector/.style={decorate, decoration={snake}, draw},
	provector/.style={decorate, decoration={snake,amplitude=2.5pt}, draw},
	antivector/.style={decorate, decoration={snake,amplitude=-2.5pt}, draw},
        smallvector/.style={decorate, decoration={snake,amplitude=1.5pt,post length=0.5mm}, draw},
    fermion/.style={draw=black, postaction={decorate},
        decoration={markings,mark=at position .55 with {\arrow[draw=black]{>}}}},
    fermionbar/.style={draw=black, postaction={decorate},
        decoration={markings,mark=at position .55 with {\arrow[draw=black]{<}}}},
    fermionnoarrow/.style={draw=black},
    gluon/.style={decorate, draw=black,
        decoration={coil,amplitude=4pt, segment length=5pt}},
    scalar/.style={dashed,draw=black, postaction={decorate},
        decoration={markings,mark=at position .55 with {\arrow[draw=black]{>}}}},
    scalarbar/.style={dashed,draw=black, postaction={decorate},
        decoration={markings,mark=at position .55 with {\arrow[draw=black]{<}}}},
    scalarnoarrow/.style={dashed,draw=black},
    electron/.style={draw=black, postaction={decorate},
        decoration={markings,mark=at position .55 with {\arrow[draw=black]{>}}}},
    bigvector/.style={decorate, decoration={snake,amplitude=4pt}, draw},
    arrow/.style={draw=black, postaction={decorate},
        decoration={markings,mark=at position 1 with {\arrow[draw=black]{>}}}},
}

\tikzstyle{block} = [draw, rectangle, 
    minimum height=3em, minimum width=6earticlem]
    
\newcommand{\boxcut}{\begin{tikzpicture}
        \begin{scope}
        \draw[-,thick] (0,0)--(1,0);
        \draw[-,double] (1,0)--(2,0);
        \draw[vector,double] (1,0)--(1,-2);
        \draw[fermion] (0,-2)--(1,-2);
        \draw[fermion] (1,-2)--(2,-2);
        \draw[-,double] (3,0)--(4,0);
        \draw[-,thick] (4,0)--(5,0);
        \draw[vector,double] (4,0)--(4,-2);
        \draw[fermion] (3,-2)--(4,-2);
        \draw[fermion] (4,-2)--(5,-2);
        \draw[->] (0.25,0.25)--(0.75,0.25);
        \draw[->] (4.35,0.25)--(4.85,0.25);
        \draw[->] (1.45,-0.5)--(1.45,-1.5);
        \draw[->] (4.35,-0.5)--(4.35,-1.5);
        \draw[dashed, blue!100!black, line width = 1.25pt] (2.5,0.75)--(2.5,-2.75);
        \node at (0,0.5) {$m_1 u_1, s_1$};
        \node at (0,-2.3) {$m_2 u_2$};
        \node at (5,-2.3) {$m_2 u_2 + q$};
        \node at (5,0.5) {$m_1 u_1 -q, s_1$};
        \node at (1.65,-1) {$\ell$};
        \node at (0.5,-1) {$|h|$};
        \node at (3.5,-1) {$|h|$};
        \node at (4.85,-1) {$q-\ell$};
        \filldraw (1,0) circle (3pt);
        \filldraw (4,0) circle (3pt);
        \end{scope}
     \end{tikzpicture}}

\newcommand{\absorptivecut}{\begin{tikzpicture}
        \begin{scope}
        \draw[-,thick] (0,0)--(2,0);
        \draw[-,double] (2,0)--(3,0);
        \draw[vector,double] (2,0)--(2,-2);
        \draw[fermion] (0,-2)--(2,-2);
        \draw[fermion] (2,-2)--(4,-2);
        \draw[-,double] (3,0)--(4,0);
        \draw[-,thick] (4,0)--(6,0);
        \draw[vector,double] (4,0)--(4,-2);
        \draw[fermion] (4,-2)--(6,-2);
        \draw[->] (0.5,0.25)--(1.5,0.25);
        \draw[->] (4.5,0.25)--(5.5,0.25);
        \draw[->] (1.65,-0.5)--(1.65,-1.5);
        \draw[->] (4.35,-0.5)--(4.35,-1.5);
        \draw[dashed, blue!100!black, line width = 1.25pt] (3,0.25)--(3,-0.25);
        \draw[dashed, blue!100!black, line width = 1.25pt] (1.75,-0.8)--(2.25,-0.8);
        \draw[dashed, blue!100!black, line width = 1.25pt] (3.75,-0.8)--(4.25,-0.8);
        \node at (1,0.5) {$m_1  u_1, s_1$};
        \node at (3,0.5) {$x, s_1'$};
        \node at (1,-2.3) {$m_2  u_2$};
        \node at (5.2,-2.3) {$m_2  u_2 + q$};
        \node at (5,0.5) {$m_1  u_1-q, s_1$};
        \node at (1.45,-1) {$\ell$};
        \node at (3,-1) {$|h|$};
        \node at (4.85,-1) {$q-\ell$};
        \filldraw (2,0) circle (3pt);
        \filldraw (4,0) circle (3pt);
        \filldraw[color=black, fill=black!20, very thick] (3, -2) ellipse (1.4 and 0.5);
        \end{scope}
    \end{tikzpicture}}

\newcommand{\absorptive}{\begin{tikzpicture}
        \begin{scope}
        \draw[-,thick] (0,0)--(2,0);
        \draw[-,double] (2,0)--(3,0);
        \draw[vector,double] (2,0)--(2,-2);
        \draw[fermion] (0,-2)--(2,-2);
        \draw[fermion] (2,-2)--(4,-2);
        \draw[-,double] (3,0)--(4,0);
        \draw[-,thick] (4,0)--(6,0);
        \draw[vector,double] (4,0)--(4,-2);
        \draw[fermion] (4,-2)--(6,-2);
        \draw[->] (0.5,0.25)--(1.5,0.25);
        \draw[->] (4.5,0.25)--(5.5,0.25);
        \draw[->] (1.65,-0.5)--(1.65,-1.5);
        \draw[->] (4.35,-0.5)--(4.35,-1.5);
        \node at (1,0.5) {$m_1  u_1, s_1$};
        \node at (3,0.5) {$x, s_1'$};
        \node at (1,-2.3) {$m_2  u_2$};
        \node at (5.2,-2.3) {$m_2  u_2 + q$};
        \node at (5,0.5) {$m_1  u_1-q, s_1$};
        \node at (1.45,-1) {$\ell$};
        \node at (3,-1) {$|h|$};
        \node at (4.85,-1) {$q-\ell$};
        \filldraw (2,0) circle (3pt);
        \filldraw (4,0) circle (3pt);
        \filldraw[color=black, fill=black!20, very thick] (3, -2) ellipse (1.4 and 0.5);
        \end{scope}
    \end{tikzpicture}}

\newcommand{\spectrumSOHOgraph}{\begin{tikzpicture}
        \begin{scope}
        \draw[-, thin] (0, 1.5)--(12, 1.5);
        \draw[-, thin] (0, 1.5)--(0, 7);
        \draw[-, thin] (0, 1.5)--(0.25, 1.5);
        \draw[-, thin] (0, 3)--(0.25, 3);
        \draw[-, thin] (0, 4.5)--(0.25, 4.5);
        \draw[-, thin] (0, 6)--(0.25, 6);
        \draw[line width = 2pt] (1, 1.55)--(3, 1.55);
        \draw[dashed, green!60!black, line width = 1.5pt] (3, 1.55)--(5, 1.55);
        \draw[dashed, green!60!black, line width = 1.5pt] (3, 1.55)--(7,3);
        \draw[dashed, draw opacity=0.3] (3, 1.55)--(7, 0);
        \draw[dashed, green!60!black, line width = 1.5pt] (3, 1.55)--(9, 4.5);
        \draw[dashed, draw opacity=0.3] (3, 1.55)--(9, -1.5);
        \draw[double, green!60!black, line width = 0.5pt, double distance = 2.5pt] (5, 1.55)--(7, 1.55);
        \draw[double, green!60!black, line width = 0.5pt, double distance = 2.5pt] (7, 3)--(9, 3);
        \draw[double, line width = 0.5pt, double distance = 2.5pt, draw opacity=0.2] (7, 0)--(9, 0);
        \draw[double, green!60!black, line width = 0.5pt, double distance = 2.5pt] (9, 4.5)--(11, 4.5);
        \draw[double, line width = 0.5pt, double distance = 2.5pt, draw opacity=0.2] (9, -1.5)--(11, -1.5);
        \node[scale = 1.7] at (2, 1) {$|\Delta s|$};
        \node[scale = 1.7] at (4, 1) {$=$};
        \node[scale = 1.7] at (6, 1) {$ 0 $};
        \node[scale = 1.7] at (8, 1) {$ 1 $};
        \node[scale = 1.7] at (10, 1) {$ 2 $};
        \node[scale = 1.7] at (-0.5, 7.5) {$s$};
        \node[scale = 1.7] at (-0.5, 1.5) {$0$};
        \node[scale = 1.7] at (-0.5, 3) {$1$};
        \node[scale = 1.7] at (-0.5, 4.5) {$2$};
        \node[scale = 1.7] at (-0.5, 6) {$3$};
        \node[scale = 1.7] at (6, 7) {$|h| = 0$};
        \node[scale = 1.7] at (6, 2) {$s' = 0$};
        \node[scale = 1.7] at (8, 3.4) {$s' = 1$};
        \node[scale = 1.7] at (10, 5) {$s' = 2$};
        \end{scope}
\end{tikzpicture}}

\newcommand{\spectrumSIHOgraph}{
\begin{tikzpicture}
        \begin{scope}
        \draw[-, thin] (0, 1.5)--(12, 1.5);
        \draw[-, thin] (0, 1.5)--(0, 7);
        \draw[-, thin] (0, 1.5)--(0.25, 1.5);
        \draw[-, thin] (0, 3)--(0.25, 3);
        \draw[-, thin] (0, 4.5)--(0.25, 4.5);
        \draw[-, thin] (0, 6)--(0.25, 6);
        \draw[line width = 2pt] (1, 3)--(3, 3);
        \draw[dashed, green!60!black, line width = 1.5pt] (3, 3)--(5, 3);
        \draw[dashed, green!60!black, line width = 1.5pt] (3, 3)--(7, 4.5);
        \draw[dashed, green!60!black, line width = 1.5pt] (3, 3)--(7, 1.55);
        \draw[dashed, green!60!black, line width = 1.5pt] (3, 3)--(9, 6);
        \draw[dashed, draw opacity=0.3] (3, 3)--(9, 0);
        \draw[-, double, green!60!black, line width = 0.5pt, double distance = 2.5pt] (5, 3)--(7, 3);
        \draw[-, double, green!60!black, line width = 0.5pt, double distance = 2.5pt] (7, 4.5)--(9, 4.5);
        \draw[-, double, green!60!black, line width = 0.5pt, double distance = 2.5pt] (7, 1.55)--(9, 1.55);
        \draw[-, double, green!60!black, line width = 0.5pt, double distance = 2.5pt] (9, 6)--(11, 6);
        \draw[-, double, line width = 0.5pt, double distance = 2.5pt, draw opacity=0.2] (9, 0)--(11, 0);
        \node[scale = 1.7] at (2, 1) {$|\Delta s|$};
        \node[scale = 1.7] at (4, 1) {$=$};
        \node[scale = 1.7] at (6, 1) {$ 0 $};
        \node[scale = 1.7] at (8, 1) {$ 1 $};
        \node[scale = 1.7] at (10, 1) {$ 2 $};
        \node[scale = 1.7] at (-0.5, 7.5) {$s$};
        \node[scale = 1.7] at (-0.5, 1.5) {$0$};
        \node[scale = 1.7] at (-0.5, 3) {$1$};
        \node[scale = 1.7] at (-0.5, 4.5) {$2$};
        \node[scale = 1.7] at (-0.5, 6) {$3$};
        \node[scale = 1.7] at (6, 7.5) {$|h| = 0$};
        \node[scale = 1.7] at (6, 3.4) {$s' = 1$};
        \node[scale = 1.7] at (8, 2) {$s' = 0$};
        \node[scale = 1.7] at (8, 4.9) {$s' = 2$};
        \node[scale = 1.7] at (10, 6.4) {$s' = 3$};
        \end{scope}
\end{tikzpicture}}

\newcommand{\spectrumSOHIgraph}{\begin{tikzpicture}
        \begin{scope}
        \draw[-, thin] (0, 1.5)--(12, 1.5);
        \draw[-, thin] (0, 1.5)--(0, 7);
        \draw[-, thin] (0, 1.5)--(0.25, 1.5);
        \draw[-, thin] (0, 3)--(0.25, 3);
        \draw[-, thin] (0, 4.5)--(0.25, 4.5);
        \draw[-, thin] (0, 6)--(0.25, 6);
        \draw[line width = 2pt] (1, 1.55)--(3, 1.55);
        \draw[dashed, red!80!black, draw opacity=0.9] (3, 1.55)--(5, 1.55);
        \draw[dashed, green!60!black, line width = 1.5pt] (3, 1.55)--(7,3);
        \draw[dashed, draw opacity=0.3] (3, 1.55)--(7, 0);
        \draw[dashed, green!60!black, line width = 1.5pt] (3, 1.55)--(9, 4.5);
        \draw[dashed, draw opacity=0.3] (3, 1.55)--(9, -1.5);
        \draw[-, double, red!80!black, line width = 0.5pt, double distance = 2.5pt, draw opacity=0.6] (5, 1.55)--(7, 1.55);
        \draw[-, double, green!60!black, line width = 0.5pt, double distance = 2.5pt] (7, 3)--(9, 3);
        \draw[-, double, line width = 0.5pt, double distance = 2.5pt, draw opacity=0.2] (7, 0)--(9, 0);
        \draw[-, double, green!60!black, line width = 0.5pt, double distance = 2.5pt] (9, 4.5)--(11, 4.5);
        \draw[-, double, line width = 0.5pt, double distance = 2.5pt, draw opacity=0.2] (9, -1.5)--(11, -1.5);
        \node[scale = 1.7] at (2, 1) {$|\Delta s|$};
        \node[scale = 1.7] at (4, 1) {$=$};
        \node[scale = 1.7] at (6, 1) {$ 0 $};
        \node[scale = 1.7] at (8, 1) {$ 1 $};
        \node[scale = 1.7] at (10, 1) {$ 2 $};
        \node[scale = 1.7] at (-0.5, 7.5) {$s$};
        \node[scale = 1.7] at (-0.5, 1.5) {$0$};
        \node[scale = 1.7] at (-0.5, 3) {$1$};
        \node[scale = 1.7] at (-0.5, 4.5) {$2$};
        \node[scale = 1.7] at (-0.5, 6) {$3$};
        \node[scale = 1.7] at (6, 6.5) {$|h| = 1$};
        \node[scale = 1.7] at (6, 2) {$s' = 0$};
        \node[scale = 1.7] at (8, 3.4) {$s' = 1$};
        \node[scale = 1.7] at (10, 5) {$s' = 2$};
        \end{scope}
\end{tikzpicture}}

\newcommand{\spectrumSIHIgraph}{\begin{tikzpicture}
        \begin{scope}
        \draw[-, thin] (0, 1.5)--(12, 1.5);
        \draw[-, thin] (0, 1.5)--(0, 7);
        \draw[-, thin] (0, 1.5)--(0.25, 1.5);
        \draw[-, thin] (0, 3)--(0.25, 3);
        \draw[-, thin] (0, 4.5)--(0.25, 4.5);
        \draw[-, thin] (0, 6)--(0.25, 6);
        \draw[line width = 2pt] (1, 3)--(3, 3);
        \draw[dashed, green!60!black, line width = 1.5pt] (3, 3)--(5, 3);
        \draw[dashed, green!60!black, line width = 1.5pt] (3, 3)--(7, 4.5);
        \draw[dashed, green!60!black, line width = 1.5pt] (3, 3)--(7, 1.55);
        \draw[dashed, green!60!black, line width = 1.5pt] (3, 3)--(9, 6);
        \draw[dashed, draw opacity=0.3] (3, 3)--(9, 0);
        \draw[-, double, green!60!black, line width = 0.5pt, double distance = 2.5pt] (5, 3)--(7, 3);
        \draw[-, double, green!60!black, line width = 0.5pt, double distance = 2.5pt] (7, 4.5)--(9, 4.5);
        \draw[-, double, green!60!black, line width = 0.5pt, double distance = 2.5pt] (7, 1.55)--(9, 1.55);
        \draw[-, double, green!60!black, line width = 0.5pt, double distance = 2.5pt] (9, 6)--(11, 6);
        \draw[-, double, line width = 0.5pt, double distance = 2.5pt, draw opacity=0.2] (9, 0)--(11, 0);
        \node[scale = 1.7] at (2, 1) {$|\Delta s|$};
        \node[scale = 1.7] at (4, 1) {$=$};
        \node[scale = 1.7] at (6, 1) {$ 0 $};
        \node[scale = 1.7] at (8, 1) {$ 1 $};
        \node[scale = 1.7] at (10, 1) {$ 2 $};
        \node[scale = 1.7] at (-0.5, 7.5) {$s$};
        \node[scale = 1.7] at (-0.5, 1.5) {$0$};
        \node[scale = 1.7] at (-0.5, 3) {$1$};
        \node[scale = 1.7] at (-0.5, 4.5) {$2$};
        \node[scale = 1.7] at (-0.5, 6) {$3$};
        \node[scale = 1.7] at (6, 6.5) {$|h| = 1$};
        \node[scale = 1.7] at (6, 3.4) {$s' = 1$};
        \node[scale = 1.7] at (8, 2) {$s' = 0$};
        \node[scale = 1.7] at (8, 4.9) {$s' = 2$};
        \node[scale = 1.7] at (10, 6.4) {$s' = 3$};
        \end{scope}
\end{tikzpicture}}

\newcommand{\spectrumSIHIIgraph}{\begin{tikzpicture}
        \begin{scope}
        \draw[-, thin] (0, 1.5)--(12, 1.5);
        \draw[-, thin] (0, 1.5)--(0, 7);
        \draw[-, thin] (0, 1.5)--(0.25, 1.5);
        \draw[-, thin] (0, 3)--(0.25, 3);
        \draw[-, thin] (0, 4.5)--(0.25, 4.5);
        \draw[-, thin] (0, 6)--(0.25, 6);
        \draw[line width = 2pt] (1, 3)--(3, 3);
        \draw[dashed, green!60!black, line width = 1.5pt] (3, 3)--(5, 3);
        \draw[dashed, green!60!black, line width = 1.5pt] (3, 3)--(7, 4.5);
        \draw[dashed, red!80!black, draw opacity=0.9] (3, 3)--(7, 1.55);
        \draw[dashed, green!60!black, line width = 1.5pt] (3, 3)--(9, 6);
        \draw[dashed, draw opacity=0.3] (3, 3)--(9, 0);
        \draw[-, double, green!60!black, line width = 0.5pt, double distance = 2.5pt] (5, 3)--(7, 3);
        \draw[-, double, green!60!black, line width = 0.5pt, double distance = 2.5pt] (7, 4.5)--(9, 4.5);
        \draw[-, double, red!80!black, line width = 0.5pt, double distance = 2.5pt, draw opacity=0.6] (7, 1.55)--(9, 1.55);
        \draw[-, double, green!60!black, line width = 0.5pt, double distance = 2.5pt] (9, 6)--(11, 6);
        \draw[-, double, line width = 0.5pt, double distance = 2.5pt, draw opacity=0.2] (9, 0)--(11, 0);
        \node[scale = 1.7] at (2, 1) {$|\Delta s|$};
        \node[scale = 1.7] at (4, 1) {$=$};
        \node[scale = 1.7] at (6, 1) {$ 0 $};
        \node[scale = 1.7] at (8, 1) {$ 1 $};
        \node[scale = 1.7] at (10, 1) {$ 2 $};
        \node[scale = 1.7] at (-0.5, 7.5) {$s$};
        \node[scale = 1.7] at (-0.5, 1.5) {$0$};
        \node[scale = 1.7] at (-0.5, 3) {$1$};
        \node[scale = 1.7] at (-0.5, 4.5) {$2$};
        \node[scale = 1.7] at (-0.5, 6) {$3$};
        \node[scale = 1.7] at (6, 6.5) {$|h| = 2$};
        \node[scale = 1.7] at (6, 3.4) {$s' = 1$};
        \node[scale = 1.7] at (8, 2) {$s' = 0$};
        \node[scale = 1.7] at (8, 4.9) {$s' = 2$};
        \node[scale = 1.7] at (10, 6.4) {$s' = 3$};
        \end{scope}
\end{tikzpicture}}

\newcommand{\spectrumSOHIIgraph}{\begin{tikzpicture}
        \begin{scope}
        \draw[-, thin] (0, 1.5)--(12, 1.5);
        \draw[-, thin] (0, 1.5)--(0, 7);
        \draw[-, thin] (0, 1.5)--(0.25, 1.5);
        \draw[-, thin] (0, 3)--(0.25, 3);
        \draw[-, thin] (0, 4.5)--(0.25, 4.5);
        \draw[-, thin] (0, 6)--(0.25, 6);
        \draw[ line width = 2pt] (1, 1.55)--(3, 1.55);
        \draw[dashed, red!80!black, draw opacity=0.9] (3, 1.55)--(5, 1.55);
        \draw[dashed, red!80!black, draw opacity=0.9] (3, 1.55)--(7,3);
        \draw[dashed, draw opacity=0.3] (3, 1.55)--(7, 0);
        \draw[dashed, green!60!black, line width = 1.5pt] (3, 1.55)--(9, 4.5);
        \draw[dashed, draw opacity=0.3] (3, 1.55)--(9, -1.5);
        \draw[-, red!80!black, double, line width = 0.5pt, double distance = 2.5pt, draw opacity=0.6] (5, 1.55)--(7, 1.55);
        \draw[-, double, red!80!black, line width = 0.5pt, double distance = 2.5pt, draw opacity=0.6] (7, 3)--(9, 3);
        \draw[-, double, line width = 0.5pt, double distance = 2.5pt, draw opacity=0.2] (7, 0)--(9, 0);
        \draw[-, double, green!60!black, line width = 0.5pt, double distance = 2.5pt] (9, 4.5)--(11, 4.5);
        \draw[-, double, line width = 0.5pt, double distance = 2.5pt, draw opacity=0.2] (9, -1.5)--(11, -1.5);
        \node[scale = 1.7] at (2, 1) {$|\Delta s|$};
        \node[scale = 1.7] at (4, 1) {$=$};
        \node[scale = 1.7] at (6, 1) {$ 0 $};
        \node[scale = 1.7] at (8, 1) {$ 1 $};
        \node[scale = 1.7] at (10, 1) {$ 2 $};
        \node[scale = 1.7] at (-0.5, 7.5) {$s$};
        \node[scale = 1.7] at (-0.5, 1.5) {$0$};
        \node[scale = 1.7] at (-0.5, 3) {$1$};
        \node[scale = 1.7] at (-0.5, 4.5) {$2$};
        \node[scale = 1.7] at (-0.5, 6) {$3$};
        \node[scale = 1.7] at (6, 6.5) {$|h| = 2$};
        \node[scale = 1.7] at (6, 2) {$s' = 0$};
        \node[scale = 1.7] at (8, 3.4) {$s' = 1$};
        \node[scale = 1.7] at (10, 5) {$s' = 2$};
        \end{scope}
\end{tikzpicture}}

\begin{document}

\title{Classical Spin Transitions and Absorptive Scattering}
\author[a]{Juan Pablo Gatica,}
\emailAdd{jpgatica3541@g.ucla.edu}
\author[b]{Callum R. T. Jones,}
\emailAdd{crtjones@arizona.edu}
\affiliation[a]{
Mani L. Bhaumik Institute for Theoretical Physics, Department of Physics and Astronomy, University of California Los Angeles, Los Angeles, CA 90095, USA
}
\affiliation[b]{
Department of Physics, University of Arizona, Tucson, Arizona 85721, USA
}

\abstract{We describe an on-shell, amplitudes-based approach to incorporating radiation absorption effects in the post-Minkowskian scattering of generic, compact, spinning bodies. Classical spinning observables are recovered by extrapolating to large spin, results calculated with finite quantum spin-$s$ particles using the properties of spin universality and Casimir interpolation. At leading-order our results give a completely general and non-redundant parametrization of absorptive observables in terms of a finite number of Wilson coefficients associated with 3-particle mass and spin-magnitude changing on-shell amplitudes. We denote these semi-fictitious microscopic processes: \textit{classical spin transitions}. Explicit results for the leading-order impulse due to the absorption of scalar, electromagnetic and gravitational radiation, for spin transitions $\Delta s = 0,\pm 1, \pm 2$ are given in a fully interpolated form up to $\mathcal{O}\left(S^2\right)$, and Casimir independent contributions given up to $\mathcal{O}\left(S^4\right)$. Our explicit results reveal some surprising universal patterns. We find that, up to identification of Wilson coefficients, the Casimir independent contributions to the impulse for spinning-up and spinning-down by the same magnitude $|\Delta s|$ are identical. For processes where the quantum $\Delta s<0$ transition is forbidden, the corresponding classical observable is suppressed in powers of $S$ by a predictable amount. Additionally we find that, while for generic non-aligned spin configurations there is a non-zero scattering angle at leading-order, for aligned spin, similar to non-spinning absorption, the scattering angle vanishes and the impulse is purely longitudinal. The formalism and results presented provide a significant extension of the amplitudes-based calculational pipeline for gravitational waveforms from binary black hole and neutron star systems beyond the point-particle approximation. 
}

\maketitle
\flushbottom

\section{Introduction}
\label{sec:intro}

The development of effective point-particle methods to simplify the two-body problem in general relativity is an important problem, both for it's own theoretical interest and more recently as part of a suite of calculational tools for making precision predictions in gravitational wave astrophysics. Complementary to expensive numerical methods, effective point-particle approaches based on the post-Newtonian (PN) \cite{Einstein:1938yz,Ohta:1973je,Goldberger:2004jt} and post-Minkowskian (PM) \cite{Bertotti:1956pxu,Kerr:1959zlt,Bertotti:1960wuq,Westpfahl:1979gu,Portilla:1980uz,Bel:1981be} perturbative expansions give compact, analytic expressions for observables associated with the early inspiral. Since the first detection of gravitational waves more than a decade ago \cite{LIGOScientific:2016aoc,LIGOScientific:2017vwq}, there has been an explosion of activity in this area. Importing modern methods from high-energy particle physics, effective field theory (EFT) and the mathematics of Feynman integrals, the state-of-the art has advanced by several orders of perturbation theory in a few short years \cite{Cheung:2018wkq,Bern:2019nnu,Bern:2021yeh,Dlapa:2021npj,Dlapa:2021vgp,Driesse:2024xad,Driesse:2024feo}. Future generations of gravitational wave observatories, both ground and space-based, are expected to increase the signal-to-noise ratio by two orders of magnitude \cite{Reitze:2019iox,Punturo:2010zz,LISA:2017pwj,ET:2025xjr}. Both for these forthcoming observational programs, and into the farther future, precision measurement requires precision theory to resolve detailed properties of the gravitational wave sources. 

These predictions must be made in the context of a systematic EFT framework that allows us to incorporate, and parametrize our ignorance of, all relevant physical effects and degrees-of-freedom. In this context, the strict point-particle approximation cannot be the end of the story. For generic macroscopic bodies, including black holes and neutron stars, low-frequency radiation can be \textit{absorbed} by microscopic degrees of freedom, leading to a change of \textit{rest-mass} and \textit{spin-magnitude}. In the past few years, there has been a renewal of interest in this subject, both from the perspective of on-shell scattering amplitude methods \cite{Aoude:2023fdm,Jones:2023ugm,Chen:2023qzo,Aoude:2024jxd,Bautista:2024emt}, and more traditional worldline EFT approaches \cite{Goldberger:2005cd,Porto:2007qi,Goldberger:2020wbx,Goldberger:2020fot}. In this paper we further develop the scattering amplitude based approach to absorption, developing a systematic EFT applicable to generic spinning bodies.

 In the conservative and radiative sectors, the use of quantum scattering amplitudes methods with spinning external states to model classical, macroscopic, spinning bodies is now mature and well-established. Beginning with the pioneering works \cite{Holstein:2008sx,Holstein:2008sw,Vaidya:2014kza}, it was shown that one can correctly recover predictions for the scattering (unbound) dynamics of classical spinning bodies from quantum scattering amplitudes with finite spin particles. In essence this requires extrapolating a sequence of quantum observables, calculated from microscopically small values of the spin: $s = \frac{1}{2}\hbar$, $\hbar$, $\frac{3}{2}\hbar$, $2\hbar$, ... to macroscopically large values.\footnote{As an explicit example, for a black hole of mass $M=35 M_{\odot}$ near the Kerr bound (e.g. the larger black hole in GW150914 \cite{LIGOScientific:2016wkq}), the spin angular momentum is approximately $s \approx \frac{G M^2}{c} \approx 10^{45} \;\text{J}\cdot \text{s} \approx 10^{79} \times \hbar $.} Remarkably, this approach is verifiably successful, for conservative and radiative observables, giving results in complete agreement with intrinsically classical calculations \cite{Holstein:2008sx,Holstein:2008sw,Maybee:2019jus,FebresCordero:2022jts,Akpinar:2025bkt,Akpinar:2025byi}. This \textit{quantum first} approach has some unique advantages; it allows the use of modern unitarity methods for the construction of loop integrands \cite{Bern:1994zx,Bern:1994cg,Bern:1995db}, and intrinsically on-shell tools like the double-copy as a means of taming the complex non-linear structure of general relativity \cite{Kawai:1985xq,Bern:2008qj,Bern:2010ue}. Moreover, this approach is naturally suited to parameterizing EFT contributions, scattering amplitudes being intrinsically gauge and field redefinition invariant.\footnote{For a (probably incomplete) list of related recent approaches to classical spinning observables see \cite{Vines:2017hyw,Guevara:2018wpp,Chung:2018kqs,Guevara:2019fsj,Chung:2019duq,Chung:2020rrz,Bern:2020buy,Kosmopoulos:2021zoq,Chen:2021kxt,Bern:2022kto,Bern:2023ity,Menezes:2022tcs,Aoude:2022thd,Aoude:2023vdk,Gatica:2024mur,Gatica:2023iws,Liu:2021zxr,Jakobsen:2021lvp,Jakobsen:2021zvh,Jakobsen:2022fcj,Jakobsen:2022zsx,Jakobsen:2023ndj,Chen:2024mmm,Bohnenblust:2024hkw,Haddad:2024ebn,Ben-Shahar:2023djm}.}

 In this paper we apply this finite quantum spin approach to the problem of absorption. This framework gives an interesting (semi-fictitious) perspective on the underlying “microscopic” process of absorption. The incoming quantum particle has spin-magnitude $s$, absorbs a (potential mode) of the mediator field, producing a (not necessarily particle-like) excited state with spin $s\pm\Delta s$ and mass $\mu^2 > m^2$. As we explain in greater detail in Section \ref{sec:review}, the leading-order (in the soft expansion) absorptive contribution to the impulse requires only the 3-particle \textit{mass/spin-magnitude changing} amplitudes of the form $\mathcal{A}_3^{(s,s\pm\Delta s,h)}$, depicted in Figure \ref{fig:3pt}.
 \begin{figure}
    \begin{center}
    \begin{tikzpicture}
        \draw[-,thick] (-1.5,0)--(0,0);
        \draw[-,double] (0,0)--(1.5,0);
        \draw[vector] (0,0)--(0,-1.5);
        \filldraw (0,0) circle (3pt);
        \node at (-7,-0.5) {
        \begin{minipage}{3cm}
        \begin{equation*}
            \mathcal{A}_3^{(s_1,s_2,h_3)}\left(p_1,p_2,p_3\right) \hspace{10mm} \equiv
        \end{equation*}
        \end{minipage}
        };
        \node[left] at (-1.5,0) {$\epsilon_1^{(s_1)}$};
        \node[right] at (1.5,0) {$\epsilon_2^{(s_2)}$};
        \node[below] at (0,-1.5) {$\epsilon_3^{(h_3)}$};
        \draw[->] (-0.3,0.2)--(-1.2,0.2);
        \draw[->] (0.3,0.2)--(1.2,0.2);
        \draw[->] (-0.3,-0.4)--(-0.3,-1.3);
        \node at (-0.75,0.5) {$p_1$};
        \node at (0.75,0.5) {$p_2$};
        \node at (-0.55,-0.75) {$p_3$};
    \end{tikzpicture}
\end{center}
    \caption{Building blocks of the leading-order absorption calculation: mass/spin-magnitude changing on-shell 3-particle amplitudes. Here parametrized in the all-outgoing convention, an incoming asymptotic state with mass $m^2$ and spin-magnitude $s_1$ absorbs a quantum of radiation (scalar, electromagnetic or gravitational); the resulting body is then in an ``excited state" with mass $\mu^2>m^2$ and spin-magnitude $s_2$, which in general differs from $s_1$. After extrapolating to the classical, large-spin regime we refer to such processes as \textit{classical spin transitions}.}
    \label{fig:3pt}
\end{figure}

We use the on-shell in-in formalism of KMOC to calculate the classical 4-momentum impulse $\Delta p_1^\mu$ \cite{Kosower:2018adc}. In \cite{Jones:2023ugm}, this formalism was used together with a general K\"all\'en-Lehmann spectral decomposition of the two-point function describing the propagation of the ``excited states", to calculate the leading absorptive impulse due to the scattering of a spinless compact body (including a Schwarzschild black hole) with a structureless second body sourcing a scalar, electromagnetic or gravitational field. This paper is largely an extension of the framework introduced in \cite{Jones:2023ugm} to incorporate spin degrees-of-freedom; but is also a refinement of it. In this work we dispense with the ultimately unnecessary off-shell concept of an invisible sector composite operator and work fully with on-shell amplitudes. Fundamentally the two approaches are mathematically identical, differing only in their conceptual apparatus. 

 We show that the calculated absorptive impulse exhibits the expected universality when extrapolated to large spin, and since there are finitely many on-shell 3-particle amplitudes $\mathcal{A}_3^{(s,s\pm\Delta s,h)}$, at this order it is possible to give a completely general parametrization of our ignorance of classical absorptive dynamics. Extending this beyond leading-order presents additional challenges. Explicitly we calculate the leading-order absorptive impulse for external spin $s=0,1,2$ (in natural units $\hbar=1$), with spin-magnitude change $\Delta s=0,\pm 1,\pm 2$. The results up to $\mathcal{O}(S^2)$ are given in Appendix \ref{app:results}, and up to $\mathcal{O}(S^4)$ in an ancillary file. The explicit results reveal some unexpected universal patterns. 

 Remarkably, the Casimir independent contributions (described in greater detail in Section \ref{subsec:spinuniandInterpolation}) for $s\rightarrow s+\Delta s$ (the \textit{ceiling}) and $s\rightarrow s-\Delta s$ (the \textit{floor}) are identical. In other words, there is an approximate symmetry between spinning-up and spinning-down by the same magnitude that is broken at leading order in the soft expansion only by contributions proportional to the spin Casimir. Similarly we observe a universal spin-suppression associated with the absence of a “floor” process $s \rightarrow s-\Delta s$. As discussed in greater detail in Section \ref{sec:spinsuppression}; the logic behind this observation is a simple consequence of the floor-ceiling symmetry and spin universality. Since, trivially, spin-magnitude is positive, if $\Delta s > s$, then the “floor” process $s \rightarrow s-\Delta s$ does not exist. This manifests as the suppression of the “ceiling” process by a predictable power of the soft scale $\lambda$. Generically, the first unsuppressed contribution appears from a calculation involving particles with external spin $s=\Delta s$ and therefore by universality must be $\mathcal{O}(S^{2\Delta s})$. The final surprise we observe is the vanishing of the leading-order absorptive scattering angle in the aligned spin limit $\mathbf{S}\propto \mathbf{L}$. This extends previous observations that the scattering angle vanishes for spinless absorption \cite{Jones:2023ugm}. Contrarily, for non-aligned spin configurations we observe that the leading-order scattering angle is generically \textit{non-zero}. Each of these observations is made for arbitrary values of the Wilson coefficients; these relations are therefore not special properties of black holes, but are universal kinematic properties and no doubt have a simple explanation that remains to be found. 

For macroscopic spinning bodies, a new physical effect becomes important, \textit{rotational superradiance} \cite{Brito:2015oca,Endlich:2016jgc}. From the point of view of the microscopic spin transitions, this corresponds to the initial state spontaneously emitting a quantum of radiation, decreasing the mass and changing the magnitude of the spin of the body. This is closely related to absorption (of positive energy modes), but is also clearly physically distinct. Our calculated absorptive contributions to the impulse are analytic in spin (excepting possible spin-dependent Wilson coefficients), whereas it has long been known that the absorption cross section for spinning black holes are non-analytic at $S^\mu \rightarrow 0$ as a consequence of superradiant contributions \cite{Page:1976df}. In Section \ref{sec:discussion} we speculate on extensions of our framework to incorporate rotational superradiance, but otherwise leave it as future work.\\
\\
The paper is organized as follows. In Section \ref{sec:review} we review the KMOC formalism and the soft limit in the context of absorptive and spin effects while focusing on the impulse as our classical observable. In Section \ref{sec:spinreview}, we go in to detail of how we use definite-spin representations to calculate spin effects in our amplitudes. We also consider unavoidable ambiguities in this scheme that come from the Casimir operator $S^2$. In Section \ref{sec:spin}, we describe the building blocks for the spin-transition amplitudes and introduce some over-arching results in our impulse calculation. In the Appendix \ref{app:3ptAmps} we provide the explicit three-point amplitudes and in Appendix \ref{app:results} we provide the explicit impulse results.\\
\\
\textbf{Notation and conventions:} Throughout this paper, unless otherwise indicated, we use natural units $\hbar = c = 1$. Lorentz covariant expressions are defined with the mostly minus metric signature $\eta_{\mu\nu} = \eta^{\mu\nu} = \text{diag}\left(+1,-1,-1,-1\right)$. For expressions involving Levi-Civita symbols we employ Schoonschip notation: $\varepsilon^{abcd} \equiv \varepsilon^{\mu\nu\rho\sigma} a_\mu b_\nu c_\rho d_\sigma$, $\varepsilon^{\mu bcd} \equiv \varepsilon^{\mu\nu\rho\sigma} b_\nu c_\rho d_\sigma$, etc. We also use the normalization for the Levi-Civita symbol $\varepsilon^{0123} = +1$. Tensor indices are symmetrized and anti-symmetrized as 
\begin{equation}
    T^{(\mu_1...\mu_n)} \equiv \frac{1}{n!}\sum_{\sigma \in S_n} T^{\mu_{\sigma(1)}...\mu_{\sigma(n)}}, \hspace{10mm} T^{[\mu_1...\mu_n]} \equiv \frac{1}{n!}\sum_{\sigma \in S_n} \text{sgn}\left(\sigma\right)T^{\mu_{\sigma(1)}...\mu_{\sigma(n)}},
\end{equation}
respectively. Loop and phase space integrals are evaluated using dimensional regularization, in $D=4-2\epsilon$ dimensions. The following shorthand is used to suppress factors of $2\pi$: $\hat{\text{d}}^{D}p_i \equiv {\text{d}^{D}p_i}/{(2 \pi)^D}$, and $\hat{\delta}(p_i^2 - m_i^2) \equiv 2 \pi \delta(p_i^2 - m_i^2)$.

\section{Classical Observables from Quantum Amplitudes}
\label{sec:review}

In this section we describe the use of the KMOC formalism \cite{Kosower:2018adc, Maybee:2019jus} to calculate absorptive contributions to the classical impulse in a 2-body scattering event. Details of the integrand construction, soft expansion and loop integration are given.

\subsection{In-in Scattering Observables}
\label{subsec:absorption}
In this paper the physical observable we study is the impulse imparted on body-1 during a 2-to-2 scattering event. Quantum mechanically this is given by the asymptotic change in the expectation value of the 4-momentum operator $\mathbb{P}^{\mu}_1$
\begin{equation}
\label{kmocbase}
    \Delta p^{\mu}_1 \equiv \langle \text{in}|S^\dagger \mathbb{P}^{\mu}_1 S | \text{in} \rangle - \langle\text{in}| \mathbb{P}^{\mu}_1 |\text{in} \rangle.
\end{equation}
Following \cite{Kosower:2018adc}, and assuming body-2 is non-spinning, we choose the in-state to be 
\begin{equation}
\label{wavepacket}
    |\text{in} \rangle \equiv \sum_{a_1=1}^{2s_1+1} \int \text{d} \Phi(p_1,m_1^2) \text{d} \Phi(p_2,m_2^2) \, \phi(p_1)\phi(p_2) \, \xi^{a_1}  \, e^{i b_1 \cdot p_1}\, e^{i b_2 \cdot p_2} \, | p_1, \{s_1,a_1\}; p_2\rangle,
\end{equation}
where $s_1$ is the \textit{spin-magnitude} and $a_1$ is an index labeling the components of the $2s_1+1$ dimensional representation of $SU(2)$, the \textit{little-group} of body-1. The state $| p_1, \{s_1,a_1\}; p_2\rangle$ is a 2-body momentum eigenstate and the minimal-uncertainty wavepackets $\phi(p_i)e^{i b_i \cdot p_i}$ are peaked around 4-momenta $p_i^\mu\approx m_i u_i^\mu$ and impact parameter $b_i^\mu$ \cite{Kosower:2018adc, Maybee:2019jus, Luna:2023uwd, Cristofoli:2021jas}. The little-group vector $\xi^{a_1}$ defines a \textit{spin-coherent} state approximating the classical spin vector $S_1^\mu$ in the correspondence limit \cite{Aoude:2021oqj}. For simplicity, in this paper body-2 is assumed to be non-spinning. Both parts of the wavefunction are normalized such that $\int \text{d}\Phi(p,m^2) |\phi(p)|^2 = \sum_a |\xi^a|^2 = 1$; where the Lorentz invariant phase space measure is defined as 
\begin{equation}
    \int \text{d}\Phi(p_i,m_i^2) \equiv \int \hat{\text{d}}^{D}p_i \, \hat{\delta}^{(+)}(p_i^2 - m_i^2).
\end{equation}
Assuming the standard unitarity relation $T^\dagger T = i(T -T^\dagger)$, where $S\equiv 1+iT$, it is useful to rewrite (\ref{kmocbase}) as the sum of so-called \textit{virtual} and \textit{real} contributions
\begin{equation}\label{vandr}
    \Delta p^{\mu}_1 = \underbrace{\langle \text{in}|  i[\mathbb{P}^{\mu}_1, T] |\text{in} \rangle}_{\text{virtual}} + \underbrace{\langle \text{in}|  T^{\dagger} [\mathbb{P}^{\mu}_1, T] |\text{in} \rangle}_{\text{real}}.
\end{equation}
We re-express the change in the observable for 2-body scattering as 
\begin{equation}\label{kmocschem}
    \Delta p^{\mu}_1 = \sum_{a_1, a_1'=1}^{2s_1+1}\int \left(\prod_{i = 1}^{2} \text{d} \Phi(p_i,m_i^2) \, \text{d} \Phi(p_i',m_i^2) \,\phi^{*}(p_i')\phi(p_i)  \, e^{-i b_i \cdot( p_i'- p_i)} \right) \xi^{*}_{a_1'} \,   \left(\mathcal{I}^{\mu}_{v}  + \mathcal{I}^{\mu}_{r}\right)^{a_1'}_{a_1}\xi^{a_1},
\end{equation}
where
\begin{align}
    \mathcal{I}^{\mu}_{v} &\equiv \langle{p_1',\{s_1,a_1'\}; p_2'}|  i [\mathbb{P}^{\mu}_1, T] |{ p_1, \{s_1,a_1\}; p_2} \rangle, \nonumber\\
    \mathcal{I}^{\mu}_{r} &\equiv \langle{p_1',\{s_1,a_1'\}; p_2'}|  T^{\dagger} [\mathbb{P}^{\mu}_1, T] |{p_1, \{s_1,a_1\}; p_2} \rangle,
\end{align}
are the virtual and real \textit{kernels} respectively. The virtual kernel is simply related to the elastic scattering amplitude
\begin{equation}
    \mathcal{I}^{\mu}_{v} = i\hat{\delta}^{(D)}\left(\sum_{i=1}^2 p_i-\sum_{i=1}^2 p'_i\right)(p_1'^\mu-p_1^\mu)\mathcal{A}_4\left(\phi^{a_1}_1(p_1)\phi_2(p_2)\rightarrow \phi_1^{a_1
'}(p_1') \phi_2(p_2')\right).
\end{equation}
The real kernel is evaluated by inserting a complete set of states between $T^\dagger$ and $[\mathbb{P}_1^\mu,T]$. Different sectors of the Hilbert space give different contributions, and at leading-order these are separately well-defined and gauge invariant. Contributions of the form
\begin{equation}
    \mathds{1} \supset \sum_{b_1=1}^{2s_1+1}\int \text{d}\Phi(r_1,m_1^2) \text{d}\Phi(r_2,m_2^2) |r_1, \{s_1,b_1\}; r_2\rangle \langle r_1, \{s_1,b_1\}; r_2|,
\end{equation}
define the \textit{conservative} sector; other contributions, including radiation modes and internal excited states, define the \textit{dissipative} sector. We are interested in contributions corresponding to excited single-particle states of body-1
\begin{equation}
    \label{absstates}
    \mathds{1} \supset \sum_{s_1',\pm}\sum_{b_1=1}^{2s_1'+1} \int_{m_{1*}^2}^\infty \text{d} \mu^2 \rho_{s_1'}^\pm(\mu^2)\int \text{d}\Phi(r_1,\mu^2) \text{d}\Phi(r_2,m_2^2) |r_1, \{s_1',b_1\}, \mu^2; r_2\rangle \langle r_1, \{s_1',b_1\}, \mu^2; r_2|.
\end{equation}
Here the excited-state in general is labeled by its Lorentz invariant quantum numbers: mass $\mu^2$, spin-magnitude $s_1'$, and (if a symmetry of the system) an intrinsic parity $\pm$. All of these may be different from body-1 in the in-state. For a given spin and parity there may be contributions of states with a range of masses, and in general we do not assume these are narrow-width particle-like excitations. Consequently we include an integration over $\mu^2$ weighted by an unknown spectral density function $\rho_{s_1'}^\pm(\mu^2)$. For macroscopic spinning bodies we expect that it should be possible to \textit{decrease} the total rest-mass by the spontaneous emission of a superradiant mode \cite{Brito:2015oca}, and so for at least some contributions $m_{1*}^2 < m_1^2$.

The excited-state contributions to the real kernel then takes the form 
\begin{align}
    \label{realkernel}
    \mathcal{I}^{\mu}_{r} &= \sum_{s_1',\pm}\sum_{b_1=1}^{2s_1'+1} \int_{m_{1*}^2}^\infty \text{d} \mu^2 \rho_{s_1'}^\pm(\mu^2)\int \text{d}\Phi(r_1,\mu^2) \text{d}\Phi(r_2,m_2^2) \nonumber\\
    &\hspace{5mm} \langle{p_1',\{s_1,a_1'\}; p_2'}|  T^{\dagger}|r_1, \{s_1',b_1\}, \mu^2; r_2\rangle \langle r_1, \{s_1',b_1\}, \mu^2; r_2| [\mathbb{P}^{\mu}_1, T] |{p_1, \{s_1,a_1\}; p_2} \rangle \nonumber\\
    &= \hat{\delta}^{(D)}\left(\sum_{i=1}^2 p_i-\sum_{i=1}^2 p'_i\right)\sum_{s_1',\pm}\sum_{b_1=1}^{2s_1'+1} \int_{m_{1*}^2}^\infty \text{d} \mu^2 \rho_{s_1'}^\pm(\mu^2)\int \text{d}\Phi(r_1,\mu^2) \text{d}\Phi(r_2,m_2^2)\hat{\delta}^{(D)}\left(\sum_{i=1}^2 p_i-\sum_{i=1}^2 r_i\right)\nonumber\\
    &\hspace{5mm} \left(r^\mu_1-p^\mu_1\right)\mathcal{A}_4^*\left(\phi^{a_1'}_1(p_1)\phi_2(p_2) \rightarrow X_{s_1',\pm}^{b_1}(r_1)\phi_2(r_2)\right) \mathcal{A}_4\left(\phi^{a_1}_1(p_1)\phi_2(p_2)\rightarrow X_{s_1',\pm}^{b_1}(r_1)\phi_2(r_2)\right).
\end{align}   
We have written the $T$-matrix elements as scattering amplitudes involving excited states, denoted generically as $X$. Since the excited $X$-states are unstable, their inclusion in the asymptotic Hilbert space (\ref{absstates}) and the associated scattering amplitudes in (\ref{realkernel}) exist only in the usual formal sense in perturbation theory, stability of these states being recovered in the limit of zero coupling \cite{Eden:1966dnq}.\footnote{In the presence of superradiant modes, even the incoming state with mass $m_1^2$ is unstable.} In the calculation of \textit{inclusive} observables, such as the impulse $\Delta p_1^\mu$, we are summing over the $X$-states, the \textit{exclusive} amplitudes $\mathcal{M}(\phi_1\phi_2 \rightarrow X \phi_2')$ should be regarded as formal devices for organizing the intermediate kinematics.   

\subsection{Soft Expansion}
\label{subsec:softexpansion}

We use the framework described in the previous subsection to calculate dissipative contributions to the impulse in a kinematic regime defined by the hierarchy of scales (in natural units $\hbar=c=1$)
\begin{equation}
    \label{softkin}
    \lambda_{C}^{(i)} \ll R^{(i)} \ll b^\mu,
\end{equation}
where $\lambda_C^{(i)} \sim \frac{1}{m_i}$ and $R^{(i)}$ are the \textit{Compton wavelength} and \textit{classical charge radius} of body-$i$ respectively, and $b^\mu\equiv b_1^\mu-b_2^\mu$ the (relative) covariant impact parameter. The classical charge radii have different definitions depending on the force mediating particle being used to probe them 
\begin{equation}\label{eq:expparams}
    R^{(i)}_{\psi} \sim \frac{g_i^2}{m_i^3}, \hspace{10mm} R_{\gamma}^{(i)} \sim \frac{e^2 Q_i^2}{m_i}, \hspace{10mm} R_h^{(i)} \sim G m_i,
\end{equation}
where $g_i$ and $Q_i$ are the scalar and electric charges of body-$i$, $e$ is the dimensionless electromagnetic gauge coupling and $G$ is Newton's constant. For bodies that carry multiple charges we will not assume a hierarchy between these scales. We make no assumptions about the relative velocity, and all of the results presented are manifestly Lorentz covariant. Together, this regime describes the scattering of (semi-)classical, relativistic point-particles at large impact parameter; in the gravitational context this is usually referred to as \textit{post-Minkowskian (PM)} scattering, and in the electromagnetic context as \textit{post-Lorentzian (PL)} scattering. 

In practice this double-hierarchy is implemented by first expanding in the couplings $g$, $e$ and $G$, the usual Feynman diagrammatic loop expansion, and then subsequently expanding the momentum space KMOC kernels $\mathcal{I}^\mu_{r,v}(q)$ in the limit of small momentum transfer ($q\sim \frac{1}{b}$)
\begin{equation}
    q^\mu \ll m_1 u_1^\mu \sim m_2 u_2^\mu,
\end{equation}
where our kinematic conventions are depicted in Figure \ref{fig:unitaritycut}. Loop integrals are expanded in this limit prior to integration using the method of regions \cite{Beneke:1997zp}. In this context the dominant long-range contribution corresponds to a single relativistic region, the \textit{soft region} defined as $\ell^\mu \sim q^\mu$, where $\ell^\mu$ is the momentum associated with an internal massless mediator. 

To construct a systematic EFT description of the scattering of compact macroscopic bodies in the kinematic regime (\ref{softkin}) we first identify the relevant degrees-of-freedom. In this problem these are: Goldstone modes arising from the spontaneous breaking of translation and rotation symmetry, soft radiation and low-energy internal degrees-of-freedom associated with absorption. The translational and rotational Goldstone modes arise from the fact that, in the absence of external fields, there is no unique minimum energy position or orientation of a non-spherically symmetric compact body. Any choice of ground state necessarily spontaneously breaks translational and rotational symmetry. In classical physics, the rotational modes are associated with a spin vector $S^\mu$ which we will assume is macroscopically large in the sense that $S \sim \frac{1}{q} $; the relation with quantum spin in the correspondence limit is described in detail in Section \ref{sec:spinreview}.

Macroscopic compact bodies are expected to exhibit a nearly continuous spectrum of excited states, corresponding to excitations of the enormous number of (possibly unknown) internal degrees of freedom. As discussed in the previous section, we can incorporate these excited states into the calculation of dissipative observables using the KMOC in-in formalism, by including the contributions of (formal) $X$-states with spins $s_1'$ and invariant mass-squared $\mu^2$ with an integral over an \textit{a priori} unknown spectral density $\rho_{s_1'}^\pm(\mu^2)$. As depicted in Figure \ref{fig:spectral}, it is useful to separate this integration into three distinct regions: 

\begin{figure}
    \centering
    \begin{tikzpicture}
        \draw[->] (0,0)--(12,0);
        \draw[-] (1,0.2)--(1,-0.2);
        \draw[-] (4,0.2)--(4,-0.2);
        \draw[-] (7,0.2)--(7,-0.2);
        \node[right] at (12,0) {$\mu^2$};
        \node[below] at (1,-0.2) {$m_{1*}^2$};
        \node[below] at (4,-0.2) {$m_1^2$};
        \node[below] at (7,-0.2) {$\Lambda$};
        \node at (2.5,0.5) {
        \begin{minipage}{5cm}
            \begin{equation*}
                \overbrace{\hspace{28mm}}^{\text{superradiant modes}}
            \end{equation*}
        \end{minipage}
        };
        \node at (5.5,0.5) {
        \begin{minipage}{5cm}
            \begin{equation*}
                \overbrace{\hspace{28mm}}^{\text{absorptive modes}}
            \end{equation*}
        \end{minipage}
        };
        \node at (9.4,0.7) {
        \begin{minipage}{5cm}
            \begin{equation*}
                \overbrace{\hspace{45mm}}^{\text{heavy (off-shell) modes}}
            \end{equation*}
        \end{minipage}
        };
    \end{tikzpicture}
    \caption{Schematic representation of contributions to the spectral integral. The mass scale $\mu^2$ is the mass of the ``excited" internal state, $m_1^2$ is the mass of the incoming body, $m_{1*}^2$ is the minimal mass that can be reached by spontaneous emission of superradiant modes. Modes with $\mu^2 > \Lambda \sim q$, are always off-shell for scattering at large impact parameter, and so can be consistently integrated out giving a contribution to conservative tidal response operators.   }
    \label{fig:spectral}
\end{figure}
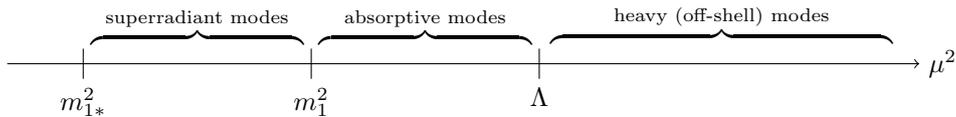

\newpage

\begin{enumerate}[(i)]
    \item $m_{1*}^2 \leq \mu^2 < m_1^2$: \hspace{4mm}\textit{superradiant modes},
    \item $m_1^2 \leq \mu^2 \leq \Lambda$: \hspace{7.5mm}\textit{absorptive modes},
    \item $\Lambda < \mu^2$: \hspace{17mm}\textit{heavy (off-shell) modes}.
\end{enumerate}
At leading-order, corresponding to diagrams of the form depicted in Figure \ref{fig:unitaritycut}, contributions from region (i) describe the spontaneous emission of a superradiant mode. We will discuss prospects for incorporating these contributions further in Section \ref{sec:discussion}. 

Absorption (of positive energy radiation), by definition, corresponds to the contributions of regions (ii) and (iii) for which the mass of the internal state exceeds the mass of the incoming body: $\mu^2 > m_1^2$. The scale $\Lambda$ is chosen (somewhat arbitrarily) to separate modes that can and cannot go on-shell during scattering in the regime of small momentum transfer. The contributions from region (iii) can be consistently integrated out to produce higher-derivative  effective (tidal response) operators; we will assume this has been parametrized as part of the calculation in the conservative sector \cite{Cheung:2020sdj,Bern:2020uwk}. For the rest of the paper, \textit{absorption} will refer to the contributions from region (ii). 

As discussed in \cite{Jones:2023ugm}, it is convenient to re-define the spectral integration in the formal $X$-state propagator as
\begin{equation}
    \label{eq:spectralintegral}
    \int_{m_1^2}^{\Lambda} \text{d} \mu^2 \, \rho(\mu^2)\,\frac{i \, \Pi ^{\alpha_1 \cdots \alpha_s}_{\,\,\beta_1 \cdots \beta_s}(k)}{k^2 - \mu^2 + i0}  \rightarrow  \frac{1}{2 m_1}\int_{0}^{\infty} \text{d} x \, \rho(x) \frac{i\Pi ^{\alpha_1 \cdots \alpha_s}_{\,\,\beta_1 \cdots \beta_s}(k)}{ k^2 -m_1^2 -2m_1 x +i0},
\end{equation}
where $\mu^2 \equiv m_1^2 +2m_1 x$, and $\Pi^{\alpha_1...\alpha_s}_{\beta_1...\beta_s}$ is the projector for a massive, symmetric-traceless spin-$s$ field. Here we have replaced the hard-cutoff $\Lambda$ with a form of analytic regularization and expand the spectral function to leading-order near the mass of the incoming state, $\mu^2 \approx m_1^2$. In what follows we will assume that this has the form of a power law\footnote{Without loss of generality we can absorb the overall coefficient of the spectral function into a redefinition of the Wilson coefficients appearing in the 3-point mass/spin changing amplitudes given in Appendix \ref{app:3ptAmps}.}
\begin{equation}
    \rho(x) \sim x^{1+\alpha}, \hspace{10mm} \text{as} \hspace{10mm} x\rightarrow 0^+.
\end{equation}
The definition of the analytic regularization is that the integral on the right-hand-side is evaluated for some range of values of $\alpha$ for which it is convergent, and then analytically continued to the physical value. In \cite{Jones:2023ugm}, these low-energy spectral functions were determined for scalar, photon and graviton absorption by a Schwarzschild black hole by matching with the known absorption cross-section. In each case, the spectral function was found to be a linear function at low-energies, meaning $\rho(x)\sim x$ as $x\rightarrow 0^+$. For the rest of this paper, we will continue to make this assumption, defining the spectral integral by continuation to $\alpha \rightarrow 0$. In principle however, this represents a further UV parameter to be determined by matching; it is straightforward to repeat our calculations for other values. If body-1 is incoming with momentum $p_1^\mu$ and absorbs radiation with momentum $q^2$, the invariant mass of the excited state is $\mu^2 \approx m_1^2 + 2p_1 \cdot q$. The $X$-states that are relevant for absorption in the soft region therefore scale as $x\sim q$. 

All together, we define the \textit{soft expansion} by the following scaling relations
\begin{align}
    \label{soft}
     l^\mu \sim q^\mu \sim x \sim b^{-1} \sim \lambda,  \hspace{10mm}
    u_1 ^\mu \sim u_2^\mu \sim \lambda^0,  \hspace{10mm}
    \frac{S_1^\mu}{m_1} &\sim \lambda^{-1}.
\end{align}
As discussed in greater detail in the following sub-section, the leading-order impulse corresponds to expanding the KMOC kernel to leading-order as $\lambda \rightarrow 0$. 

\subsection{Leading Absorptive Impulse}
\label{subsec:absampint}

To calculate the leading contribution to the absorptive impulse, we need the corresponding contribution to the one-loop amplitude, 
\begin{equation}\label{eq:absamplitude}
     \mathcal{A}_4^{(s_1, s_1', |h|)} = \int_{0}^{\infty} \text{d} x \,\rho(x) \int \hat{\text{d}}^{D} \ell  \hspace{5mm}\vcenter{\hbox{\scalebox{1.0}{\absorptive}}},
\end{equation}
where $\rho(x)$ is the spectral density function, $s_1, s_1'$ are the spin of the external and internal states, respectively, and $|h|$ is the helicity of the massless mediator. 
\begin{figure}
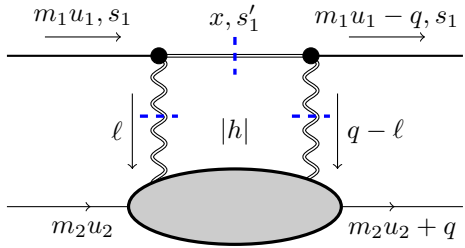

    \centering
    \absorptivecut
    \caption{Unitarity cut required to reconstruct the integrand for the leading-order absorptive impulse. Pinched mediator contributions are scaleless in the soft region of $\ell$-integration and pinched $X$-state contributions are scaleless in the soft region of $x$-integration corresponding to absorptive modes. }
    \label{fig:unitaritycut}
\end{figure}

We construct the integrand using generalized unitarity by sewing the cut depicted in Figure \ref{fig:unitaritycut}. The required input are the relevant mass/spin-changing 3-point amplitudes and, since body-2 is assumed to be non-spinning, the usual conservative scalar Compton amplitudes. On the cut $X$-state line we insert physical state projectors; for example, for $s_1'=1$ and $s_1'=2$ we use
\begin{align}
    \label{completeness1}
    \Pi^{\alpha_1}_{\,\,\beta_1}(\mu^2, k) &= \delta^{\alpha_1}_{\,\,\beta_1} - \frac{k^{\alpha_1} k_{\beta_1}}{\mu^2}, \\
    \label{completeness2}
    \Pi^{\alpha_1 \alpha_2}_{\,\,\beta_1 \beta_2}(\mu^2, k) &= \frac{1}{2} \Pi^{\alpha_1}_{\,\,\beta_1}(\mu^2, k) \Pi^{\alpha_2}_{\,\,\beta_2}(\mu^2, k) + \frac{1}{2} \Pi^{\alpha_1}_{\,\,\beta_2}(\mu^2, k) \Pi^{\alpha_2}_{\,\,\beta_1}(\mu^2, k) - \frac{1}{3} \Pi^{\alpha_1 \alpha_2}(\mu^2, k) \Pi_{\beta_1\beta_2}(\mu^2, k),
\end{align}
respectively.  For any choice of $(s_1,s_1',h)$ there are strictly finitely many independent mass/spin-changing 3-point amplitudes and therefore we can give a completely general parametrization; the explicit amplitudes used are given in Appendix \ref{app:3ptAmps}.\footnote{Of course the integrand can also be calculated using standard Feynman diagrammatics beginning with an off-shell effective action for the massive spinning states: Proca for $s=1$ \cite{Proca:1936fbw}, Fierz-Pauli for $s=2$ \cite{Fierz:1939ix}, and the higher-spin Singh-Hagen construction for $s>2$ \cite{Singh:1974qz}.}

Unlike the corresponding one-loop contributions to the conservative impulse \cite{Kosower:2018adc}, when calculating the leading-order absorptive impulse there are no super-classical contributions to the amplitude; the classical observable is given by the leading-order term in the soft expansion. As discussed in detail in \cite{Jones:2023ugm}, the \textit{real part} of the virtual kernel contribution vanishes after calculating $x$- and $\ell$-integrals due to the manifest $\ell \;\leftrightarrow \; q -\ell$ symmetry of the integrand
\begin{equation}
    \mathcal{I}_v^{\mu } \biggr\vert_{\ell \;\leftrightarrow \; q -\ell} = \mathcal{I}_v^{\mu }.
\end{equation}
The final observable therefore depends only on the \textit{imaginary part} of the amplitude (\ref{eq:absamplitude}). All together, our master formula for the calculation of the leading-order absorptive impulse simplifies to:
\begin{align}
\label{absimpulseDiagram}
     \left(\Delta p_1^\mu\right)_{|h|}^{s\rightarrow s'} = & \frac{1}{4 m_1 m_2} \int_{0}^{\infty} \text{d}x \, x^{1+\alpha}  \int \hat{\text{d}}^{D}q \, \hat{\delta}(u_1 \cdot q ) \, \hat{\delta}(u_2 \cdot q ) \, e^{-i b \cdot q} \nonumber\\ 
    & \times \int \hat{\text{d}}^{D}\ell \, \left(\frac{2 \ell^{\mu} - q^{\mu} }{2}\right) \, \,  \vcenter{\hbox{\scalebox{0.95}{\boxcut}}}.
\end{align}
After we expand in the soft limit, we make a vNV decomposition of tensor integrals by making the replacement
\begin{equation} 
\label{vecdecomp}
    \ell^{\mu} \rightarrow \frac{1}{2}q^{\mu} - x \Check{u}_1^{\mu} + \left(\frac{\tilde{n} \cdot \ell}{\tilde{n}^2}\right) \tilde{n}^{\mu} +\ell^\mu_{[-2\epsilon]},
\end{equation}
where
\begin{equation}
    \label{vectorbasis1}
    \Check{u}_{1}^{\mu} = \frac{u_{1}^{\mu} - y \,u_{2}^{\mu}}{1 - y^2},\hspace{10mm} \Check{u}_{2}^{\mu} = \frac{u_{2}^{\mu} - y \,u_{1}^{\mu}}{1 - y^2},\hspace{10mm} y = u_1 \cdot u_2, \hspace{10mm} \tilde{n}^{\mu} \equiv \varepsilon^{\mu u_1 u_2 q}.
\end{equation}
The master integrals we require (\ref{eq:masterint}) turn out to be finite in the cases of interest, so we can ignore the $[-2\epsilon]$-dimensional component of $\ell^\mu$ when making the reduction. In (\ref{vecdecomp}) we have used the replacements $u_1 \cdot \ell \rightarrow - x$ and $q \cdot \ell \rightarrow \frac{1}{2}q^2$; the corresponding expressions differ only by integrals that are scaleless in the soft region (\ref{soft}). 

Note that odd factors of $\tilde{n} \cdot \ell$ are odd under the $\ell \;\leftrightarrow \; q -\ell$ symmetry and therefore vanish upon integration; so we should only consider even powers of this term which are further decomposed by the replacement
\begin{equation}
    \frac{(\tilde{n}\cdot \ell)^2}{\tilde{n}^2} \rightarrow \frac{x^2}{y^2-1} - \frac{q^2}{4}.
\end{equation}
After tensor reduction, and dropping terms that vanish due to symmetry, the remaining integrals are of the form
\begin{equation}
    \label{eq:masterint}
    I_{k,n}[q^{\mu_1}...q^{\mu_r}] \equiv \int_0^\infty\mathrm{d}x\,x^k\int\hat{\mathrm{d}}^{4}q\, \hat{\delta}( u_2\cdot q)\hat{\delta}(u_1\cdot q)
    e^{i q\cdot b}q^{\mu_1}...q^{\mu_r}\left(-q^2\right)^{n}\int\hat{\mathrm{d}}^{4}\ell\frac{\hat{\delta}( u_1\cdot \ell +x)\hat{\delta}(u_2\cdot \ell)}{\ell^2(q-\ell)^2}.
\end{equation}
As shown in Appendix E of \cite{Jones:2023ugm}, the scalar integrals evaluate to
\begin{equation}
    I_{k,n}[1]=\frac{(-1)^n 4^{n-2} 
   \Gamma \left(\frac{k+1}{2}\right)^3 
   \left(\frac{k+1}{2}\right)_n^2}{\pi ^{3/2} \Gamma
   \left(\frac{k}{2}+1\right)}\frac{\left(y^2-1\right)^{\frac{k-1}{2}}}{(-b^2)^{\frac{2n+1+k}{2}}}.
\end{equation}
Tensor integrals can be generated by taking appropriate transverse $b$-derivatives 
\begin{equation}\label{eq:qvectobdivs}
    I_{k,n}[q^{\mu_1}...q^{\mu_r}] = \prod_{j=1}^r \left(-i\;\Xi^{\mu_j \nu_j} \frac{\partial}{\partial b^{\nu_j}}\right)I_{k,n}[1],
\end{equation}
where the projector onto the space transverse to $\{u_1^\mu,u_2^\mu\}$ is \cite{Maybee:2019jus}
\begin{equation}
    \Xi^{\mu \nu} \equiv \eta^{\mu\nu} - u_1^\mu \Check{u}_1^\nu - u_2^\mu \Check{u}_2^\nu.
\end{equation}
It is convenient to decompose the final impulse into scalar contributions 
\begin{equation}
\label{eq:impform}
     \left(\Delta p_1^{\mu}\right)^{s\rightarrow s'}_{|h|} = \left(u_1 \cdot \Delta p_1\right)^{s\rightarrow s'}_{|h|}\Check{u}_1^{\mu} + \frac{\left(b \cdot \Delta p_1\right)^{s\rightarrow s'}_{|h|}}{b^2} b^{\mu} + \frac{\left(n \cdot \Delta p_1\right)^{s\rightarrow s'}_{|h|}}{(y^2 - 1) \, b^2} n^{\mu},
\end{equation}
where 
\begin{equation}
    \label{vectorbasis2}
    n^\mu \equiv \varepsilon^{\mu u_1 u_2 b}.
\end{equation}
Since we are calculating the absorptive contribution at leading-order, there are no other forms of dissipation including mixed absorptive-radiative effects (diagrams including both excited $X$-states and radiation lines) or double-absorptive effects (diagrams including excited $X$-states for both body-1 and body-2). Hence to this order we can assume overall conservation of kinetic momentum $\Delta p_2^\mu = -\Delta p_1^\mu$ and mass-conservation for body-2, $\Delta m_2 =0$. Combined, it follows that $p_2\cdot \Delta p_1 =0$, and hence $\Delta p_1^\mu$ cannot have a non-zero component in the direction of the basis vector $\Check{u}_2^\mu$. 

As we will see, for leading-order absorption not all of these components are independent; the relations between them are discussed in greater detail in Section \ref{sec:angle}.

\section{Classical Spin from Finite Representations}
\label{sec:spinreview}
In this section, we will define the spin vector and describe the method for expressing quantum polarization tensors in terms of the classical spin vector. We will also describe our strategy for resolving Casimir ambiguities associated with finite-spin representations, which follows the method of \cite{Akpinar:2025bkt, Akpinar:2024meg}.

\subsection{Classical Spin Vector}
\label{subsec:spinvector}

As explained in detail in \cite{Maybee:2019jus}, in the large spin correspondence limit $s_1\gg \hbar$, the \textit{classical spin vector} is identified with the spin-coherent state expectation value of the Pauli-Lubanski operator
\begin{equation}
    \langle \mathbb{W}^\mu \rangle \mapsto m_1 S_1^\mu, \hspace{10mm} \mathbb{W}^\mu = \frac{1}{2} \varepsilon^{\mu\nu\rho\sigma} \mathbb{P}_{1\nu} \mathbb{M}_{\rho\sigma}.
\end{equation}
The spin tensor is then given by
\begin{equation}
    S^{\mu\nu}_1 = - \frac{1}{m_1} \varepsilon^{\mu \nu \rho \sigma} p_{1\, \rho} S_{1 \, \sigma},
\end{equation}
where in this section we use $p_1^\mu \equiv m_1 u_1^\mu$. This definition of the spin tensor automatically imposes the \textit{covariant} spin supplementary condition (SSC)\footnote{Several interesting recent works have discussed the possibility of relaxing the SSC, associating additional degrees of freedom with the spinning body \cite{Bern:2023ity, Alaverdian:2024spu, Alaverdian:2025jtw}.}
\begin{equation}
    \label{eq:SSC}
    p_{1\mu} S_1^{\mu\nu} = 0,
\end{equation}
In practice we first calculate the \textit{classical spin tensor} by the replacement prescription \cite{Maybee:2019jus,Bern:2020buy}
\begin{equation}
\label{spintensordef}
     \epsilon_{\mu_1 \cdots \mu_{s_1}}(p_1) \left[ \mathit{M}^{\mu \nu}\right]^{\mu_1 \cdots \mu_{s_1}}_{\,\,\,\,\nu_1 \cdots \nu_{s_1}} \epsilon^{\nu_1 \cdots \nu_{s_1}}(p_1) \mapsto S^{\mu \nu}_1,
\end{equation}
where 
\begin{equation}
    \left[ \mathit{M}^{\mu \nu}\right]^{\mu_1 \cdots \mu_{s}}_{\,\,\,\,\nu_1 \cdots \nu_{s}} =  2 i s \, \delta^{[\mu}_{(\nu_1} \eta^{\nu] (\mu_1} \delta^{\mu_2}_{\nu_2} \cdots \delta^{\mu_{s})}_{\nu_{s})},
\end{equation}
are the generators of the spin-$s$ representation of the Lorentz algebra
\begin{equation}
    \left[M^{\mu\nu},M^{\rho\sigma}\right] = i\left(\eta^{\nu\rho}M^{\mu\sigma}-\eta^{\mu\rho}M^{\nu\sigma}-\eta^{\nu\sigma}M^{\mu\rho}+\eta^{\mu\sigma}M^{\nu\rho}\right).
\end{equation}
Here we define
\begin{equation}
    \epsilon^{\nu_1 \cdots \nu_{s_1}}(p_1) \equiv \sum_{a_1=1}^{2s_1+1}\xi^{a_1} \epsilon^{\nu_1 \cdots \nu_{s_1}}(p_1,a_1),
\end{equation}
where $\xi_{a}$ is the little-group vector, introduced in Section \ref{subsec:absorption}, defining a projection onto the described spin-coherent state. For products of spin vectors, we first symmetrize over the pair of free indices of the Lorentz generator in (\ref{spintensordef}), the replacement prescription then generalizes as \cite{Bern:2020buy}
\begin{equation}
     \frac{1}{n!} \epsilon_{\alpha_1 \cdots \alpha_{s_1}}(p_1)\,\text{Sym} \left[\textit{M}^{\,\mu_1 \nu_1} \cdots \mathit{M}^{\mu_{n} \nu_{n}}\right]^{\alpha_1 \cdots \alpha_{s_1}}_{\,\,\,\, \beta_1 \cdots \beta_{s_1}} \, \epsilon^{\beta_1 \cdots \beta_{s_1}}(p_1) \mapsto S^{\mu_1 \nu_1}_1 \cdots S_1^{\mu_{n} \nu_{n}}.
\end{equation}
We define the $\text{Sym}[\cdots]$ operator in the following way
\begin{equation}
    \text{Sym}\left[\textit{M}^{\mu_1 \nu_1} \cdots \mathit{M}^{\mu_{n} \nu_{n}}\right]^{\alpha_1 \cdots \alpha_{s_1}}_{\,\,\,\, \beta_1 \cdots \beta_{s_1}} \equiv \left[\sum_{\sigma \in S_{n}} \prod_{i = 1}^{n} \mathit{M}^{\mu_{\sigma(i)}\nu_{\sigma(i)}}\right]^{\alpha_1 \cdots \alpha_{s_1}}_{\,\,\,\, \beta_1 \cdots \beta_{s_1}},
\end{equation}
where $S_n$ is the symmetric group of $n$ elements. We provide a couple of examples
\begin{align}
    & \text{Sym}[\mathit{M}^{\mu_1 \nu_1} \mathit{M}^{\mu_2 \nu_2}]^{\alpha_1 \cdots \alpha_{s_1}}_{\,\,\,\, \beta_1 \cdots \beta_{s_1}} = [\mathit{M}^{\mu_1 \nu_1} \mathit{M}^{\mu_2 \nu_2}]^{\alpha_1 \cdots \alpha_{s_1}}_{\,\,\,\, \beta_1 \cdots \beta_{s_1}} + [\mathit{M}^{\mu_2 \nu_2} \mathit{M}^{\mu_1 \nu_1}]^{\alpha_1 \cdots \alpha_{s_1}}_{\,\,\,\, \beta_1 \cdots \beta_{s_1}}, \\
    &\nonumber \\
    & \text{Sym}[\mathit{M}^{\mu_1 \nu_1} \mathit{M}^{\mu_2 \nu_2} \mathit{M}^{\mu_3 \nu_3}]^{\alpha_1 \cdots \alpha_{s_1}}_{\,\,\,\, \beta_1 \cdots \beta_{s_1}} = [\mathit{M}^{\mu_1 \nu_1} \mathit{M}^{\mu_2 \nu_2}\mathit{M}^{\mu_3 \nu_3}]^{\alpha_1 \cdots \alpha_{s_1}}_{\,\,\,\, \beta_1 \cdots \beta_{s_1}} + [\mathit{M}^{\mu_2 \nu_2} \mathit{M}^{\mu_1 \nu_1} \mathit{M}^{\mu_3 \nu_3}]^{\alpha_1 \cdots \alpha_{s_1}}_{\,\,\,\, \beta_1 \cdots \beta_{s_1}} \nonumber \\
    & \hspace{ 4cm}+ [\mathit{M}^{\mu_3 \nu_3} \mathit{M}^{\mu_2 \nu_2}\mathit{M}^{\mu_1 \nu_1}]^{\alpha_1 \cdots \alpha_{s_1}}_{\,\,\,\, \beta_1 \cdots \beta_{s_1}} + [\mathit{M}^{\mu_1 \nu_1} \mathit{M}^{\mu_3 \nu_3} \mathit{M}^{\mu_2 \nu_2}]^{\alpha_1 \cdots \alpha_{s_1}}_{\,\,\,\, \beta_1 \cdots \beta_{s_1}} \nonumber \\
    & \hspace{ 4cm}+ [\mathit{M}^{\mu_2 \nu_2} \mathit{M}^{\mu_3 \nu_3}\mathit{M}^{\mu_1 \nu_1}]^{\alpha_1 \cdots \alpha_{s_1}}_{\,\,\,\, \beta_1 \cdots \beta_{s_1}} + [\mathit{M}^{\mu_3 \nu_3} \mathit{M}^{\mu_1 \nu_1} \mathit{M}^{\mu_2 \nu_2}]^{\alpha_1 \cdots \alpha_{s_1}}_{\,\,\,\, \beta_1 \cdots \beta_{s_1}},
\end{align}
where ordering matters because these are matrix valued elements. We explain our notation for the product of matrices
\begin{equation}
    \left[A_1 A_2 \cdots A_{n-1} A_n \right]^{\alpha_1 \cdots \alpha_s}_{\,\,\,\, \beta_1 \cdots \beta_s} \equiv \left[A_1\right]^{\alpha_1 \cdots \alpha_s}_{\,\,\,\,\gamma_1 \cdots \gamma_s} \left[A_2\right]^{\gamma_1 \cdots \gamma_s}_{\,\,\,\, \delta_1 \cdots \delta_s} \cdots \left[A_{n-1}\right]^{\eta_1 \cdots \eta_s}_{\,\,\,\,\zeta_1 \cdots \zeta_s} \left[A_{n}\right]^{\zeta_1 \cdots \zeta_s}_{\,\,\,\,\beta_1 \cdots \beta_s},
\end{equation}
for some matrix $A_i$. With the spin variable well defined, we can proceed with obtaining spin information from our amplitudes. 

In general, polarization tensors appear in the integrand (\ref{absimpulseDiagram}) in the form 
\begin{equation}
    \epsilon^{*}_{\mu_1 \cdots \mu_{s_1}}(p_1 - q) \mathcal{N}^{\mu_1 \cdots \mu_{s_1}}_{\nu_1 \cdots \nu_{s_1}}\left(u_1,u_2,q,l\right) \epsilon^{\nu_1\cdots \nu_{s_1}}(p_1),
\end{equation}
where $\mathcal{N}$ is some tensor numerator constructed from the available Lorentz covariant objects. The first step to rewriting this in terms of classical spin vectors is to expand  $\epsilon^{*}_{\mu_1 \cdots \mu_{s_1}}(p_1 - q)$ in the soft limit $q\ll p_1$. The procedure for the perturbative expansion of the polarization tensors can be found in \cite{Bern:2023ity, Gatica:2023iws, Chung:2018kqs}. In our work, we choose the reference momentum\footnote{For spinning fields, the polarization tensor $\epsilon(p',\sigma)$ is defined by applying a \textit{standard Lorentz boost}, $p'= U(p';p)\cdot p$, to a basis of fiducial polarization tensors $\epsilon(p,\sigma)$ defined with respect to an arbitrary reference momentum $p$: $\epsilon(p',\sigma) \equiv U(p';p)\cdot \epsilon(p,\sigma)$ \cite{Weinberg:1995mt}. The choice of reference momentum represents an arbitrary \textit{scheme dependence} of spinning observables. When comparing results between references, care must be taken to ensure the results are calculated in a compatible scheme. For a detailed discussion there are numerous sources \cite{Bern:2023ity, Gatica:2023iws, Chung:2018kqs}.}
to be $p_1^\mu$, which allows us to use the following resummed expression \cite{Gatica:2023iws,Cangemi:2023ysz}
\begin{equation}\label{polexpansionresum}
    \epsilon^{*}_{\alpha(s_1)}(p_1 - q)\,\epsilon_{\gamma_{(s_1)}}(p_1 ) =  \text{exp}\left[\frac{i q_{\mu}p_{1\,\nu}}{m_1^2}  \frac{\text{arcsin}\left(\sqrt{\frac{-q^2}{4 m_1^2}}\right)}{\sqrt{\frac{-q^2}{4 m_1^2}}\sqrt{ 1 + \frac{q^2}{4 m_1^2}}}\mathit{M}^{\mu \nu}\right]^{\,\,\,\,\,\,\beta(s_1)}_{\alpha(s_1)} \epsilon^{*}_{\beta(s_1)}(p_1) \, \epsilon_{\gamma_{(s_1)}}(p_1 ) .
\end{equation}
After expanding  (\ref{polexpansionresum}), we need to replace the product of polarization tensors with spin vectors. To do this, we construct a system of equations relating the polarization tensors to different powers of spin vectors. We start by taking the unordered outer product of $2 s_1$ Lorentz generators with the polarization tensors, and plugging in for the spin-$s_1$ representation. We then construct a system of equations by successively symmetrizing and anti-symmetrizing the generators and identifying these combinations with the appropriate spin vector structure. For example, for a spin-$1$ representation we start with
\begin{align}
   &\epsilon^{*\,\mu_2}(p_1) \epsilon^{\mu_1}(p_1) -\Pi^{\mu_1 \mu_2}(m_1^2, p_1)\nonumber \\
   &= \left(\frac{-1}{2 m_1}\right)^2 \varepsilon^{\mu_1 \nu_1 \rho_1 \sigma_1} \varepsilon^{\mu_2 \nu_2 \rho_2 \sigma_2} p_{1 \, \nu_1}p_{1 \, \nu_2}   \epsilon^{*}_{\alpha_1}(p_1) \left(\mathit{M}_{\rho_1 \sigma_1}\right)^{\alpha_1}_{\,\,\,\, \beta} \left(\mathit{M}_{\rho_2 \sigma_2}\right)^{\beta}_{\,\,\,\, \alpha_2}\epsilon^{\alpha_2}(p_1).
\end{align}
Then we symmetrize and anti-symmetrize over the $\mu_1, \mu_2$ indices to get a system of equations
\begin{align}
    \epsilon^{*\,\mu_2}(p_1) \epsilon^{\mu_1}(p_1) +  \epsilon^{*\,\mu_1}(p_1) \epsilon^{\mu_2}(p_1)  &\mapsto 2S^{\mu_1}_1 S^{\mu_2}_1 +2\Pi^{\mu_1 \mu_2}(m_1^2, p_1), \\
     \epsilon^{*\,\mu_2}(p_1) \epsilon^{\mu_1}(p_1) - \epsilon^{*\,\mu_1}(p_1) \epsilon^{\mu_2}(p_1) &\mapsto    -\frac{i}{m_1} \varepsilon^{\mu_1 \mu_2 p_1 S_1}.
\end{align}
Adding these together gives our desired replacement rule
\begin{equation}
\label{spin1reppoltovec}
    \epsilon^{\mu_1}(p_1) \epsilon^{*\,\mu_2}(p_1) \mapsto S^{\mu_1}_1 S^{\mu_2}_1 + \Pi^{\mu_1 \mu_2}(m_1^2, p_1)-\frac{i}{2 m_1} \varepsilon^{\mu_1 \mu_2 p_1
    S_1} .
\end{equation}
With  (\ref{polexpansionresum}) and (\ref{spin1reppoltovec}), we can reliably expand the universal polarization product in our amplitudes in the soft limit to any desired order for the spin-1 representation. 

Following the same procedure for the spin-2 representation, we find
\begin{align}\label{eq:spin2epstospinvec}
    \epsilon^{*\,\mu_1 \mu_2}(p_1)\,\epsilon^{\mu_3 \mu_4}(p_1) \mapsto &\, \frac{2}{3} \Pi^{\mu_1 \mu_2}\Pi^{\mu_3 \mu_4}  + \frac{1}{6} S^{\mu_1}_1 S^{\mu_2}_1 S^{\mu_3}_1 S^{\mu_4}_1 \nonumber \\
     &+ \frac{7 i }{72 m_1} \left[\Pi^{\mu_1 \mu_3} \varepsilon^{\mu_2 \mu_4 p_1 S_1} +  \Pi^{\mu_1 \mu_4} \varepsilon^{\mu_2 \mu_3 p_1 S_1} \right. \nonumber \\
     & \left. \hspace{2cm}+ \Pi^{\mu_2 \mu_3} \varepsilon^{\mu_1 \mu_4 p_1 S_1}+ \Pi^{\mu_2 \mu_4} \varepsilon^{\mu_1 \mu_3 p_1 S_1}\right] \nonumber \\
     &-\frac{3}{36} S_1^2 \left[\Pi^{\mu_1 \mu_4} \Pi^{\mu_2 \mu_3} + \Pi^{\mu_1 \mu_3} \Pi^{\mu_2 \mu_4} - 2 \,\Pi^{\mu_1 \mu_2} \Pi^{\mu_3 \mu_4}\right] \nonumber \\
     &+ \frac{1}{36} \left[\Pi^{\mu_1 \mu_2} S^{\mu_3}_1 S^{\mu_4}_1 + \Pi^{\mu_3 \mu_4} S^{\mu_1}_1 S^{\mu_2}_1 + 7 \, \Pi^{\mu_1 \mu_3} S^{\mu_2}_1 S^{\mu_4}_1 \right. \nonumber \\
      &\left. \hspace{2cm}+  7 \, \Pi^{\mu_1 \mu_4} S^{\mu_2}_1 S^{\mu_3}_1+ 7 \, \Pi^{\mu_2 \mu_3} S^{\mu_1}_2 S^{\mu_4}_1+ 7 \, \Pi^{\mu_2 \mu_4} S^{\mu_1}_1 S^{\mu_3}_1\right] \nonumber \\
       &+ \frac{i }{24 m_1} \left[S^{\mu_1}_1 S^{\mu_3}_1 \varepsilon^{\mu_2 \mu_4 p_1 S_1} +  S^{\mu_1}_1 S^{\mu_4}_1 \varepsilon^{\mu_2 \mu_3 p_1 S_1} \right. \nonumber \\
       &\left. \hspace{2cm}+ S^{\mu_2}_1 S^{\mu_3}_1 \varepsilon^{\mu_1 \mu_4 p_1 S_1}+ S^{\mu_2}_1 S^{\mu_4}_1 \varepsilon^{\mu_1 \mu_3 p_1 S_1}\right], 
\end{align}
where now we can calculate up to $\mathcal{O}\left(S^{4}_1\right)$. We also suppress the arguments of the projectors, such that $\Pi^{\alpha \beta} \equiv \Pi^{\alpha \beta}(m_1^2, p_1)$. 

At this point, we are able to take the universal product of polarization tensors, $\epsilon^{\mu_1 \cdots \mu_{s_1}}(p_1 - q)\epsilon^{\nu_1 \cdots \nu_{s_1}}(p_1)$, and translate it to a function of spin vectors and momenta. If we were only interested in leading order contributions, we would expand (\ref{polexpansionresum}) to $\mathcal{O}(q^{2 s_1})$; this is because for a spin-$s_1$ representation we can access spin vector contributions up to $\mathcal{O}(S_1^{2s_1})$. Therefore, in the soft limit we would be expanding up to $\mathcal{O}(\lambda^{0})$. However, as well will see in the next section, we will need to expand to higher contributions in the soft limit to resolve ambiguities associated with the spin Casimir.  

\subsection{Spin Universality and Interpolation}
\label{subsec:spinuniandInterpolation}

As described, the general form of the leading-order absorptive impulse is (\ref{eq:impform}); we can generically parametrize the coefficients of the components as an expansion in the available spin structures\footnote{In the next section, we will comment on the structure of (\ref{eq:impform}); specifically, the additional components when compared to the non-spinning case \cite{Jones:2023ugm}.}, 
\begin{align}
\label{eq:cdef}
    \frac{\left(v \cdot \Delta p_1\right)^{s\rightarrow s'}_{|h|}}{v^2} &=   \frac{\sqrt{y^2 - 1}}{ \sqrt{-b^2}} \sum_{j_1,j_2,j_3} (c_{v})^{(s,s',|h|)}_{j_1, j_2, j_3} \left(b \cdot S_1\right)^{j_1} \left(n \cdot S_1\right)^{j_2} \left(\sqrt{-b^2} \, u_2 \cdot S_1\right)^{j_3},
\end{align}
where $v \in \left\{u_1,b,n\right\}$. For each scalar component, at $\mathcal{O}\left(S_1^k\right)$, the total number of $(c_i)_{j_1,j_2,j_3}^{(s,s',|h|)}$ coefficients (with $j_1+j_2+j_3=k$) is equal to the $k$-th triangular number: $\frac{1}{2}\left(k+2\right)\left(k+1\right)$. 

Following the method described in the previous subsection, the $(c_i)_{j_1,j_2,j_3}^{(s,s',|h|)}$ coefficients, for $j_1+j_2+j_3\leq 2s_1$, are determined from a calculation with a finite spin-$s_1$ external state. This means the coefficients up to $\mathcal{O}\left(S^2\right)$ are determined by an external spin-1 calculation, up to $\mathcal{O}\left(S^4\right)$ are determined by an external spin-2 calculation and so on. A basic requirement is then that this procedure gives self-consistent results as the external spin is increased. For example, the coefficients up to $\mathcal{O}\left(S^2\right)$ for the spin-1 and spin-2 calculations should agree, this is the property of \textit{spin universality}. Since each finite spin calculation is given in terms of a different set of Wilson coefficients (the coefficients parameterizing the structures of $\mathcal{A}_3^{(s_1,s_2,h_3)}$), agreement here means that there should exist a \textit{bijection} between Wilson coefficients. We denote equality up to such a bijective map by the symbol $\cong$. The existence of such a map is extremely over-constrained, in general only a very small number of Wilson coefficients will appear in a given calculation (usually $\leq 4$), but this map must ensure equality between every possible spin structure up to $\mathcal{O}\left(S^{2s_1}\right)$ for all possible values of the relative velocity $y$.

When calculating classical observables with a fixed spin representation, special care must be taken for the Casimir operator
\begin{equation}\label{casimirdef}
    S^{\mu}_{1} S_{1 \, \mu} \equiv S_1^2 = - s_1 (s_1 + 1).
\end{equation}
In a finite spin calculation, this combination is a finite number; this creates an \textit{ambiguity} in the definition of the map ($\mapsto$) described in Section \ref{subsec:spinvector} to promote matrix elements of Lorentz generators $M^{\mu\nu}$ to products of classical spin tensors. Moreover, this ambiguity leads to a complicated mixing of orders in the soft expansion; as seen in (\ref{casimirdef}), where the left-hand side scales as $\mathcal{O}(\lambda^{-2})$ while the right-hand side is $\mathcal{O}(\lambda^{0})$. Equivalently, terms that are naively $\mathcal{O}\left(\lambda^2\right)$ suppressed can be \textit{promoted} by multiplying them by factors of $S^2_1$. A systematic solution to the problem was described in \cite{Akpinar:2025bkt, Akpinar:2024meg} referred to there as \textit{spin interpolation}.\footnote{See also \cite{Cangemi:2022abk} for a closely related systematic interpolation method for finite spin observables.}

The Casimir ambiguity can be generically parametrized by introducing \textit{ambiguity parameters} $(\beta_{v})^{(s,s',|h|)}_{j_1, j_2, j_3}$ as:
\begin{align}
    \frac{\left(v \cdot \Delta p_1\right)^{s\rightarrow s'}_{|h|}}{v^2} \rightarrow &\;\frac{\left(v \cdot \Delta p_1\right)^{s\rightarrow s'}_{|h|}}{v^2} \\
    &+   S_1^2\left[\frac{\sqrt{y^2 - 1}}{ \sqrt{-b^2}}  \sum_{j_1,j_2,j_3} (\beta_{v})^{(s,s',|h|)}_{j_1, j_2, j_3} \left(b \cdot S_1\right)^{j_1} \left(n \cdot S_1\right)^{j_2} \left(\sqrt{-b^2} \, u_2 \cdot S_1\right)^{j_3}\right].\nonumber
\end{align}
At $\mathcal{O}(S_1^k)$, for $k\geq 2$ there are $\frac{1}{2}k\left(k-1\right)$ such parameters. Expanding the spin vector into our chosen basis
\begin{equation}
\label{eq:spinvectobasis}
    S^{\mu}_1 = \left(\frac{b \cdot S_1}{b^2}\right) b^{\mu} + \left(\frac{n \cdot S_1 }{b^2(y^2-1)}\right)n^{\mu} + (u_2 \cdot S_1) \Check{u}_2^{\mu},
\end{equation}
the Casimir takes the form
\begin{equation}
    \label{eq:spinsq}
    S_1^2 = \frac{(b \cdot S_1)^2}{b^2}+ \frac{(n \cdot S_1)^2}{b^2 \,(y^2 - 1)} + \frac{\left(\sqrt{-b^2} \, u_2 \cdot S_1\right)^2}{b^2(y^2-1)}.
\end{equation}
Introducing the Casimir ambiguity parameters is then seen to be equivalent to a linear shift of the $(c_{v})^{(s,s',|h|)}_{j_1, j_2, j_3}$ coefficients in (\ref{eq:cdef}). For example, at $\mathcal{O}(S_1^2)$ the shift takes the explicit form
\begin{flalign}
   &(c_{v})_{2, 0, 0}^{(s, \,s',\,|h|)} \rightarrow (c_{v})_{2, 0, 0}^{(s, \,s',\,|h|)}+ \frac{(\beta_v)_{0, 0, 0}^{(s, s', |h|)}}{b^2}, \\ \
   &(c_{v})_{0, 2, 0}^{(s, \,s',\,|h|)} \rightarrow (c_{v})_{0, 2, 0}^{(s, \,s',\,|h|)} + \frac{(\beta_v)_{0, 0, 0}^{(s, s', |h|)}}{b^2 (y^2 - 1)}, \\ 
   &(c_{v})_{0, 0, 2}^{(s, \,s',\,|h|)} \rightarrow (c_{v})_{0, 0, 2}^{(s, \,s',\,|h|)}  + \frac{(\beta_v)_{0, 0, 0}^{(s, s', |h|)}}{b^2(y^2-1)},
\end{flalign}
the ``off-diagonal" coefficients $(c_v)^{(s,s',|h|)}_{0,1,1}$, $(c_v)^{(s,s',|h|)}_{1,0,1}$, and $(c_v)^{(s,s',|h|)}_{1,1,0}$ are unaffected by the Casimir ambiguity and do not shift. Since at this order there are 6 spin structures in total ($c$-coefficients) and a 1-parameter ambiguity ($\beta$-coefficients) we expect that there should be 5 linear combinations of coefficients that are unaffected by the shift. That is, we expect that there are 5 linear combinations that can be unambiguously reconstructed \textit{without} Casimir interpolation. Trivially, 3 of these correspond to the ``off-diagonal" coefficients, the remaining 2 non-trivial combinations are easily found
\begin{align}
    \label{eq:dcoeffS2}
    (d_v)^{(s, s', h)}_{2, 1} &\equiv (c_v)^{(s,s',|h|)}_{0,0,2}-\frac{(c_v)^{(s,s',|h|)}_{2,0,0}}{y^2-1}, \\
    (d_v)^{(s, s', h)}_{2, 2} &\equiv (c_v)^{(s,s',|h|)}_{0,2,0}-\frac{(c_v)^{(s,s',|h|)}_{2,0,0}}{y^2-1}.
\end{align}
More generally, at $\mathcal{O}(S_1^k)$ there are $2k+1$ such Casimir independent (ambiguity free) combinations of coefficients. 

The interpolation procedure described in \cite{Akpinar:2025bkt, Akpinar:2024meg}, is a means of fixing the ambiguity parameters by imposing universality at \textit{subleading order} in the soft expansion. A simple, illustrative example is to consider the calculation of the $1\rightarrow 2$ transition for scalar absorption; this has the schematic form 
\begin{align}
    \left(\Delta p_1^\mu\right)_{0}^{1\rightarrow 2} &\sim \underbrace{\frac{S_1^2}{|b|^5}}_{\lambda^3} + \underbrace{\frac{S_1^1}{|b|^5}+\frac{S_1^2}{|b|^6}}_{\lambda^4}+ \underbrace{\frac{S_1^0}{|b|^5}+\frac{S_1^1}{|b|^6}+\frac{S_1^2}{|b|^7}}_{\lambda^5} +...
\end{align}
For an external spin-1 particle the spin Casimir takes the value $S_1^2 = -2$. Introducing the Casimir ambiguity parameter $\beta$ is a kind creative way of adding zero
\begin{equation}
\label{interpolationex}
    \left(\Delta p_1^\mu\right)_{0}^{1\rightarrow 2}\biggr\vert_{S_1^2 b^{-5}} = \left(\Delta p_1^\mu\right)_{0}^{1\rightarrow 2}\biggr\vert_{S_1^2 b^{-5}} + \beta S_1^2 \left(\left(\Delta p_1^\mu\right)_{0}^{1\rightarrow 2}\biggr\vert_{S_1^0 b^{-5}}\right) + (1+2\beta) \left(\left(\Delta p_1^\mu\right)_{0}^{1\rightarrow 2}\biggr\vert_{S_1^0 b^{-5}}\right)+...
\end{equation}
We then compare to a calculation for the $0\rightarrow 1$ transition; as discussed further in Section \ref{sec:spinsuppression} this has the leading-order scaling $\left(\Delta p_1^\mu\right)_{0}^{0\rightarrow 1} \sim S_1^0 b^{-5} \sim \lambda^{-5}$. Imposing universality, that these two calculations agree at $\mathcal{O}\left(S_1^0\right)$ but at \textit{all orders} in $\lambda$, gives an additional matching condition
\begin{equation}
    \left(\Delta p_1^\mu\right)_{0}^{0\rightarrow 1} \biggr\vert_{S_1^0 b^{-5}} \cong (1+2\beta) \left(\left(\Delta p_1^\mu\right)_{0}^{1\rightarrow 2}\biggr\vert_{S_1^0 b^{-5}}\right),
\end{equation}
that can be used to determine $\beta$. Inserting this value into (\ref{interpolationex}) gives the final \textit{interpolated} observable. In practice the interpolation step can be carried out at the level of the loop integrand. We will use the following decomposition of the amplitude
\begin{equation}
    \mathcal{A}_4^{(s, s', h)} = \frac{1}{4 m_1 m_2}\int \hat{\text{d}}^{4}\ell \,\hat{\delta}(u_2 \cdot \ell) \frac{\mathcal{N}^{(s, s', h)}}{\ell^2 \, (q-l)^2[u_1 \cdot \ell + x-i0]}.
\end{equation}
The relevant parts of the numerator of the $1\rightarrow 2$ calculation has the form
\begin{align}
    \tilde{\mathcal{N}}^{(1, 2, 0)}\vert_{ S_1^2 q^2} &= - \frac{g^2 } {6}\left(g_{1}^{(1, 2, 0)}\right)^2 \left(\ell \cdot S_1\right) \left((q-\ell)\cdot S_1\right), \\
    \tilde{\mathcal{N}}^{(1, 2, 0)}\vert_{ S_1^0 q^2} &= -\frac{g^2 } {3}\left(g_{1}^{(1, 2, 0)}\right)^2 \left(2x^2+q^2\right),
\end{align}
where $\tilde{\mathcal{N}}$ denotes that these are the expressions before interpolation. Similarly the contribution to the $0\rightarrow 1$ process is calculated to be
\begin{equation}
    \mathcal{N}^{(0, 1, 0)}\vert_{ S_1^0 q^2} = \frac{g^2 \left(g_{1}^{(0, 1, 0)}\right)^2}{2} \left(2 x^2 + q^2\right).
\end{equation}
Matching the $\mathcal{O}\left(q^2 S_1^0\right)$ contributions of $1\rightarrow 2$ and $0\rightarrow 1$ determines the Casimir promotion. The interpolated numerator is given by 
\begin{align}
    \mathcal{N}^{(1, 2, 0)}\vert_{ S_1^2 q^2} &= \tilde{\mathcal{N}}^{(1, 2, 0)}\vert_{q^2 S_1^2} + S_1^2 \left(f^{(1, 2, 0)}_1 x^2 + f^{(1, 2, 0)}_2 q^2\right), \nonumber\\
    \mathcal{N}^{(1, 2, 0)}\vert_{ S_1^0 q^2} &= \frac{g^2 \left(g_{1}^{(0, 1, 0)}\right)^2}{2} \left(2 x^2 + q^2\right),
\end{align}
where
\begin{equation}
    f^{(1, 2, 0)}_1 = - \frac{g^2}{6} \left(3 \left(g_1^{(0, 1, 0)}\right)^2 + 2\left(g_1^{(1, 2, 0)}\right)^2\right), \hspace{0.3cm} f^{(1, 2, 0)}_2 = - \frac{g^2}{12} \left(3 \left(g_1^{(0, 1, 0)}\right)^2 + 2\left(g_1^{(1, 2, 0)}\right)^2\right).
\end{equation}
At $\mathcal{O}(S_1^2)$ we have carried out the Casimir interpolation procedure described above, determining the value of the ambiguity parameters $(\beta_{v})^{(s,s',|h|)}_{0,0,0}$, the results are given in Appendix \ref{app:results}. At higher-order in the spin expansion, the interpolation procedure becomes more intricate, involving promotion from several sub-leading orders in the soft expansion, we leave this to future work. 

On the basis of the above discussion however, we can still extract \textit{physically meaningful} information from un-interpolated results by studying the Casimir independent combinations of coefficients. Results calculated with external states up to spin-2, up to $\mathcal{O}(S_1^4)$ in the spin expansion, are reported in an ancillary file. The explicit definitions of the Casimir independent combinations we use are recorded below. At $\mathcal{O}(S_1^3)$ the following 6 non-trivial combinations of coefficients are Casimir independent
{\allowdisplaybreaks\begin{flalign}
    \label{eq:dcoeffS3}
    (d_v)^{(s, s', h)}_{3, 1} &\equiv (c_v)^{(s,s',|h|)}_{0,0,3}-\frac{(c_v)^{(s,s',|h|)}_{2,0,1}}{y^2-1}, \\
    (d_v)^{(s, s', h)}_{3, 2} &\equiv (c_v)^{(s,s',|h|)}_{0,1,2}-\frac{(c_v)^{(s,s',|h|)}_{2,1,0}}{y^2-1}, \\
    (d_v)^{(s, s', h)}_{3, 3} &\equiv (c_v)^{(s,s',|h|)}_{0,2,1}-\frac{(c_v)^{(s,s',|h|)}_{2,0,1}}{y^2-1}, \\
    (d_v)^{(s, s', h)}_{3, 4} &\equiv (c_v)^{(s,s',|h|)}_{0,3,0}-\frac{(c_v)^{(s,s',|h|)}_{2,1,0}}{y^2-1}, \\
    (d_v)^{(s, s', h)}_{3, 5} &\equiv (c_v)^{(s,s',|h|)}_{1,0,2}-\frac{(c_v)^{(s,s',|h|)}_{3,0,0}}{y^2-1}, \\
    (d_v)^{(s, s', h)}_{3, 6} &\equiv (c_v)^{(s,s',|h|)}_{1,2,0}-\frac{(c_v)^{(s,s',|h|)}_{3,0,0}}{y^2-1}, 
\end{flalign}}
and the coefficient $(c_v)^{(s,s',|h|)}_{1,1,1}$ is individually Casimir independent. At $\mathcal{O}(S_1^4)$ the following 9 non-trivial combinations of coefficients are Casimir independent
{\allowdisplaybreaks\begin{flalign}
    \label{eq:dcoeffS4}
    (d_v)^{(s, s', h)}_{4, 1} &\equiv (c_v)^{(s,s',|h|)}_{0,0,4}-\frac{(c_v)^{(s,s',|h|)}_{2,0,2}}{y^2-1}+\frac{(c_v)^{(s,s',|h|)}_{4,0,0}}{\left(y^2-1\right)^2}, \\
    (d_v)^{(s, s', h)}_{4, 2} &\equiv (c_v)^{(s,s',|h|)}_{0,1,3}-\frac{(c_v)^{(s,s',|h|)}_{2,1,1}}{y^2-1}, \\
    (d_v)^{(s, s', h)}_{4, 3} &\equiv (c_v)^{(s,s',|h|)}_{0,2,2}-(c_v)^{(s,s',|h|)}_{0,4,0}-\frac{(c_v)^{(s,s',|h|)}_{2,0,2}}{y^2-1}+\frac{(c_v)^{(s,s',|h|)}_{4,0,0}}{\left(y^2-1\right)^2}, \\
    (d_v)^{(s, s', h)}_{4, 4} &\equiv (c_v)^{(s,s',|h|)}_{0,3,1}-\frac{(c_v)^{(s,s',|h|)}_{2,1,1}}{y^2-1}, \\
    (d_v)^{(s, s', h)}_{4, 5} &\equiv (c_v)^{(s,s',|h|)}_{1,0,3}-\frac{(c_v)^{(s,s',|h|)}_{3,0,1}}{y^2-1}, \\
    (d_v)^{(s, s', h)}_{4, 6} &\equiv (c_v)^{(s,s',|h|)}_{1,1,2}-\frac{(c_v)^{(s,s',|h|)}_{3,1,0}}{y^2-1}, \\
    (d_v)^{(s, s', h)}_{4, 7} &\equiv (c_v)^{(s,s',|h|)}_{1,2,1}-\frac{(c_v)^{(s,s',|h|)}_{3,0,1}}{y^2-1}, \\
    (d_v)^{(s, s', h)}_{4, 8} &\equiv (c_v)^{(s,s',|h|)}_{1,3,0}-\frac{(c_v)^{(s,s',|h|)}_{3,1,0}}{y^2-1}, \\
    (d_v)^{(s, s', h)}_{4, 9} &\equiv (c_v)^{(s,s',|h|)}_{0,4,0}-\frac{(c_v)^{(s,s',|h|)}_{2,2,0}}{y^2-1}+\frac{(c_v)^{(s,s',|h|)}_{4,0,0}}{(y^2-1)^2}.
\end{flalign}}
At $\mathcal{O}\left(S_1^5\right)$ there are 11 Casimir independent combinations, at $\mathcal{O}\left(S_1^6\right)$ there are 13 and so on. 

Using these combinations we are able to explicitly verify the emergence of spin universality at $\mathcal{O}(S_1^2)$ by comparing the $d$-coefficients corresponding to the transitions $1\rightarrow 2$ and $2\rightarrow 3$ for $h=0$ and $1$\footnote{As explained in greater detail in Section \ref{sec:spinsuppression}, for $\Delta s=1$ transitions due to graviton absorption, $h=2$, we only expect universality for external spins $s\geq 2$. Therefore the first non-trivial test of universality would be to compare $2\rightarrow 3$ with $3\rightarrow 4$, we leave this calculation for future work.}. 
\newpage

\section{Results}
\label{sec:spin}

\begin{figure}[t!]
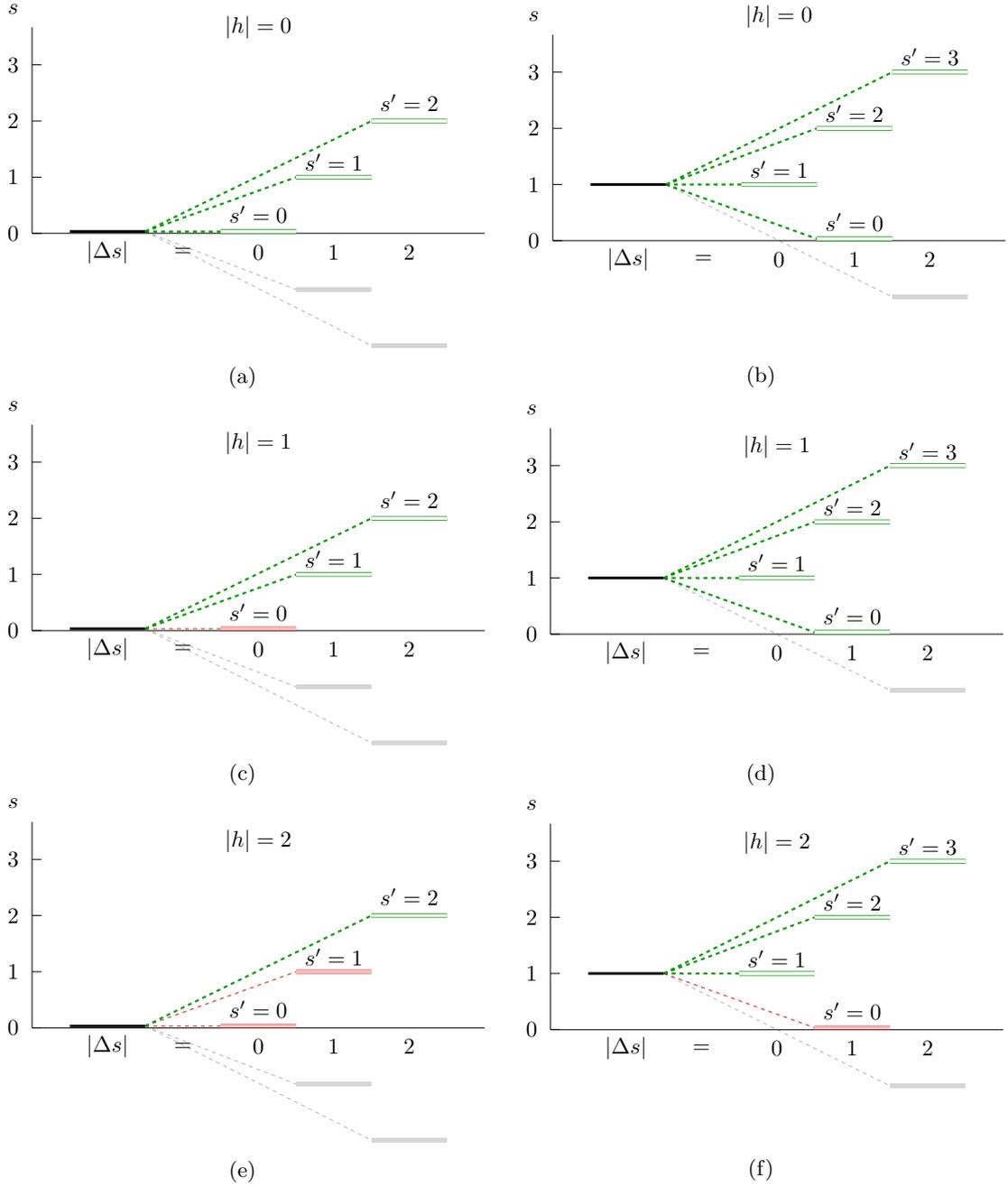

    \centering
    \begin{subfigure}[c]{0.5\textwidth}
        \centering
        \scalebox{0.55}{\spectrumSOHOgraph}
        \caption{}
        \label{fig:spectragraphS0H0}
    \end{subfigure}%
    \begin{subfigure}[c]{0.5\textwidth}
        \centering
        \raisebox{20pt}{\scalebox{0.55}{\spectrumSIHOgraph}}
        \caption{}
        \label{fig:spectragraphS1H0}
    \end{subfigure}
    \begin{subfigure}{0.5\textwidth}
        \centering
        \scalebox{0.55}{\spectrumSOHIgraph}
        \caption{}
        \label{fig:spectragraphS0H1}
    \end{subfigure}%
    \begin{subfigure}{0.5\textwidth}
        \centering
        \raisebox{22pt}{\scalebox{0.55}{\spectrumSIHIgraph}}
        \caption{}
        \label{fig:spectragraphS1H1}
    \end{subfigure}
    \begin{subfigure}[c]{0.5\textwidth}
        \centering
        \scalebox{0.55}{\spectrumSOHIIgraph}
        \caption{}
        \label{fig:spectragraphS0H2}
    \end{subfigure}%
    \begin{subfigure}[c]{0.5\textwidth}
        \centering
        \raisebox{22pt}{\scalebox{0.55}{\spectrumSIHIIgraph}}
        \caption{}
        \label{fig:spectragraphS1H2}
    \end{subfigure}
\caption{Calculated quantum spin transition spectrum of external spin-0 and spin-1 states. The bold line represents the external (asymptotic) mass-$m$ state, while the double line represents an internal (excited) mass-$\mu$ state. \color{gray}Gray \color{black} transitions are trivially forbidden because $s'\geq0$ by definition. The \color{red!80!black}red \color{black} transitions are non-trivially forbidden because there are no corresponding, physically allowable, 3-point on-shell amplitudes. The \color{green!60!black}green \color{black} transitions are allowed. For a fixed $|\Delta s|$ we refer to transitions with $\Delta s > 0$ as the \textit{ceiling} and those with $\Delta s < 0$ as the \textit{floor}.  }
    \label{fig:spectragraph}
\end{figure}

Here we summarize our calculation method and some structural aspects of the results. We use the spin transitions spectra in Figure \ref{fig:spectragraph} to guide the explanation of our results. We provide the spin-interpolated impulse for external spin-$0$ and external spin-$1$ for spin transitions of $\Delta s = 0, \pm 1, \pm2$ with scalar, photon, and graviton exchange. We also draw conclusions from the non-interpolated spin-$2$ calculation by considering Casimir-ambiguity-independent coefficients, as explained in Section \ref{subsec:spinuniandInterpolation}. Therefore, our observations are informed by information valid to $\mathcal{O}(S_1^4)$.

\subsection{Summary of Calculation}
\label{subsec:3pt}

\noindent\textit{1. Construct the integrand.} We construct the one-loop integrand by the method of generalized unitarity. We sew together the cut depicted in Figure \ref{fig:unitaritycut}, using minimal-coupling Compton amplitudes and the absorptive 3-point amplitudes in Appendix \ref{app:3ptAmps} via physical state projectors for the scalar, photon, and graviton and completeness relations (\ref{completeness1}) and (\ref{completeness2}) for the internal $X$-states. For the spectral mass function, we choose $\rho(x) \sim x^k$ with $k=1$; as discussed this may change for different physical bodies.

\vspace{0.5cm}
\noindent \textit{2. Replace polarization tensors with spin vectors.} After sewing together the integrand, we replace the universal product of polarization tensors with (\ref{polexpansionresum}). At this point, we can follow the procedure of Section \ref{subsec:spinvector} and replace the polarization tensors with spin vectors and projectors. 

\vspace{0.5cm}
\noindent \textit{3. Soft expansion of the integrand.} Now that the integrand is fully in terms of momenta and spin vectors, we can perform the soft expansion.  As explained in Section \ref{subsec:spinuniandInterpolation}, we will need to expand to $2^{s_1}$ extra orders of $\lambda$ in order to be able to carry out the interpolation procedure. The leading order contribution should reduce to the sum of box and cross-box contributions, which leads to the cuts in the impulse integrand of (\ref{absimpulseDiagram}). 

\vspace{0.5cm}
\noindent \textit{4. Spin interpolation.} With the integrand expanded in the soft limit, we can perform the spin interpolation procedure laid out in Section \ref{subsec:spinuniandInterpolation}. This should only be done if there is a corresponding transition that has a lower external spin-$s_1'$ field whose contribution should be entirely contained by the spin-$s_1$ calculation, as dictated by spin universality. Once the interpolation is complete, we have the unambiguous integrand, which may carry Wilson coefficients associated with lower external spin 3-point amplitudes. We can alternatively not interpolate the integrand but instead consider a non-fixed correction to their result proportional to appropriate powers of $S_1^2$. Then, after integrating,  we must construct the Casimir-ambiguity-independent $d$-coefficients.

\vspace{0.5cm}
\noindent \textit{5. Integration.} Now that we have the unambiguous integrand, the final step now is to vNV decompose tensor integrals (\ref{vecdecomp}), and integrate using the master formula (\ref{eq:masterint}). After projecting all leftover vectors in basis $\left\{b^{\mu}, u_1^{\mu}, u_2^{\mu}, n^{\mu}\right\}$, we have successfully calculated the impulse. This also means that we replace all factors of $S_{1}^{2}$ with (\ref{eq:spinsq}).\\ 
\\
At this point we arrive at the final impulse, which we parameterize into components as in (\ref{eq:impform}), and organize the coefficients those components in (\ref{eq:cdef}). We provide the results for the interpolated impulses in Appendix \ref{app:results}, up to $\mathcal{O}(S_1^2)$. For the non-interpolated results, we provide the combination of coefficients that are immune to the Casimir ambiguity discussed in Section \ref{subsec:spinuniandInterpolation} in an ancillary file. 

\subsection{Floor-Ceiling Symmetry}

We have observed that in all cases the Casimir independent contributions to the absorptive impulse satisfy the following relation
\begin{equation}
    \label{eq:floorceilingd}
    \left(d_v\right)_{k,i}^{(s,s+\Delta s,|h|)} \cong \left(d_v\right)_{k,i}^{(s,s-\Delta s,|h|)},
\end{equation}
where as explained in Section \ref{subsec:spinuniandInterpolation}, $\cong$ means that there exists a bijective map between Wilson coefficients. Equivalently, we have observed that the impulse associated with a \textit{decrease} of the spin-magnitude by $\Delta s$ (the \textit{floor} process) is identical to the impulse associated with an \textit{increase} of spin-magnitude by $\Delta s$ (the \textit{ceiling} process) modulo contributions proportional to the spin Casimir:
\begin{equation}
    \label{eq:floorceiling}
    \boxed{(\Delta p_1^\mu)_{|h|}^{s\rightarrow s+\Delta s} \cong (\Delta p_1^\mu)_{|h|}^{s\rightarrow s-\Delta s} + S_1^2 \times \left[...\right].}
\end{equation}
A simple illustrative example is to compare the $d_u$-coefficients for the observables $\left(\Delta p_1^\mu\right)^{2\rightarrow 1}_2$ and $\left(\Delta p_1^\mu\right)^{2\rightarrow 3}_2$. For the ``floor" we calculate (in the notation (\ref{eq:dcoeffS4}))
{\allowdisplaybreaks\begin{flalign}
    (d_u)_{4,1}^{(2,1,2)} &= \frac{15 \left(h_1^{(2,1,2)}\right)^2 \kappa ^2 m_2^2 \left(105 y^4-98 y^2+9\right)}{8388608 (-b^2)^5 \left(y^2-1\right)^2} \\
    (d_u)_{4,2}^{(2,1,2)} &= 0 \\
    (d_u)_{4,3}^{(2,1,2)} &= \frac{105 \kappa ^2 m_2^2 \left(\left(h_1^{(2,1,2)}\right)^2 \left(22 y^2-21\right)+20 \left(g_1^{(2,1,2)}\right)^2 \left(y^2-1\right)\right)}{8388608 (-b^2)^5 \left(y^2-1\right)^2} \\
    (d_u)_{4,4}^{(2,1,2)} &= 0 \\
    (d_u)_{4,5}^{(2,1,2)} &= 0 \\
    (d_u)_{4,6}^{(2,1,2)} &= \frac{105 \left(h_1^{(2,1,2)}\right) \left(g_1^{(2,1,2)}\right) \kappa ^2 m_2^2 y \left(7-15 y^2\right)}{2097152 (-b^2)^5 \left(y^2-1\right)} \\
    (d_u)_{4,7}^{(2,1,2)} &= 0 \\
    (d_u)_{4,8}^{(2,1,2)} &= \frac{1575 \left(h_1^{(2,1,2)}\right) \left(g_1^{(2,1,2)}\right) \kappa ^2 m_2^2 y \left(1-2 y^2\right)}{2097152 (-b^2)^5 \left(y^2-1\right)} \\
    (d_u)_{4,9}^{(2,1,2)} &= -\frac{1575 \kappa ^2 m_2^2 \left(\left(h_1^{(2,1,2)}\right)^2 \left(1-2 y^2\right)^2-4 \left(g_1^{(2,1,2)}\right)^2 y^2 \left(y^2-1\right)\right)}{8388608 (-b^2)^5 \left(y^2-1\right)},
\end{flalign}}
and for the ``ceiling"
{\allowdisplaybreaks\begin{flalign}
    (d_u)_{4,1}^{(2,3,2)} &= \frac{3 \left(h_{10}^{(2,3,2)}\right)^2 \kappa ^2 m_2^2 \left(105 y^4-98 y^2+9\right)}{16777216 (-b^2)^5 \left(y^2-1\right)^2} \\
    (d_u)_{4,2}^{(2,3,2)} &= 0 \\
    (d_u)_{4,3}^{(2,3,2)} &= \frac{21 \kappa ^2 m_2^2 \left(\left(h_{10}^{(2,3,2)}\right)^2 \left(22 y^2-21\right)+20 \left(g_{10}^{(2,3,2)}\right)^2 \left(y^2-1\right)\right)}{16777216 (-b^2)^5 \left(y^2-1\right)^2} \\
    (d_u)_{4,4}^{(2,3,2)} &= 0 \\
    (d_u)_{4,5}^{(2,3,2)} &= 0 \\
    (d_u)_{4,6}^{(2,3,2)} &= \frac{21 \left(h_{10}^{(2,3,2)}\right) \left(g_{10}^{(2,3,2)}\right) \kappa ^2 m_2^2 y \left(7-15 y^2\right)}{4194304 (-b^2)^5 \left(y^2-1\right)} \\
    (d_u)_{4,7}^{(2,3,2)} &= 0 \\
    (d_u)_{4,8}^{(2,3,2)} &= \frac{315 \left(h_{10}^{(2,3,2)}\right) \left(g_{10}^{(2,3,2)}\right) \kappa ^2 m_2^2 y \left(1-2 y^2\right)}{4194304 (-b^2)^5 \left(y^2-1\right)} \\
    (d_u)_{4,9}^{(2,3,2)} &= -\frac{315 \kappa ^2 m_2^2 \left(\left(h_{10}^{(2,3,2)}\right)^2 \left(1-2 y^2\right)^2-4 \left(g_{10}^{(2,3,2)}\right)^2 y^2 \left(y^2-1\right)\right)}{16777216 (-b^2)^5 \left(y^2-1\right)}.
\end{flalign}}
Each set of observables is a function of only two Wilson coefficients $\{g_1^{(2,1,2)},h_1^{(2,1,2)}\}$ and $\{g_{10}^{(2,3,2)},h_{10}^{(2,3,2)}\}$ respectively (the definitions of the corresponding 3-point amplitudes are provided in the ancillary file). The claimed bijection in this case takes the simple form 
\begin{equation}
    \label{bijection}
    g_{10}^{(2,3,2)} \cong -\sqrt{10} g_1^{(2,1,2)}, \hspace{10mm} h_{10}^{(2,3,2)} \cong -\sqrt{10} h_1^{(2,1,2)}.
\end{equation}
In the above example, not only do each of the 5 non-zero pairs of coefficients need to match, but they need to match at \textit{all values} of the relative velocity $y$. The matching condition 
\begin{equation}
    (d_u)_{4,1}^{(2,1,2)} \cong (d_u)_{4,1}^{(2,3,2)},
\end{equation}
is then actually 3 separate conditions, one each for the numerator coefficients of $y^0$, $y^2$ and $y^4$. Moreover, there are also the matching conditions for the $d_b$- and $d_n$-coefficients, each of these depend on the same set of 4 Wilson coefficients, and can be verified to be related by the bijection (\ref{bijection}). Overall the problem of finding such a map is \textit{highly} over-constrained, and its existence correspondingly highly non-trivial. We have explicitly verified that the floor-ceiling symmetry obtains in the cases: $(s,\pm \Delta s) = (1,\pm 1)$, $(2,\pm 1)$ and $(2, \pm 2)$ for scalar and photon absorption and $(2,\pm 1)$ and $(2, \pm 2)$ for graviton absorption. 

A naively surprising aspect of the existence of this symmetry concerns the counting of Wilson coefficients. As reviewed in Appendix \ref{app:3ptAmps}, as the external spin $s'$ increases the number of available on-shell 3-point amplitudes also increases. For the floor-ceiling symmetry to be possible it must be the case that, at least in the soft expansion, the impulse depends on a small subset of possible structures; as we discuss in greater detail in the subsequent subsection this is indeed the case. 

There also exists special values of the interpolation correction that extends this symmetry to include mappings of lower spin Wilson coefficients that come from the interpolation procedure to the Wilson coefficients of the corrected amplitude. This may hint at the existence of \textit{special} compact, spinning bodies for which the symmetry is an exact statement, including Casimir contributions. 

\subsection{Spin Suppression}
\label{sec:spinsuppression}

At long-distances, $|b|\rightarrow \infty$, the absorptive impulse is dominated by contributions with the slowest falloff in $1/|b|$. In the construction of the integrand using generalized unitarity, as described in Section \ref{subsec:absampint}, the largest contributions arise from a small subset of the 3-particle mass/spin-magnitude changing interactions that dominate in the soft limit. 

In the kinematic conventions of Figure \ref{fig:3pt}, in the even- and odd-parity sectors, if $\Delta s \geq |h_3|$, there are unique, \textit{soft-dominant} mass-changing 3-point amplitudes of the form\footnote{The factors $\left((p_1 \cdot \epsilon_3) (\epsilon_2 \cdot p_3) - (p_1\cdot p_3) \, (\epsilon_2 \cdot \epsilon_3)\right)^{h_3}$ and $\varepsilon^{p_1 p_3 \epsilon_2 \epsilon_3}\left((p_1 \cdot \epsilon_3) (\epsilon_2 \cdot p_3) - (p_1\cdot p_3) \, (\epsilon_2 \cdot \epsilon_3)\right)^{h_3-1}$ correspond to the on-shell matrix elements of the field strength tensor and its dual respectively: $F_{\mu\nu}$, $\tilde{F}_{\mu\nu}$ for photon absorption, $|h_3|=1$, and the Riemann tensor $R_{\mu\nu\rho\sigma}$, $\tilde{R}_{\mu\nu\rho\sigma}$ for graviton absorption, $|h_3|=2$. From an off-shell effective action perspective, the soft scaling of this 3-point interaction is a consequence of the fact that the mass/spin-magnitude changing interactions are always \textit{non-minimal} with the matter fields coupling to the mediator fields via the field strength. }
\begin{align}
    \label{eq:softA3}
    \mathcal{A}_3^{(s_1,s_2,|h_3|)} &=  g\left((p_1 \cdot \epsilon_3) (\epsilon_2 \cdot p_3) - (p_1\cdot p_3) \, (\epsilon_2 \cdot \epsilon_3)\right)^{h_3} (p_3 \cdot \epsilon_2)^{\Delta s - {h_3}} (\epsilon_1 \cdot \epsilon_2)^{s_1}, \\
    \label{eq:softA3t}
    \tilde{\mathcal{A}}_3^{(s_1,s_2,|h_3|)} &= h \,\varepsilon^{p_1 p_3 \epsilon_2 \epsilon_3}\left((p_1 \cdot \epsilon_3) (\epsilon_2 \cdot p_3) - (p_1\cdot p_3) \, (\epsilon_2 \cdot \epsilon_3)\right)^{h_3-1} (p_3 \cdot \epsilon_2)^{\Delta s - {h_3}} (\epsilon_1 \cdot \epsilon_2)^{s_1},
\end{align}
where $\Delta s \equiv s_2-s_1$ and $p^\mu_3\sim q^\mu$ is the soft momentum. These interactions manifestly scale as $\mathcal{A}_3^{(s_1,s_2,h_3)} \sim q^{\Delta s}$, and as a consequence, the slowest falloff contribution to the impulse is fixed to be
\begin{equation}
    \left(\Delta p_1^\mu\right)_{|h|}^{s\rightarrow s\pm\Delta s} \sim \frac{1}{|b|^{3+2\Delta s}}.
\end{equation}
If we were calculating the impulse for fixed (quantum) spin $s$, then these contributions would be the most important. However, as discussed in Section \ref{subsec:softexpansion}, in the soft expansion of the impulse, the classical spin vector scales like an inverse power of the soft scale $S_1 \sim \lambda^{-1}$, and therefore terms of the above form with additional powers of $S_1$ appear at \textit{lower} order in the expansion. Since we can only reconstruct terms in the spin expansion up to $\mathcal{O}(S_1^{2s})$ from calculations with finite external spin-$s$, the true leading-order soft contribution may not appear until $s$ reaches a \textit{critical} value. A simple illustration of this is to consider the case $\Delta s=1$ for $h=0$, the explicit results for integer spins are recorded in Appendix \ref{app:results} and in the ancillary file, for this discussion we only need the schematic form:
\begin{align}
    \left(\Delta p_1^\mu\right)_0^{0\rightarrow 1} &\sim \underbrace{\frac{S_1^0}{|b|^5}}_{\lambda^5} + ...\\
    \left(\Delta p_1^\mu\right)_0^{\frac{1}{2}\rightarrow \frac{3}{2}} &\sim \underbrace{\frac{S_1^1}{|b|^5}}_{\lambda^4} + \underbrace{\frac{S_1^0}{|b|^5} + \frac{S_1^1}{|b|^6}}_{\lambda^5} +...\\
    \left(\Delta p_1^\mu\right)_0^{1\rightarrow 2} &\sim \underbrace{\frac{S_1^2}{|b|^5}}_{\lambda^3} + \underbrace{\frac{S_1^1}{|b|^5}+\frac{S_1^2}{|b|^6}}_{\lambda^4}+ \underbrace{\frac{S_1^0}{|b|^5}+\frac{S_1^1}{|b|^6}+\frac{S_1^2}{|b|^7}}_{\lambda^5} +...\\
    \left(\Delta p_1^\mu\right)_0^{\frac{3}{2}\rightarrow \frac{5}{2}} &\sim \underbrace{\frac{S_1^2}{|b|^5} +\frac{S_1^3}{|b|^6}}_{\lambda^3} + \underbrace{\frac{S_1^1}{|b|^5}+\frac{S_1^2}{|b|^6}+\frac{S_1^3}{|b|^7}}_{\lambda^4}+ \underbrace{\frac{S_1^0}{|b|^5}+\frac{S_1^1}{|b|^6}+\frac{S_1^2}{|b|^7}+\frac{S_1^3}{|b|^{8}}}_{\lambda^5} +... \\
    \left(\Delta p_1^\mu\right)_0^{2\rightarrow 3} &\sim \underbrace{\frac{S_1^2}{|b|^5} +\frac{S_1^3}{|b|^6} +\frac{S_1^4}{|b|^7}}_{\lambda^3} + \underbrace{\frac{S_1^1}{|b|^5}+\frac{S_1^2}{|b|^6}+\frac{S_1^3}{|b|^{7}}+\frac{S_1^4}{|b|^8}}_{\lambda^4}+ \underbrace{\frac{S_1^0}{|b|^5}+...+\frac{S_1^4}{|b|^{9}}}_{\lambda^5} +...
\end{align}
In this example the critical value of the external spin is $s=1$. From calculations at $s=0,\frac{1}{2}$ one would incorrectly surmise that the leading soft scaling is either $\mathcal{O}(\lambda^5)$ or $\mathcal{O}(\lambda^4)$. For spins $s\geq 1$ the leading order soft contribution is \textit{stabilized} at $\mathcal{O}(\lambda^3)$, the true leading-order value. Higher-spins give us access to higher-order terms in the spin expansion, but the leading soft scaling is determined. In this example the leading-order contribution (in the soft and spin expansions) is $\sim S_1^2 b^{-5}\sim \lambda^3$. There cannot be a contribution at this order in $\lambda$ with fewer powers of spin: if there was an $S_1^0$ contribution it would necessarily have the form $\sim S_1^0 b^{-3}$. Universality requires that such a contribution also appear in the $0\rightarrow 1$ calculation, which in turn requires a 3-point on-shell amplitude with scaling  $\mathcal{A}_3^{(0,1,0)}\sim q^0$, but as previously discussed this does not exist.

The critical value of $s$ is given by $\Delta s$, and therefore the leading-order contribution for a given $\Delta s$ is predicted to be $\left(\Delta p_1^\mu\right)_{|h|}^{s\rightarrow s\pm \Delta s} \sim S_1^{2\Delta s}$; we term this universal behavior \textit{spin suppression}. The above discussion applies only if $\Delta s \geq |h|$; when this inequality is violated additional care is needed. 

For the Casimir independent contributions, the appearance of a critical $s$ value can be understood in another way by appealing to the notion of floor-ceiling symmetry described in the previous subsection. The critical value of $s$ corresponds to the lowest value for which the floor process, $s \rightarrow s -\Delta s$ exists. In the above example, $s=\frac{1}{2}$ is below the critical value for $\Delta s = 1$ because the corresponding floor process $\frac{1}{2}\rightarrow -\frac{1}{2}$ trivially does not exist. Naively, the process $\Delta s\rightarrow 0$ is the lowest spin calculation for which the floor exists, giving the critical value $s=\Delta s$ as claimed above. However, as explained in detail in Appendix \ref{app:3ptAmps}, if $\Delta s < |h_3|$, the 3-point amplitude $\mathcal{A}_3^{(\Delta s,0,h)}$ \textit{does not exist}. For $\Delta s < |h_3|$ the lowest value of $s$ for which the $s\rightarrow s-\Delta s$ process exists for a mediator of spin $|h_3|$ is $s=\frac{1}{2}(|h_3|+\Delta s)$, and therefore we predict 
\begin{equation}
    \label{eq:spinsupp}
    \boxed{\left(\Delta p_1^\mu\right)_{|h|}^{s\rightarrow s\pm \Delta s} \sim \frac{1}{|b|^{3+2\Delta s}}\begin{cases}
        S_1^{2\Delta s }, &\text{if} \hspace{3mm} \Delta s \geq |h|,\\
        S_1^{\Delta s + |h|}, &\text{if} \hspace{3mm} \Delta s < |h|.
    \end{cases}}
\end{equation}
Such an exceptional case is illustrated in Figure \ref{fig:spectragraphS1H2}, the graviton absorption process $1\rightarrow 2$ has no corresponding floor $1\rightarrow 0$ because, non-trivially, the floor amplitude does not exist. In this case the critical spin value is $s=2$ and so, as verified in the $d$-coefficients tabulated in the ancillary file, the spin suppression of the leading soft contribution is $\mathcal{O}\left(S_1^4\right)$.

\subsection{Scattering Angles and Mass Shift}
\label{sec:angle}

It is enlightening to give a more direct physical interpretation to the scalar components of the impulse (\ref{eq:impform}). As we explain in this section, they are related to the center-of-mass (COM) frame scattering angle(s) and the change in rest-mass. The latter is a Lorentz invariant quantity, naturally given by
\begin{equation}\label{eq:massshiftdef}
    \left(\Delta m_1\right)^{s \rightarrow s'}_{|h|} = \left(u_1 \cdot \Delta p_1\right)^{s \rightarrow s'}_{|h|}.
\end{equation}
As explained in \cite{Jones:2023ugm}, for non-spinning external states, $s=0$ the impulse is purely longitudinal $\Delta p_1^\mu \propto \Check{u}_1^\mu$, with the overall coefficient completely determined change in rest-mass. 

For spinning external states, in general there will be transverse contributions that will produce non-forward scattering. In the COM frame the kinematic variables take the form
\begin{equation}
    p_1^\mu = \left(E_1,\mathbf{p}\right), \hspace{10mm} p_2^\mu = \left(E_2,-\mathbf{p}\right), \hspace{10mm} b^\mu = \left(0,\mathbf{b}\right)\hspace{10mm} n^\mu = \left(0,\mathbf{n}\right),
\end{equation}
where $\mathbf{p}\cdot \mathbf{b} = \mathbf{p}\cdot \mathbf{n} = \mathbf{b}\cdot \mathbf{n}=0$, $\mathbf{b}^2 = -b^2$, $\mathbf{n}^2 = -n^2 = -b^2(y^2-1)$ and $E_i \equiv \sqrt{\mathbf{p}^2+m_i^2}$. For later reference we also define the orbital angular momentum as $\mathbf{L} = \mathbf{b} \times \mathbf{p}$. 
\begin{figure}
    \centering
    \begin{tikzpicture}[scale=0.5]
        \filldraw[color=black!5] (-10,-4) -- (-5,4) -- (10,4) -- (5,-4) -- cycle;
        \filldraw[color=black!10] (-12,-2) -- (-3,2) -- (12,2) -- (3,-2) -- cycle;
        \draw[-,color=gray] (-10,-4)--(-5,4);
        \draw[-,color=gray] (5,-4)--(10,4);
        \draw[-,color=gray] (-5,4)--(10,4);
        \draw[-,color=gray] (-10,-4)--(5,-4);
        \draw[-,color=gray] (-3,2)--(-7.5,0);
        \draw[-,color=gray] (3,-2)--(12,2);
        \draw[-,color=gray] (-12,-2)--(3,-2);
        \draw[-,color=gray] (-3,2)--(12,2);
        \draw[-,color=gray] (-7,0)--(-7+0.456906,0+0.203069)--(-7.5+0.456906,0+0.203069);
        \draw[-,color=gray] (-7.5,0.5)--(-7.5+0.456906,0.5+0.203069)--(-7.5+0.456906,0.203069);
        \draw[-,color=gray] (4,1.5)--(4-0.485071,1.5-0.121268)--(4-0.485071,1-0.121268);
        \draw (1,0) arc (0:35:1);
        \draw (8.41381,0.406138) arc (23.9625:57:1);
        \draw[->,thick,color=red] (-7.5,0)--(-12,-2);
        \draw[->,thick,color=violet] (-7.5,0)--(0,0);
        \draw[-,style=dashed,color=gray] (0,0)--(7.5,0);
        \draw[->,thick,color=blue] (0,0)--(4,3);
        \draw[-,style=dashed,color=gray] (0,0)--(4,1);
        \draw[-,style=dashed,color=gray] (4,1)--(4,3);
        \draw[->,thick,color=teal] (-7.5,0)--(-7.5,4);
        \node at (1.3,0.4) {$\chi$};
        \node at (8.7,1) {$\varphi$};
        \node at (-3.75,-0.75) {\color{violet}$\mathbf{p}$};
        \node at (-8.2,2) {\color{teal}$\mathbf{L}$};
        \node at (-10,-0.2) {\color{red}$\mathbf{b}$};
        \node at (2,2.5) {\color{blue}$\mathbf{p}\;'$};
    \end{tikzpicture}
    \caption{The \textit{polar} $\chi$, and \textit{azimuthal} $\varphi$, scattering angles defined in the COM frame in terms of the incoming 3-momentum $\color{violet}\mathbf{p}\color{black}$, outgoing 3-momentum $\color{blue}\mathbf{p}\,'\color{black}$ and impact parameter $\color{red}\mathbf{b}\color{black}$. The azimuthal angle $\varphi$ measures the degree of \textit{non-planarity} of the scattering; in the limit of aligned spin ($\mathbf{S}_1 \propto \color{teal}\mathbf{L}\color{black}$) the scattering is planar ($\color{blue}\mathbf{p}\,'\color{black}$ lies in the plane spanned by $\color{violet}\mathbf{p}\color{black}$ and $\color{red}\mathbf{b}\color{black}$) and consequently $\varphi=0$.}
    \label{fig:angles}
\end{figure}
The outgoing momentum is given by $\mathbf{p}' \equiv \mathbf{p} + \Delta \mathbf{p}$ and therefore the \textit{polar} scattering angle is defined as
\begin{equation}
    \cos(\chi) = \frac{\mathbf{p}\cdot \mathbf{p}'}{|\mathbf{p}| |\mathbf{p}'|}.
\end{equation}
By some elementary kinematics (care must be taken since the mass of the scattering body is non-constant) this can be expressed in terms of covariant quantities 
\begin{equation}\label{eq:scatteringangle}
    \chi^{s \rightarrow s'}_{|h|} = \frac{\sqrt{ m_1^2 + m_2^2 + 2 m_1 m_2 y}}{m_1 m_2 \sqrt{y^2 - 1}} \sqrt{-  \left(\Delta p_1^2\right)^{s \rightarrow s'}_{|h|} - \frac{\left(\Delta m_1^2\right)^{s \rightarrow s'}_{|h|}}{y^2 - 1}},
\end{equation}
where we ignore higher-order in $\lambda$ contributions. By squaring (\ref{eq:impform}) we can write this in an equivalent form
\begin{equation}\label{eq:scatteringangle2LO}
    \chi^{s \rightarrow s'}_{|h|} = \frac{\sqrt{ m_1^2 + m_2^2 + 2 m_1 m_2 y}}{m_1 m_2 \sqrt{y^2 - 1}} \sqrt{-  \frac{\left(\left(b \cdot \Delta p_1\right)^{s \rightarrow s'}_{|h|}\right)^2}{b^2} - \frac{\left(\left(n \cdot \Delta p_1\right)^{s \rightarrow s'}_{|h|}\right)^2}{b^2 (y^2 - 1)}}.
\end{equation}
When including spin in 2-body scattering, we expect that, in general, the scattering is \textit{non-planar}. The \textit{azimuthal} scattering angle measures the non-planarity of the scattering, as illustrated in Figure \ref{fig:angles}. More precisely it is defined as
\begin{equation}
    \cos(\varphi) = \frac{\mathbf{b}\cdot \mathbf{p}'}{|\mathbf{b}||\mathbf{p}'|}\left(1-\frac{(\mathbf{p}\cdot \mathbf{p}')^2}{|\mathbf{p}|^2|\mathbf{p}'|^2}\right)^{-1/2},
\end{equation}
we can again express it in terms of covariant quantities 
\begin{equation}\label{eq:azimuthangle}
    \varphi^{s \rightarrow s'}_{|h|} = \frac{1}{\sqrt{y^2 - 1}} \frac{\left(n\cdot \Delta p_1\right)^{s \rightarrow s'}_{|h|}}{\left(b\cdot \Delta p_1\right)^{s \rightarrow s'}_{|h|}}.
\end{equation}
Note that  $\varphi^{s \rightarrow s'}_{|h|}$ has an ambiguous $|b| \rightarrow \infty$ limit; this corresponds to forward scattering, or  $\chi^{s \rightarrow s'}_{|h|} = 0$, and as expected in spherical coordinates, in this limit the azimuthal angle is undefined.

From (\ref{eq:scatteringangle2LO}) and (\ref{eq:azimuthangle}), it is clear that, in order for the scattering event to be non-planar, the absorptive impulse needs to have components in the $b^{\mu}$ and $n^{\mu}$ directions. The non-longitudinal components of the absorptive impulse come from the vNV decomposition of the loop momentum (\ref{vecdecomp}); in particular, from the vector prefactor in the KMOC integrand (\ref{absimpulseDiagram}) upon decomposition
\begin{equation}
    \int \hat{\text{d}}^{D}l \left(\frac{2 \ell^{\mu } - q^{\mu}}{2}\right) \rightarrow \int \hat{\text{d}}^{D}l \left( \left(\frac{\tilde{n} \cdot \ell }{\tilde{n}^2}\right) \, \tilde{n}^{\mu} - x \check{u}_1^{\mu}\right),
\end{equation}
where the terms of the integrand proportional to $\tilde{n}^{\mu}$ Fourier transform to terms that are proportional to combinations of $b^{\mu}$ and $n^{\mu}$. For example, at $\mathcal{O}(S^2)$, we encounter the following
\begin{align}
    &\frac{\tilde{n}^{\mu} (\tilde{n}\cdot S_1)}{(-q^2)^{\alpha+2}}  \xrightarrow[]{\text{FT}}  \left[b^{\mu} (b \cdot S_1) - n^{\mu}  (n\cdot S_1) \left(\frac{2\alpha+1}{y^2 - 1}\right)\right](-b^2)^{-\alpha} , \nonumber \\
    & \frac{\tilde{n}^{\mu} (q\cdot S_1)}{(-q^2)^{\alpha+2}} \xrightarrow[]{\text{FT}}  \left[b^{\mu} (n \cdot S_1) + n^{\mu} (b\cdot S_1) (2\alpha + 1)\right](-b^2)^{-\alpha},
\end{align}
where we decompose the spin vector in the same basis as the impulse (\ref{eq:spinvectobasis}), and suppress an overall coefficient for clarity. This also explains why many of the coefficients $c_{b}, c_{n}$ are related to each other, as can be seen in Appendix \ref{app:results} and the ancillary file. We also find that the space of spin monomials at each order in spin is not spanned by the spin monomials in any of the contributions to the impulse, i.e. several of the $(c_{i})_{j_1, j_2, j_3}^{(s, s', |h|)}$ are zero. This reduction of structures is \textit{a priori} unexpected and must be kinematic in origin, since this appears to be a universal feature for all possible choices of Wilson coefficients.

Our most interesting observation concerns the \textit{aligned spin} limit. This corresponds to the special case where the spin angular momentum 3-vector $\mathbf{S}_1$ is collinear with the orbital angular momentum 3-vector $\mathbf{L}$. In this limit, by symmetry the scattering is necessarily planar, $\varphi^{s \rightarrow s'}_{|h|}=0$. Moreover, since $\mathbf{p}\cdot \mathbf{L} = \mathbf{b}\cdot \mathbf{L}=0$ it follows that
\begin{equation}
    b\cdot S_1, \; u_2\cdot S_1 \xrightarrow[]{\mathbf{S}_1 \propto \mathbf{L}} 0,
\end{equation}
and therefore only $c$-coefficients of the form $(c_v)^{(s,s',h)}_{0,i,0}$, for $v\in \{b,n\}$, give a non-zero transverse contribution to the impulse. Interestingly, we find that all such coefficients are \textit{zero} and therefore the scattering angle (\ref{eq:scatteringangle2LO}) vanishes in the aligned spin limit at leading order\footnote{Since (\ref{eq:scatteringangle2LO}) expresses the polar scattering angle as a sum-of-squares with the same sign, it is clear that the scattering angle vanishes \textit{if and only if} $b\cdot \Delta p_1 = n\cdot \Delta p_1 =0$.} 
\begin{equation}
\label{eq:alignedspinSA}
    \boxed{\chi^{s \rightarrow s'}_{|h|} \xrightarrow[]{\mathbf{S}_1 \propto \mathbf{L}} 0.}
\end{equation}
Conversely, when away from the aligned spin limit, we find that the leading-order absorptive scattering angle is \textit{non-zero}. This is easy to see by studying the coefficients in Appendix \ref{app:results}; up to $\mathcal{O}(S_1^2)$, the relevant $c_b$ and $c_n$ contributions always contain terms proportional to $b \cdot S_1$ and $u_2 \cdot S_1$, which vanish in the spin-aligned case. At $\mathcal{O}(S_1^2)$ this property can be observed before spin interpolation; as discussed, at $\mathcal{O}(S_1^0)$ the impulse is always longitudinal, and so therefore are the Casimir promoted contributions (\ref{interpolationex}).

In addition to the scattering angles, we have also calculated the leading order mass-shift, $\left(\Delta m_1\right)^{s \rightarrow s'}_{|h|}$, for spin transitions $\Delta s= 0, \pm 1, \pm 2$ up to $\mathcal{O}(S_1^2)$ in the spin expansion. When comparing to the results for Schwarzschild black holes in \cite{Jones:2023ugm}, we find that at first we have too many free coefficients in our results
\begin{equation}
    \left(\Delta m_1\right)_{2}^{0\rightarrow 2} = - \frac{225 \kappa^2 m_1^2 m_2^2 }{2^{21} } \frac{\sqrt{y^2 - 1}}{ \sqrt{-b^2}} \frac{\left(g_{1}^{(0, 2, 2)}\right)^2 \left(21 y^4 - 14 y^2 + 9\right) + 7 \left(h_{1}^{(0, 2, 2)}\right)^2 \left(3 y^4 - 2 y^2 -1 \right)}{\left(-b^2\right)^3}.
\end{equation}
Imposing the condition 
\begin{equation}
    g_{1}^{(0, 2, 2)} \overset{!}{=} h_{1}^{(0, 2, 2)},
\end{equation}
gives the correct polynomial structure, $21y^4 -14y^2 +1$, for a Schwarzschild black hole. In \cite{Jones:2023ugm}, following \cite{Goldberger:2005cd}, this was interpreted as a manifestation of Chandrasekhar duality \cite{Chandrasekhar:1975nkd}. It is striking that this hidden symmetry has such a simple implementation: the equality of Wilson coefficients for parity-even and parity-odd 3-point couplings. Since this is also expected to be a symmetry of the Kerr black hole, it is conceivable that the analogous relation between $g^{(s, s', 2)}$ and $h^{(s, s', 2)}$ couplings should be imposed for $s>0$. 

\newpage

\section{Discussion}
\label{sec:discussion}

In this paper we explored the effects of scalar, electromagnetic and gravitational radiation absorption and spin-magnitude change in two-body post-Minkowskian scattering. We modeled the spin dependence of the absorbing body by using states of finite quantum spin, extrapolating results to the large-spin correspondence limit. We have calculated the classical absorptive impulse up to $\mathcal{O}(S^2)$ for fully Casimir interpolated results and up to $\mathcal{O}(S^4)$ for Casimir independent contributions. The results exhibit the expected spin universality as well as a number of other, apparently universal, features. Our main results are: the observation of a surprising ``floor-ceiling symmetry" (\ref{eq:floorceiling}) relating the Casimir independent contributions from the spin transitions $s\rightarrow s \pm \Delta s$; a universal suppression of the leading-order soft contribution by a predictable power of the classical spin vector (\ref{eq:spinsupp}); and the vanishing of the leading-order scattering angle in the aligned spin limit (\ref{eq:alignedspinSA}). There are numerous ways our analysis could be extended.

It would be desirable to extend the calculation to external spinning states with $s>2$. Such a calculation is technically more involved, but conceptually straightforward using the methods described in this paper. The calculation of the Casimir independent $d$-coefficients for the $3\rightarrow 4$ transition for graviton absorption would provide a non-trivial check of the emergence of spin universality in gravitational absorptive scattering. Also clearly desirable is to calculate fully Casimir interpolated observables beyond $\mathcal{O}(S^2)$. As explained in Section \ref{subsec:spinuniandInterpolation}, this requires a more intricate matching calculation with terms promoted from several subleading orders in the soft expansion. A systematic approach to this was described in \cite{Akpinar:2025bkt, Akpinar:2024meg} , it would be interesting to adapt this to the present context. In this paper we have used the extrapolation of observables calculated for finite quantum spin states because this gives a clear conceptual picture of transitions between different values of the spin-magnitude. In principle however, it should be possible to obtain the same results using other approaches, either scattering with arbitrary spin particles \cite{Guevara:2018wpp,Guevara:2019fsj,Bern:2020buy,Kosmopoulos:2021zoq,Bern:2022kto,Aoude:2022thd,Aoude:2023vdk} or worldline-based approaches \cite{Liu:2021zxr,Jakobsen:2021lvp,Jakobsen:2021zvh,Jakobsen:2022fcj,Jakobsen:2022zsx,Jakobsen:2023ndj,Haddad:2024ebn,Ben-Shahar:2023djm}; such approaches may give an improvement in computational efficiency.

Recently there have been developments of a formalism for incorporating ``conservative spin-magnitude change" by relaxing the spin supplementary condition (\ref{eq:SSC}) \cite{Bern:2023ity,Alaverdian:2024spu,Alaverdian:2025jtw}. In these works additional degrees-of-freedom are introduced, associated with the \textit{boost vector} $\mathbf{K}$, with different spin-magnitude from the external states. Unlike our framework however, these new degrees of freedom have the \textit{same rest mass} as the incoming external state. In the language of Section \ref{subsec:absorption}, these $\mathbf{K}$-states are stable, particle-like excitations with spectral density $\rho(\mu^2) \sim \delta(\mu^2-m_1^2)$. This is to be compared to the smooth distribution of excited $X$-states, the existence of which are necessary to match the low-energy black hole absorption cross-section \cite{Page:1976df}, and as discussed in Section \ref{subsec:softexpansion} (and in greater detail in \cite{Jones:2023ugm}) are expected to have a power-law spectral density $\rho(\mu^2) \sim (\mu^2-m_1^2)^k$ in the soft region. Because of this key difference, there is no reason to ``trace-out" the $\mathbf{K}$-states in an in-in calculation as we do in this paper for the $X$-states, and so they are declared to be part of the conservative sector. It is clearly of some importance to gain a better understanding of the relation between these two different approaches to spin-magnitude change and their mutual relevance to developing a systematic EFT of black hole and neutron star scattering.

Finally, there is the important problem of determining the Wilson coefficients in the context of an explicit matching calculation, in particular matching to the Kerr black hole solution. In an interesting recent paper, the absorptive mass-shift for Kerr scattering was calculated \textit{without} EFT matching \cite{Bautista:2024emt}. Rather the authors extracted the Kerr Compton amplitude, including absorptive and superradiant effects, directly from the Teukolsky equation \cite{Teukolsky:1973ha}, and used generalized unitarity to sew this into a 2-to-2 scattering amplitude. Unfortunately, this strategy seems difficult to generalize to the problem of scattering neutron stars, for which the full theory solution is unknown. The approach of this paper is complementary, rather than calculate observables in the context of a specific UV completion we have endeavored to construct a maximally general EFT, providing a parametrization of our ignorance of spinning absorptive scattering. 

It would therefore be of great interest to revisit the calculation of Kerr absorption from the point-of-view of the EFT construction presented in this paper, but this poses a particular challenge. The results of \cite{Bautista:2024emt} incorporate \textit{all} physical effects, in the language of Section \ref{subsec:softexpansion} they include the contributions of all ``excited" states, including superradiant modes with $m_{1*}^2 \leq \mu^2 < m_1^2$. If superradiant contributions cannot be cleanly separated then we will need to extend the current framework to incorporate them.\footnote{For some observables, such as the partial wave absorption cross-section $\sigma_{l,m}$, the total superradiant contribution vanishes when summed over the magnetic quantum number $m$ \cite{Porto:2007qi}. It seems implausible that this would always be the case and so an extension of the presented formalism to incorporate them remains a well-motivated problem.} The most obvious way to do this is to extend the lower integration bound on the spectral integral (\ref{eq:spectralintegral}) as
\begin{equation*}
    \int_0^\infty \text{d}x \; x^{k}\; \left(...\right) \hspace{3mm}\rightarrow \hspace{3mm} \int_{x_*}^\infty \text{d}x \; x^k\; \left(...\right), 
\end{equation*}
where $m_{1*}^2 = m_1^2 + 2m_1 x_*$. To match the low-energy Kerr absorption cross-section \cite{Page:1976df} requires a lower-bound of the form $x_* \sim \Omega_{\text{H}} \sim |S_1| \equiv \sqrt{-S_{1\mu}S_1^\mu}$; where $\Omega_{\text{H}}$ is the angular velocity of the event horizon. In a finite spin calculation, introducing additional explicit dependence on the classical spin vector in the integration bounds in this way seems \textit{ad hoc}; dependence on the spin vector should emerge in the form of spin universality in the limit of large spin. What organizing principle is to be used to determine how Wilson coefficients should be introduced in $x_{*}$? Moreover, the spin dependence of $x_{*}$ for Kerr is \textit{non-analytic} in the limit $S_1^\mu \rightarrow 0$, in contrast to the spin dependence of the absorptive contributions which is manifestly analytic; further complicating the question of what spin-structures are to be included/excluded in the EFT parametrization. What may be needed is a better understanding of the emergence of (classical) rotational superradiance in the large spin correspondence limit. This should provide a \textit{rationale} for such non-analytic spin dependence and a pathway to a generic EFT parametrization of superradiant effects. We leave the development of this extended framework to future work.

\vspace{3mm}
\noindent \textbf{Acknowledgment}

\vspace{3mm}
We would like to thank Dogan Akpinar, Yilber Fabian Bautista, Henrik Johansson, Michael Ruf, Matteo Sergola and Mao Zeng for helpful discussions and useful feedback on the first version of the paper. JPG is supported in part by the U.S. Department of Energy (DOE) under Award Number
DE-SC0009937 and the Mani L. Bhaumik Institute for Theoretical Physics. CRTJ is supported by funding from the University of Arizona.

\newpage

\appendix

\begingroup
\allowdisplaybreaks

\section{Three-Point Amplitudes}
\label{app:3ptAmps}

\subsection*{Counting Independent Amplitudes}

For larger values of $s_i$, due to the proliferation of kinematic and dimensionality constraints, it becomes non-trivial to construct a minimal basis of truly independent on-shell 3-point amplitudes. A convenient way to count the number of independent 3-point amplitudes in $D=4$ is to make use of the massive spinor formalism\footnote{This section is a review of Section 4.2.1 of \cite{Arkani-Hamed:2017jhn}. The method presented in that reference is completely correct, but the final quoted formula is only applicable if $|h_3|\leq |s_1-s_2|$. For completeness, in this Appendix we rederive the general case.}. As explained in \cite{Arkani-Hamed:2017jhn}, after solving all relevant constraints, the general form of a 3-point amplitude with two massive particles (of unequal mass) and one massless particle is
\begin{equation}
    \mathcal{A}_3^{(s_1,s_2,h_3)} = \boldsymbol{\lambda}^{\alpha_1}_1...\boldsymbol{\lambda}^{\alpha_{2s_1}}_1 \boldsymbol{\lambda}^{\beta_{1}}_2...\boldsymbol{\lambda}^{\beta_{2s_2}}_2 \sum_{i=1}^{N(s_1,s_2,h_3)} g_i \left(u^{s_1+s_2+h_3} v^{s_1+s_2-h_3}\right)^{(i)}_{\alpha_1...\alpha_{2s_1}\beta_1...\beta_{2s_2}},
\end{equation}
where 
\begin{equation}
    u_\alpha \equiv \lambda_{3\alpha}, \hspace{10mm} v_\alpha \equiv \frac{p_{1\alpha\dot{\beta}}}{m_1} \overline{\lambda}_3^{\dot{\beta}},\hspace{10mm} \boldsymbol{\lambda}_i \equiv z_{iI}\lambda_{i\alpha}^I, \hspace{10mm} z_{iI} z_i^I = 0.
\end{equation}
Details on the definitions of massive spinors are given in the original reference \cite{Arkani-Hamed:2017jhn}. The number of independent 3-point amplitudes is equal to the number of distinct terms that can appear in the above sum. This is equivalent to a simple counting problem: given a set containing the letter $u$, $s_1+s_2+h_3$ times, and the letter $v$, $s_1+s_2-h_3$ times\footnote{If $|h_3|>s_1+s_2$ then there are no structures we can write down with the correct helicity weights for all of the particles; the problem is only non-trivial for $|h_3|\leq s_1+s_2$.}, how many different ways can that set be partitioned into two disjoint subsets of length $2s_1$ and $2s_2$? The solution is given by
\begin{equation}
    \label{counting3pt}
    N\left(s_1,s_2,h_3\right) = 
    \begin{cases}
    s_1+s_2-|s_1-s_2|+1 & \text{if} \hspace{5mm}|h_3|\leq |s_1-s_2|, \\
    s_1+s_2-|h_3|+1 & \text{if} \hspace{5mm}|s_1-s_2|<|h_3|\leq s_1+s_2, \\
    0 & \text{if} \hspace{5mm} s_1+s_2 < |h_3|.
    \end{cases}
\end{equation}
This counting of on-shell amplitudes must agree with the counting of cubic Lagrangian operators modulo equations of motion and total derivatives. It is useful to classify the latter into \textit{parity-even} (without a Levi-Civita symbol) and \textit{parity-odd} (with a Levi-Civita symbol). Since parity changes the sign of the helicity for a massless particle, we identify 
\begin{equation}
    \mathcal{A}_3^{(s_1,s_2,h_3)}\biggr\vert_{\lambda_3 \leftrightarrow \overline{\lambda}_3} = \mathcal{A}_3^{(s_1,s_2,-h_3)}, \hspace{10mm} \tilde{\mathcal{A}}_3^{(s_1,s_2,h_3)}\biggr\vert_{\lambda_3 \leftrightarrow \overline{\lambda}_3} = -\tilde{\mathcal{A}}_3^{(s_1,s_2,-h_3)},
\end{equation}
with the $+$ sign for parity even operators and the $-$ sign for parity odd. The formula (\ref{counting3pt}) counts the number of on-shell amplitudes for a given helicity, and so for $|h_3|>0$, $N\left(s_1,s_2,h_3\right)$ is separately the number of parity even operators and the number of parity odd operators; the total number of operators for a given massless spin $|h_3|$ is $2N\left(s_1,s_2,h_3\right)$. For $h_3=0$ (massless scalar mediators) $N\left(s_1,s_2,0\right)$ counts the total number of operators, parity even plus parity odd. We will use this counting to verify that we have correctly constructed a complete basis of 3-point interactions. 

Below we explicitly enumerate the on-shell 3-point amplitudes used in the calculation of the absorptive impulse. We use the all-outgoing convention, with momenta and spins/helicities labeled as in Figure \ref{fig:3pt}. Without loss of generality, we represent the traceless-symmetric polarization tensor for the graviton as a product $\epsilon^{\mu\nu}_i = \epsilon^{\mu}_i \epsilon^{\nu}_i$, where $\epsilon_i^2=0$.

\subsection*{Scalar Absorption}

Parity even interactions:
\begin{flalign}
    \mathcal{A}_3^{(0,0,0)} &= g_1^{(0,0,0)}, &&\\
    \mathcal{A}_3^{(0,1,0)} &= -i g_1^{(0,1,0)}\left(p_3\cdot \epsilon_2\right),&&\\
    \mathcal{A}_3^{(0,2,0)} &= -g_1^{(0,2,0)}\left(p_3\cdot \epsilon_2\right)^2,&&\\
    \mathcal{A}_3^{(1,0,0)} &= -i g_1^{(1,0,0)}\left(p_3\cdot \epsilon_1\right),&&\\
    \mathcal{A}_3^{(1,1,0)} &= g_1^{(1,1,0)}(\epsilon_1\cdot \epsilon_2)-g_2^{(1,1,0)}(p_3\cdot \epsilon_1)(p_3\cdot \epsilon_2),&&\\
    \mathcal{A}_3^{(1,2,0)} &= -ig_1^{(1,2,0)} (p_3\cdot \epsilon_2)(\epsilon_1\cdot \epsilon_2) + ig_2^{(1,2,0)}(p_3\cdot \epsilon_1)(p_3\cdot \epsilon_2)^2,&&\\
    \mathcal{A}_3^{(1,3,0)} &= -g_1^{(1,3,0)} (p_3\cdot \epsilon_2)^2(\epsilon_1\cdot \epsilon_2) + g_2^{(1,3,0)}(p_3\cdot \epsilon_1)(p_3\cdot \epsilon_2)^3.&&
\end{flalign}
\\
Parity odd interactions:
\begin{flalign}
    \tilde{\mathcal{A}}_3^{(1,1,0)} &= h_1^{(1,1,0)}\varepsilon^{p_1 p_3 \epsilon_1 \epsilon_2},&&\\
    \tilde{\mathcal{A}}_3^{(1,2,0)} &= ih_1^{(1,2,0)}(p_3\cdot \epsilon_2)\varepsilon^{p_1 p_3 \epsilon_1 \epsilon_2},&&\\
    \tilde{\mathcal{A}}_3^{(1,3,0)} &= -h_1^{(1,3,0)}(p_3\cdot \epsilon_2)^2\varepsilon^{p_1 p_3 \epsilon_1 \epsilon_2}.&&
\end{flalign}

\subsection*{Photon Absorption}

Parity even interactions:
\begin{flalign}
    \mathcal{A}_3^{(0,1,\pm 1)} &= g_1^{(0,1,1)}\left[(p_1\cdot \epsilon_3)(p_3\cdot \epsilon_2)-\frac{1}{2}(\mu^2-m_1^2)(\epsilon_2\cdot \epsilon_3)\right],&&\\
    \mathcal{A}_3^{(0,2,\pm 1)} &= -ig_1^{(0,2,1)}(p_3\cdot \epsilon_2)\left[(p_1\cdot \epsilon_3)(p_3\cdot \epsilon_2)-\frac{1}{2}(\mu^2-m_1^2)(\epsilon_2\cdot \epsilon_3)\right],&&\\
    \mathcal{A}_3^{(1,0,\pm 1)} &= g_1^{(1,0,1)}\left[(p_2\cdot \epsilon_3)(p_3\cdot \epsilon_1)+\frac{1}{2}(\mu^2+m_1^2)(\epsilon_1\cdot \epsilon_3)\right],&&\\
    \mathcal{A}_3^{(1,1,\pm 1)} &= ig_1^{(1,1,1)}\left[(p_3\cdot \epsilon_2)(\epsilon_1\cdot \epsilon_3)-(p_3\cdot \epsilon_1)(\epsilon_2\cdot \epsilon_3)\right],\nonumber&&\\
    &\hspace{5mm} +ig_2^{(1,1,1)}\left[(p_3\cdot \epsilon_2)(\epsilon_1\cdot \epsilon_3)-(p_3\cdot \epsilon_1)(\epsilon_2\cdot \epsilon_3)\right]&&\\
    \mathcal{A}_3^{(1,2,\pm 1)} &= g_1^{(1,2,1)} (p_3\cdot \epsilon_2) \left[(p_3\cdot \epsilon_2) (\epsilon_1\cdot \epsilon_3)-(p_3\cdot \epsilon_1) (\epsilon_2\cdot \epsilon_3)\right]\nonumber&&\\
    &\hspace{5mm}+g_2^{(1,2,1)} (\epsilon_1\cdot \epsilon_2) \left[(p_1\cdot \epsilon_3) (p_3\cdot \epsilon_2)-\frac{1}{2}(\mu^2-m_1^2) (\epsilon_2\cdot \epsilon_3)\right]\nonumber&&\\
    &\hspace{5mm}+g_3^{(1,2,1)} (p_3\cdot \epsilon_1) (p_3\cdot \epsilon_2) \left[\frac{1}{2}(\mu^2-m_1^2) (\epsilon_2\cdot \epsilon_3)-(p_1\cdot \epsilon_3) (p_3\cdot \epsilon_2)\right],&&\\
    \mathcal{A}_3^{(1,3,\pm 1)} &= i g_1^{(1,3,1)} (p_3\cdot \epsilon_2)^{2} \left[(p_3\cdot \epsilon_1) (\epsilon_2\cdot \epsilon_3)-(p_3\cdot \epsilon_2) (\epsilon_1\cdot \epsilon_3)\right]\nonumber&&\\
    &\hspace{5mm} -i g_2^{(1,3,1)} (p_3\cdot \epsilon_2) (\epsilon_1\cdot \epsilon_2) \left[(p_1\cdot \epsilon_3) (p_3\cdot \epsilon_2)-\frac{1}{2}(\mu^2-m_1^2) (\epsilon_2\cdot \epsilon_3)\right]\nonumber&&\\
    &\hspace{5mm} +i g_3^{(1,3,1)} (p_3\cdot \epsilon_1) (p_3\cdot \epsilon_2)^{2} \left[(p_1\cdot \epsilon_3) (p_3\cdot \epsilon_2)-\frac{1}{2}(\mu^2-m_1^2) (\epsilon_2\cdot \epsilon_3)\right].&&
\end{flalign}
\\
Parity odd interactions:
\begin{flalign}
    \tilde{\mathcal{A}}_3^{(0,1,\pm 1)} &= h_1^{(0,1,1)}\varepsilon^{p_1 p_3 \epsilon_2 \epsilon_3}, &&\\
    \tilde{\mathcal{A}}_3^{(0,2,\pm 1)} &= -ih_1^{(0,2,1)}\left(p_3\cdot \epsilon_2\right)\varepsilon^{p_1 p_3 \epsilon_2 \epsilon_3}, &&\\
    \tilde{\mathcal{A}}_3^{(1,0,\pm 1)} &= h_1^{(1,0,1)}\varepsilon^{p_2 p_3 \epsilon_1 \epsilon_3}, &&\\
    \tilde{\mathcal{A}}_3^{(1,1,\pm 1)} &= -i h_1^{(1,1,1)} \varepsilon^{p_3 \epsilon_1 \epsilon_2 \epsilon_3}+i h_2^{(1,1,1)} (p_3\cdot \epsilon_2)\varepsilon^{p_2 p_3 \epsilon_1 \epsilon_3} , &&\\
    \tilde{\mathcal{A}}_3^{(1,2,\pm 1)} &= h_1^{(1,2,1)} (\epsilon_1\cdot \epsilon_2) \varepsilon^{p_1 p_3 \epsilon_2 \epsilon_3}-h_2^{(1,2,1)} (p_3\cdot \epsilon_1) (p_3\cdot \epsilon_2) \varepsilon^{p_1 p_3 \epsilon_2 \epsilon_3}\nonumber &&\\
    &\hspace{5mm}+h_3^{(1,2,1)} (p_3\cdot \epsilon_2)^{2}\varepsilon^{p_2 p_3 \epsilon_1 \epsilon_3}, &&\\
    \tilde{\mathcal{A}}_3^{(1,3,\pm 1)} &= i h_1^{(1,3,1)} (p_3\cdot \epsilon_2)^{2} \varepsilon^{p_3\epsilon_1\epsilon_2\epsilon_3}-i h_2^{(1,3,1)} (p_3\cdot \epsilon_2) (\epsilon_1\cdot \epsilon_2) \varepsilon^{p_1 p_3 \epsilon_2 \epsilon_3}\nonumber &&\\
    &\hspace{5mm}+i h_3^{(1,3,1)} (p_3\cdot \epsilon_1) (p_3\cdot \epsilon_2)^{2} \varepsilon^{p_1 p_3 \epsilon_2 \epsilon_3}.&&
\end{flalign}

\subsection*{Graviton Absorption}

Parity even interactions:
\begin{flalign}
    \mathcal{A}_3^{(0,2,\pm 2)} &= -\frac{1}{2} g_1^{(0,2,2)} \left[(p_1\cdot \epsilon_3) (p_3\cdot \epsilon_2)-\frac{1}{2}(\mu^2-m_1^2) (\epsilon_2\cdot \epsilon_3)\right]^2, &&\\
    \mathcal{A}_3^{(1,1,\pm 2)} &= -\frac{1}{2} g_1^{(1,1,2)} \left[(p_1\cdot \epsilon_3) (p_3\cdot \epsilon_1)-\frac{1}{2}(\mu^2-m_1^2) (\epsilon_1\cdot \epsilon_3)\right] \nonumber &&\\
    &\hspace{30mm}\times\left[(p_2\cdot \epsilon_3) (p_3\cdot \epsilon_2)+\frac{1}{2}(\mu^2+m_1^2) (\epsilon_2\cdot \epsilon_3)\right], &&\\
    \mathcal{A}_3^{(1,2,\pm 2)} &= -\frac{1}{2} i g_1^{(1,2,2)} \left[(p_1\cdot \epsilon_3) (p_3\cdot \epsilon_2)-\frac{1}{2}(\mu^2-m_1^2) (\epsilon_2\cdot \epsilon_3)\right]\nonumber&&\\
    &\hspace{40mm}\times \left[(p_3\cdot \epsilon_2) (\epsilon_1\cdot \epsilon_3)-(p_3\cdot \epsilon_1) (\epsilon_2\cdot \epsilon_3)\right]\nonumber&&\\
    &\hspace{5mm}+\frac{1}{2} i g_2^{(1,2,2)} (p_3\cdot \epsilon_2) \left[(p_1\cdot \epsilon_3) (p_3\cdot \epsilon_1)-\frac{1}{2}(\mu^2-m_1^2) (\epsilon_1\cdot \epsilon_3)\right] \nonumber&&\\
    &\hspace{40mm}\times \left[(p_2\cdot \epsilon_3) (p_3\cdot \epsilon_2)+\frac{1}{2}(\mu^2+m_1^2) (\epsilon_2\cdot \epsilon_3)\right],&&\\
    \mathcal{A}_3^{(1,3,\pm 2)} &= -\frac{1}{2} g_1^{(1,3,2)} (p_3\cdot \epsilon_2) \left[(p_1\cdot \epsilon_3) (p_3\cdot \epsilon_2)-\frac{1}{2}(\mu^2-m_1^2) (\epsilon_2\cdot \epsilon_3)\right] \nonumber&&\\
    &\hspace{40mm}\times \left[(p_3\cdot \epsilon_2) (\epsilon_1\cdot \epsilon_3)-(p_3\cdot \epsilon_1) (\epsilon_2\cdot \epsilon_3)\right]\nonumber&&\\
    &\hspace{5mm}+\frac{1}{2} g_2^{(1,3,2)} (p_3\cdot \epsilon_2)^{2} \left[(p_1\cdot \epsilon_3) (p_3\cdot \epsilon_1)-\frac{1}{2}(\mu^2-m_1^2) (\epsilon_1\cdot \epsilon_3)\right] \nonumber&&\\
    &\hspace{40mm}\times \left[(p_2\cdot \epsilon_3) (p_3\cdot \epsilon_2)+\frac{1}{2}(\mu^2+m_1^2) (\epsilon_2\cdot \epsilon_3)\right]\nonumber&&\\
    &\hspace{5mm}-\frac{1}{2} g_3^{(1,3,2)} (\epsilon_1\cdot \epsilon_2) \left[(p_1\cdot \epsilon_3) (p_3\cdot \epsilon_2)-\frac{1}{2}(\mu^2-m_1^2) (\epsilon_2\cdot \epsilon_3)\right]^2.&&
\end{flalign}
\\
Parity odd interactions:
\begin{flalign}
    \tilde{\mathcal{A}}_3^{(0,2,\pm 2)} &= \frac{1}{2} h_1^{(0,2,2)} \left[(p_1\cdot \epsilon_3) (p_3\cdot \epsilon_2)-\frac{1}{2}(\mu^2-m_1^2) (\epsilon_2\cdot \epsilon_3)\right] \varepsilon^{p_1 p_3 \epsilon_2 \epsilon_3},&&\\
    \tilde{\mathcal{A}}_3^{(1,1,\pm 2)} &= -\frac{1}{2} h_1^{(1,1,2)} \left[(p_2\cdot \epsilon_3) (p_3\cdot \epsilon_2)+\frac{1}{2}(\mu^2+m_1^2) (\epsilon_2\cdot \epsilon_3)\right] \varepsilon^{p_1 p_3 \epsilon_1 \epsilon_3},&&\\
    \tilde{\mathcal{A}}_3^{(1,2,\pm 2)} &= \frac{i}{2}  h_1^{(1,2,2)} \left[(p_3\cdot \epsilon_1) (\epsilon_2\cdot \epsilon_3)-(p_3\cdot \epsilon_2) (\epsilon_1\cdot \epsilon_3)\right] \varepsilon^{p_1 p_3 \epsilon_2 \epsilon_3}\nonumber&&\\
    &\hspace{5mm}+\frac{i}{2}  h_2^{(1,2,2)} (p_3\cdot \epsilon_2) \left[(p_2\cdot \epsilon_3) (p_3\cdot \epsilon_2)+\frac{1}{2}(\mu^2+m_1^2) (\epsilon_2\cdot \epsilon_3)\right] \varepsilon^{p_1 p_3 \epsilon_1 \epsilon_3},&&\\
    \tilde{\mathcal{A}}_3^{(1,3,\pm 2)} &= \frac{i}{2} h_1^{(1,3,2)} (p_3\cdot \epsilon_2) \left[(p_3\cdot \epsilon_2) (\epsilon_1\cdot \epsilon_3)-(p_3\cdot \epsilon_1) (\epsilon_2\cdot \epsilon_3)\right] \varepsilon^{p_1 p_3 \epsilon_2 \epsilon_3}\nonumber&&\\
    &\hspace{5mm}+\frac{i}{2} h_2^{(1,3,2)} (\epsilon_1\cdot \epsilon_2) \left[(p_1\cdot \epsilon_3) (p_3\cdot \epsilon_2)-\frac{1}{2}(\mu^2-m_1^2) (\epsilon_2\cdot \epsilon_3)\right] \varepsilon^{p_1 p_3 \epsilon_2 \epsilon_3}\nonumber&&\\
    &\hspace{5mm}-\frac{i}{2} h_3^{(1,3,2)} (p_3\cdot \epsilon_2)^{2} \left[(p_2\cdot \epsilon_3) (p_3\cdot \epsilon_2)+\frac{1}{2}(\mu^2+m_1^2) (\epsilon_2\cdot \epsilon_3)\right] \varepsilon^{p_1 p_3 \epsilon_1 \epsilon_3}.&&
\end{flalign}

\section{Impulse Results}
\label{app:results} 

Below we give the Casimir interpolated impulse coefficients up to $\mathcal{O}(S_1^2)$ in terms of the Wilson coefficients appearing in the 3-point amplitudes constructed in Appendix \ref{app:3ptAmps}; coefficients which are not explicitly enumerated are zero. 
We organize each set of coefficients first by massless mediator and then by exchange channel, labeled as $\left(s, s', |h|\right)$. The definition of the coefficients is given in (\ref{eq:impform}) and (\ref{eq:cdef}).

\subsection*{Scalar Absorption}
\label{subsec:scalarabsresults}

\begin{itemize}[leftmargin= 0 pt]
    \item \scalebox{1.1}{$\left(0, 0, 0\right):$}
    \begin{flalign}
    & (c_u)^{(0, 0, 0)}_{0,0,0} = -\frac{g^2}{2^{11}  m_1^2 m_2^2 }\frac{ \left(g_1^{(0, 0, 0)}\right)^2}{ (-b^2)  }. &&
\end{flalign}
\end{itemize}

\begin{itemize}[leftmargin= 0 pt]
    \item \scalebox{1.1}{$\left(0, 1, 0\right):$}
    \begin{flalign}
    & (c_u)^{(0, 1, 0)}_{0,0,0} = - \frac{9 g^2}{2^{15}  m_1^2 m_2^2 }\frac{ (3 y^2 + 5) \,  \left(g_1^{(0, 1, 0)}\right)^2}{ (-b^2)^2  }. &&
\end{flalign}
\end{itemize}

\begin{itemize}[leftmargin= 0 pt]
    \item \scalebox{1.1}{$\left(0, 2, 0\right):$}
    \begin{flalign}
     & (c_u)^{(0, 2, 0)}_{0,0,0} = - \frac{75 g^2}{2^{17}  m_1^2 m_2^2 }\frac{ (5 y^4 + 26 y^2 + 17) \,  \left(g_1^{(0, 2, 0)}\right)^2}{ (-b^2)^3  }. &&
\end{flalign}
\end{itemize}

\begin{itemize}[leftmargin= 0 pt]
    \item \scalebox{1.1}{$\left(1, 0, 0\right):$}
    \begin{flalign}
    & (c_u)^{(1, 0, 0)}_{2,0,0} = - \frac{45 g^2}{2^{15}  m_1^2 m_2^2 }\frac{ \left(g_1^{(1, 0, 0)}\right)^2}{ (-b^2)^3  }, &&\\
    & (c_u)^{(1, 0, 0)}_{0,0,2} = -\frac{27 g^2}{2^{15}  (y^2 - 1) m_1^2 m_2^2 }\frac{ y^2 \,  \left(g_1^{(1, 0, 0)}\right)^2}{ (-b^2)^3  }, &&\\
    & (c_b)^{(1, 0, 0)}_{1,0,1} = \frac{3 g^2}{2^{14}  (y^2 - 1) m_1^2 m_2^2 }\frac{ y \,  \left(g_1^{(1, 0, 0)}\right)^2}{ (-b^2)^3  }, &&\\
    & (c_n)^{(1, 0, 0)}_{0,1,1} = -\frac{4}{ (y^2 - 1)}(c_b)^{(1, 0, 0)}_{1,0,1}.&&
\end{flalign}
\end{itemize}

\begin{itemize}[leftmargin= 0 pt]
    \item \scalebox{1.1}{$\left(1, 1, 0\right):$}\\
    \\
    For this transition there is a particular combination of Wilson coefficients,
\begin{equation}
    \left(\Tilde{g}_2^{(1, 1, 0)}\right)^2 :=  \left(g_1^{(1, 1, 0)}\right)^2 +  2 m_1^2\left(g_1^{(1, 1, 0)}\right) \left(g_2^{(1, 1, 0)}\right),
\end{equation}
that is more natural for the impulse.
\begin{flalign}
    & (c_u)^{(1, 1, 0)}_{0,0,0} =  \frac{g^2}{2^{11}  m_1^2 m_2^2 }\frac{ \left(g_1^{(1, 1, 0)}\right)^2}{ \sqrt{-b^2}  }, && \\
    & (c_u)^{(1, 1, 0)}_{1,0,0} = \frac{3 g^2}{2^{11}  m_1 m_2^2 }\frac{  \left(g_1^{(1, 1, 0)} h_1^{(1, 1, 0)}\right)}{ (-b^2)^2  }, && \\
    & (c_u)^{(1, 1, 0)}_{2,0,0} = -\frac{3 g^2}{2^{16}  m_1^4 m_2^2 }\frac{ (9 y^2 - 43) \,  \left(\Tilde{g}_2^{(1, 1, 0)}\right)^2 - 30 m_1^4 \left(h_1^{(1, 1, 0)}\right)^2 }{ (-b^2)^3  }, && \\
    & (c_u)^{(1, 1, 0)}_{0,2,0} = -\frac{3 g^2}{2^{16} (y^2 - 1) m_1^4 m_2^2 }\frac{ (9 y^2 + 7) \,  \left(\Tilde{g}_2^{(1, 1, 0)}\right)^2  }{ (-b^2)^3  }, && \\
    & (c_u)^{(1, 1, 0)}_{0,0,2} =  -\frac{27 g^2}{2^{16} (y^2 - 1) m_1^4 m_2^2 }\frac{ (3 y^2 -1) \,  \left(\Tilde{g}_2^{(1, 1, 0)}\right)^2 - 2 y^2 m_1^4 \left(h_1^{(1, 1, 0)}\right)^2 }{ (-b^2)^3  }, && \\
    & (c_b)^{(1, 1, 0)}_{1,0,1} =  \frac{3 y g^2}{2^{14} (y^2 - 1) m_1^4 m_2^2 }\frac{ \left(\Tilde{g}_2^{(1, 1, 0)}\right)^2 - m_1^4 \left(h_1^{(1, 1, 0)}\right)^2 }{ (-b^2)^3  }, && \\
    & (c_n)^{(1, 1, 0)}_{0,1,1} = - \frac{4}{ (y^2 - 1)} (c_b)^{(1, 1, 0)}_{1,0,1}. &&
\end{flalign}
\end{itemize}

\begin{itemize}[leftmargin = 0pt]
    \item \scalebox{1.1}{$\left(1, 2, 0\right):$}
    \begin{flalign}
    & (c_u)^{(1, 2, 0)}_{2,0,0} = -\frac{9 g^2}{2^{16}  m_1^2 m_2^2 }\frac{ (3 y^2 + 5) \,  \left(g_1^{(1, 2, 0)}\right)^2 + (2 y^2 +5) \,  \left(g_1^{(0, 1, 0)}\right)^2}{ (-b^2)^3  }, && \\
    & (c_u)^{(1, 2, 0)}_{0,2,0} = - \frac{3 g^2 (3 y^2+5)}{2^{16}  (y^2 - 1) m_1^2 m_2^2 }\frac{ 2 \,  \left(g_1^{(1, 2, 0)}\right)^2 + 3 \,  \left(g_1^{(0, 1, 0)}\right)^2}{ (-b^2)^3  }, && \\
    & (c_u)^{(1, 2, 0)}_{0,0,2} = -\frac{3 g^2}{2^{16}  (y^2 - 1) m_1^2 m_2^2 }\frac{  (9 y^2 + 10) \,  \left(g_1^{(1, 2, 0)}\right)^2 + 3 (3 y^2 + 5) \,  \left(g_1^{(0, 1, 0)}\right)^2}{ (-b^2)^3  }, && \\
    & (c_b)^{(1, 2, 0)}_{1,0,1} =  \frac{y g^2}{2^{15}  (y^2 - 1) m_1^2 m_2^2 }\frac{ \left(g_1^{(1, 2, 0)}\right)^2}{ (-b^2)^3  }, && \\
    & (c_n)^{(1, 2, 0)}_{0,1,1} = - \frac{4}{ (y^2 - 1)}  (c_b)^{(1, 2, 0)}_{1,0,1}. &&
\end{flalign}
For equivalence between (1,0,0) and (1,2,0) there exists the map of Wilson coefficients
    \begin{equation}
        g_1^{(1, 0, 0)} \rightarrow 1/6\,  g_1^{(1, 2, 0)}, \hspace{2cm}
        \left(g_1^{(0, 1, 0)} \right)^2 \rightarrow  - 2/3 \left( g_1^{(1, 2, 0)}\right)^2,
    \end{equation}
    which is determined by spin universality and floor-ceiling symmetry.
\end{itemize}

\begin{itemize}[leftmargin = 0pt]
    \item \scalebox{1.1}{$\left(1, 3, 0\right):$}
    \begin{flalign}
    & (c_u)^{(1, 3, 0)}_{2,0,0} = \frac{3 g^2}{2^{18}  m_1^2 m_2^2 }\frac{ (75 y^4 + 152 y^2 + 101) \,  \left(g_1^{(1, 3, 0)}\right)^2 +25(5 y^4 + 26 y^2 + 17) \,  \left(g_1^{(0, 2, 0)}\right)^2}{ (-b^2)^4  }, && \\
    & (c_u)^{(1, 3, 0)}_{0,2,0} =  \frac{3 g^2}{2^{18} (y^2 - 1)  m_1^2 m_2^2 }\frac{ (75 y^4 + 558 y^2 + 199) \,  \left(g_1^{(1, 3, 0)}\right)^2 +25(5 y^4 + 26 y^2 + 17) \,  \left(g_1^{(0, 2, 0)}\right)^2}{ (-b^2)^4  }, && \\
    & (c_u)^{(1, 3, 0)}_{0,0,2} = \frac{15 g^2}{2^{18} (y^2 - 1)  m_1^2 m_2^2 }\frac{ (5 y^2(y^2 + 8) + 59) \,  \left(g_1^{(1, 3, 0)}\right)^2 + 5(5 y^4 + 26 y^2 + 17) \,  \left(g_1^{(0, 2, 0)}\right)^2}{ (-b^2)^4  }, && \\
    & (c_b)^{(1, 3, 0)}_{1,0,1} =  \frac{3 y g^2}{2^{16} (y^2 - 1)  m_1^2 m_2^2 }\frac{ ( y^2 + 11) \,  \left(g_1^{(1, 3, 0)}\right)^2 }{ (-b^2)^4  }, && \\
    & (c_n)^{(1, 3, 0)}_{0,1,1} =   - \frac{6}{ (y^2 - 1)} (c_b)^{(1, 3, 0)}_{1,0,1}. &&
\end{flalign}
\end{itemize}

\subsection*{Photon Absorption}

\begin{itemize}[leftmargin = 0pt]
    \item \scalebox{1.1}{$\left(0, 1, 1\right):$}
    \begin{flalign}
         & (c_u)^{(0, 1, 1)}_{0,0,0} = - \frac{9 e^2}{2^{13}  }\frac{ (5 y^2 + 3) \, \left(g_1^{(0,1, 1)}\right)^2 + 5 ( y^2 - 1) \, \left(h_1^{(0,1, 1)}\right)^2}{ (-b^2)^2  }. &&
    \end{flalign}
\end{itemize}

\begin{itemize}[leftmargin = 0pt]
    \item \scalebox{1.1}{$\left(0, 2, 1\right):$}
    \begin{flalign}
    & (c_u)^{(0, 2, 1)}_{0,0,0} =  -\frac{225 e^2}{2^{17}  m_1^2 m_2^2 }\frac{ (7 y^4 + 50 y^2 + 7) \, \left(g_1^{(0, 2, 1)}\right)^2 +  7(  y^4 + 2 y^2 -3) \, \left(h_1^{(0, 2, 1)}\right)^2}{ (-b^2)^3 }. && 
\end{flalign}
\end{itemize}
 

\begin{itemize}[leftmargin = 0pt]
    \item \scalebox{1.1}{$\left(1, 0, 1\right):$}
    \begin{flalign}
    & (c_u)^{(1, 0, 1)}_{2,0,0} = -\frac{45 e^2}{2^{13}  }\frac{ y^2 \left(g_1^{(1, 0, 1)}\right)^2 - (y^2-1) \left(h_1^{(1, 0, 1)}\right)^2}{ (-b^2)^3  }, && \\
    & (c_u)^{(1, 0, 1)}_{1,1,0} = \frac{45 y e^2}{2^{12}  }\frac{  \left(g_1^{(1, 0, 1)}\right) \left(h_1^{(1, 0, 1)}\right)}{ (-b^2)^3  }, && \\
    & (c_u)^{(1, 0, 1)}_{0,0,2} =  - \frac{9 e^2}{2^{13} (y^2 - 1)  }\frac{ 3 \left(g_1^{(1, 0, 1)}\right)^2 - 5 (y^2 - 1) \left(h_1^{(1, 0, 1)}\right)^2}{ (-b^2)^3  }, && \\
    & (c_b)^{(1, 0, 1)}_{1,0,1} =  \frac{3 y e^2}{2^{12} (y^2 - 1)  }\frac{  \left(g_1^{(1, 0, 1)}\right)^2}{ (-b^2)^3  }, && \\
    & (c_b)^{(1, 0, 1)}_{0,1,1} = -\frac{3  e^2}{2^{12} (y^2 - 1)  }\frac{ \left(g_1^{(1, 0, 1)}\right) \left(h_1^{(1, 0, 1)}\right)}{ (-b^2)^3  }, && \\
    & (c_n)^{(1, 0, 1)}_{1,0,1} = 4 \, (c_b)^{(1, 0, 1)}_{0,1,1}, && \\
    & (c_n)^{(1, 0, 1)}_{0,1,1} = - \frac{4}{ (y^2 - 1)} (c_b)^{(1, 0, 1)}_{1,0,1}. &&
\end{flalign}
\end{itemize}


\begin{itemize}[leftmargin = 0pt]
    \item \scalebox{1.1}{$\left(1, 1, 1\right):$}
    \begin{flalign}
        & (c_u)^{(1, 1, 1)}_{2,0,0} = -\frac{45 e^2}{2^{13}  m_1^2}\frac{ ( y^2 - 1) \,  \left(g_1^{(1, 1, 1)}\right)^2 - y^2 \left(h_1^{(1, 1, 1)}\right)^2 }{ (-b^2)^3  }, && \\
        & (c_u)^{(1, 1, 1)}_{1,1,0} = \frac{45 y e^2}{2^{12}  m_1^2}\frac{   \left(g_1^{(1, 1, 1)}\right) \left(h_1^{(1, 1, 1)}\right) }{ (-b^2)^3  }, && \\
        & (c_u)^{(1, 1, 1)}_{0,0,2} =  -\frac{9 e^2}{2^{13} (y^2 - 1) m_1^2}\frac{ 5 (y^2 - 1 ) \,  \left(g_1^{(1, 1, 1)}\right)^2 - 3 \left(h_1^{(1, 1, 1)}\right)^2 }{ (-b^2)^3  }, && \\
        & (c_b)^{(1, 1, 1)}_{1,0,1} =   -\frac{3 y e^2}{2^{12} (y^2 - 1) m_1^2}\frac{  \left(h_1^{(1, 1, 1)}\right)^2 }{ (-b^2)^3  }, && \\
        & (c_b)^{(1, 1, 1)}_{0,1,1} = -\frac{3 e^2}{2^{12}  m_1^2}\frac{  \left(g_1^{(1, 1, 1)}\right) \left(h_1^{(1, 1, 1)}\right) }{ (-b^2)^3  }, && \\
        & (c_n)^{(1, 1, 1)}_{1,0,1} = 4 \, (c_b)^{(1, 1, 1)}_{0,1,1},  && \\
        & (c_n)^{(1, 1, 1)}_{0,1,1} = - \frac{4}{ (y^2 - 1)} \, (c_b)^{(1, 1, 1)}_{1,0,1}. &&
    \end{flalign}
\end{itemize}

\begin{itemize}[leftmargin = 0pt]
    \item \scalebox{1.1}{$\left(1, 2, 1\right):$}
    \begin{flalign}
    & (c_u)^{(1, 2, 1)}_{2,0,0} =  -\frac{9 e^2}{2^{14}  } \frac{1}{ (-b^2)^3  }\left( (5 y^2 + 2) \,  \left(g_1^{(1, 2, 1)}\right)^2 + (5 y^2 + 3) \,  \left(g_1^{(0, 1, 1)}\right)^2  \right. && \\
    & \left. \hspace{4cm} + 5 (y^2 - 1) \,  \left(h_1^{(0, 1, 1)}\right)^2\right), && \\
    & (c_u)^{(1, 2, 1)}_{1,1,0} =  \frac{15 y e^2}{2^{13}  }\frac{ \left(g_1^{(1, 2, 1)}\right)^2 \left(h_1^{(1, 2, 1)}\right)^2}{ (-b^2)^3  },  && \\
    & (c_u)^{(1, 2, 1)}_{0,2,0} =  - \frac{3 e^2}{2^{14} (y^2 - 1) } \frac{1}{(-b^2)^3}\left( 2 (5 y^2 + 3) \,  \left(g_1^{(1, 2, 1)}\right)^2 + 3(5 y^2 + 3) \,  \left(g_1^{(0, 1, 1)}\right)^2  \right. && \\
    & \left. \hspace{5cm} + 5 (y^2 - 1) \,  \left(h_1^{(1, 2, 1)}\right)^2 + 15 (y^2 - 1) \,  \left(h_1^{(0, 1, 1)}\right)^2\right), && \\
    & (c_u)^{(1, 2, 1)}_{0,0,2} =  - \frac{3 e^2}{2^{14} (y^2 - 1) } \frac{1}{ (-b^2)^3  }\left( (10 y^2 + 9) \,  \left(g_1^{(1, 2, 1)}\right)^2 + 3(5 y^2 + 3) \,  \left(g_1^{(0, 1, 1)}\right)^2 \right. && \\
    & \left. \hspace{5cm} + 15 (y^2 - 1) \,  \left(h_1^{(0, 1, 1)}\right)^2\right), && \\
    & (c_b)^{(1, 2, 1)}_{1,0,1} =  \frac{y e^2}{2^{13} (y^2 - 1) }\frac{   \left(g_1^{(1, 2, 1)}\right)^2  }{ (-b^2)^3  }, && \\
    & (c_b)^{(1, 2, 1)}_{0,1,1} =- \frac{ y e^2}{2^{13} (y^2 - 1) }\frac{ \left(g_1^{(1, 2, 1)}\right)\left(h_1^{(1, 2, 1)}\right) }{ (-b^2)^3  }, &&  \\
    & (c_n)^{(1, 2, 1)}_{1,0,1} = 4\,  (c_b)^{(1, 2, 1)}_{0,1,1}, && \\
    & (c_n)^{(1, 2, 1)}_{0,1,1} = -\frac{4}{ (y^2 - 1)} (c_b)^{(1, 2, 1)}_{1,0,1}. &&
\end{flalign}
For equivalence between (1,0,1) and (1,2,1) there exists the map of Wilson coefficients
    \begin{align}
        &\left(g_1^{(1, 0, 1)}\right) \rightarrow 1/\sqrt{6} \left(g_1^{(1, 2, 1)}\right), \hspace{2cm}\left(g_1^{(0, 1, 1)}\right)^2 \rightarrow -2/3 \left(g_1^{(1, 2, 1)}\right)^2, \nonumber\\
        \\
         &\left(h_1^{(1, 0, 1)}\right) \rightarrow 1/\sqrt{6} \left(h_1^{(1, 2, 1)}\right), \hspace{2cm} \left(h_1^{(0, 1, 1)}\right)^2 \rightarrow -1/3 \left(h_1^{(1, 2, 1)}\right)^2 , \nonumber
    \end{align}
    which is determined by spin universality and floor-ceiling symmetry.
\end{itemize}

\begin{itemize}[leftmargin = 0pt]
    \item \scalebox{1.1}{$\left(1, 3, 1\right):$}
    \begin{flalign}
    & (c_u)^{(1, 3, 1)}_{2,0,0} = \frac{3 e^2}{2^{18}  } \frac{1}{(-b^2)^4} \left(  (175 y^4 + 962 y^2 + 175) \,  \left(g_1^{(1, 3, 1)}\right)^2 + 75( 7 y^2 +1)(y^2 + 1) \,  \left(g_1^{(0, 2, 1)}\right)^2 \right. \nonumber && \\
    & \left. \hspace{3cm} + \,   \,  35( 11 y^4 + 4 y^2 -15)\left( h_1^{(1, 3, 1)}\right)^2 + 525 (y^4 + 2 y^2 - 3) \,  \left(h_1^{(0, 2, 1)}\right)^2  \right), && \\
    & (c_u)^{(1, 3, 1)}_{1,1,0} =    \frac{63 y e^2}{2^{16}  } \frac{ (5 y^2 + 19) \,  \left(g_1^{(1, 3, 1)}\right)\,  \left(h_1^{(1, 3, 1)}\right) }{ (-b^2)^4  }, && \\
    & (c_u)^{(1, 3, 1)}_{0,2,0} =  \frac{3 e^2}{2^{18} (y^2 - 1) } \frac{1}{ (-b^2)^4  } \nonumber && \\
    & \hspace{2cm} \left(  (385 y^4 + 2558 y^2 + 385) \,  \left(g_1^{(1, 3, 1)}\right)^2 + 75(7 y^2 + 1) (y^2 + 7) \,  \left(g_1^{(0, 2, 1)}\right)^2  \right. \nonumber && \\
    & \left. \hspace{3cm} + 175 (y^4 + 2 y^2 - 3) \left( \,  \left(h_1^{(1, 3, 1)}\right)^2 + 3 \,  \left(h_1^{(0, 2, 1)}\right)^2\right) \right), && \\
    & (c_u)^{(1, 3, 1)}_{0,0,2} =  \frac{15 e^2}{2^{18} (y^2 - 1) } \frac{1}{ (-b^2)^4  } \nonumber && \\
    & \hspace{2cm} \left( (35 y^4 + 346 y^2  + 35) \,  \left(g_1^{(1, 3, 1)}\right)^2 + 15(7 y^2 + 1)(y^2 + 7) \,  \left(g_1^{(0, 2, 1)}\right)^2  \right.  \nonumber && \\
    & \left. \hspace{3cm} + 7 (5 y^4 + 28 y^2 -33) \,  \left(h_1^{(1, 3, 1)}\right)^2 + 105 (y^4 + 2 y^2 -3) \,  \left(h_1^{(0, 2, 1)}\right)^2\right), && \\
    & (c_b)^{(1, 3, 1)}_{1,0,1} =  \frac{9  y e^2}{2^{16} (y^2 - 1) }\frac{ 8 (y^2 + 1) \,  \left(g_1^{(1, 3, 1)}\right)^2  + 7 (y^2 - 1) \,  \left(h_1^{(1, 3, 1)}\right)^2 }{ (-b^2)^4  }, && \\
    & (c_b)^{(1, 3, 1)}_{0,1,1} =   - \frac{9   e^2}{2^{16} (y^2 - 1) }\frac{ ( y^2 + 7) \,  \left(g_1^{(1, 3, 1)}\right)\left(h_1^{(1, 3, 1)}\right) }{ (-b^2)^4  }, && \\
    & (c_n)^{(1, 3, 1)}_{1,0,1} =  6\, (c_b)^{(1, 3, 1)}_{0,1,1},  && \\
    & (c_n)^{(1, 3, 1)}_{0,1,1} = -\frac{6}{ (y^2 - 1)} (c_b)^{(1, 3, 1)}_{1,0,1}.  && 
\end{flalign}
\end{itemize}

\subsection*{Graviton Absorption}

\begin{itemize}[leftmargin = 0pt]
    \item \scalebox{1.1}{$\left(0, 2, 2\right):$}
    \begin{flalign}
    & (c_u)^{(0, 2, 2)}_{0,0,0} =  - \frac{225 \kappa^2 m_1^2 m_2^2 }{2^{21} } \,\,\,\frac{\left(g_{1}^{(0, 2, 2)}\right)^2 \left(21 y^4 - 14 y^2 + 9\right) + 7 \left(h_{1}^{(0, 2, 2)}\right)^2 \left(3 y^4 - 2 y^2 -1 \right)}{\left(-b^2\right)^3}. &&
\end{flalign}
\end{itemize}

\begin{itemize}[leftmargin = 0pt]
    \item \scalebox{1.1}{$\left(1, 1, 2\right):$}
    \begin{flalign}
    & (c_u)^{(1, 1, 2)}_{2,0,0} =   -\frac{45 \kappa^2 m_1^2 m_2^2 }{2^{22} } \,\, \frac{\left(g_{1}^{(1, 1, 2)}\right)^2 \left(80 y^4 + 29 y^2 - 17\right) - 5 \left(h_{1}^{(1, 1, 2)}\right)^2 \left(16 y^4 - 9 y^2 -7 \right)}{\left(-b^2\right)^4},  && \\
    & (c_u)^{(1, 1, 2)}_{1,1,0} = \frac{945 y \kappa^2 m_1^2 m_2^2 }{2^{19} } \,\, \frac{\left(g_{1}^{(1, 1, 2)}\right)  \left(h_{1}^{(1, 1, 2)}\right) }{\left(-b^2\right)^4}, && \\
    & (c_u)^{(1, 1, 2)}_{0,2,0} = -\frac{45 \kappa^2 m_1^2 m_2^2 }{2^{19} (y^2 - 1) } \,\, \frac{\left(g_{1}^{(1, 1, 2)}\right)^2 \left(10 y^4 - 13 y^2 + 4\right) - 10 y^2 \left(h_{1}^{(1, 1, 2)}\right)^2 \left(y^2 -1 \right)}{\left(-b^2\right)^4}, && \\
    & (c_u)^{(1, 1, 2)}_{0,0,2} =  \frac{225 y^2  \kappa^2 m_1^2 m_2^2 }{2^{22} (y^2 - 1) } \,\, \frac{\left(g_{1}^{(1, 1, 2)}\right)^2 \left( 5 y^2 - 17\right) +37 \left(h_{1}^{(1, 1, 2)}\right)^2 \left( y^2 - 1 \right)}{\left(-b^2\right)^4}, && \\
    & (c_b)^{(1, 1, 2)}_{1,0,1} =  \frac{45 y  \kappa^2 m_1^2 m_2^2 }{2^{21} (y^2 - 1) } \,\, \frac{\left(g_{1}^{(1, 1, 2)}\right)^2 \left( 7 y^2 - 3\right) + 7 \left(h_{1}^{(1, 1, 2)}\right)^2 \left( y^2 - 1 \right)}{\left(-b^2\right)^4}, && \\
    & (c_b)^{(1, 1, 2)}_{0,1,1} = -\frac{45  \kappa^2 m_1^2 m_2^2 }{2^{19} (y^2 - 1)} \,\, \frac{\left(g_{1}^{(1, 1, 2)}\right)  \left(h_{1}^{(1, 1, 2)}\right) }{\left(-b^2\right)^4}, && \\
    & (c_n)^{(1, 1, 2)}_{1,0,1} = 6  (c_b)^{(1, 1, 2)}_{0,1,1}, && \\
    & (c_n)^{(1, 1, 2)}_{0,1,1} =  -\frac{6}{ (y^2 - 1)}  (c_b)^{(1, 1, 2)}_{1,0,1}. &&
\end{flalign}
\end{itemize}

\begin{itemize}[leftmargin = 0pt]
    \item \scalebox{1.1}{$\left(1, 2, 2\right):$}
    \begin{flalign}
    & (c_u)^{(1, 2, 2)}_{2,0,0} = -\frac{45 \kappa^2 m_2^2 }{2^{23} } \,\, \frac{ 5 \left(g_{1}^{(1, 2, 2)}\right)^2 \left(16 y^4 + 5 y^2 -21\right) + \left(h_{1}^{(1, 2, 2)}\right)^2 \left(80 y^4 - 407 y^2 +51 \right)}{\left(-b^2\right)^4}, && \\
    & (c_u)^{(1, 2, 2)}_{1,1,0} = \frac{2835 y \kappa^2  m_2^2 }{2^{20} } \,\, \frac{\left(g_{1}^{(1, 2, 2)}\right)  \left(h_{1}^{(1, 2, 2)}\right) }{\left(-b^2\right)^4}, && \\
    & (c_u)^{(1, 2, 2)}_{0,2,0} = -\frac{45 \kappa^2 m_2^2 }{2^{20} (y^2 - 1) } \,\, \frac{ 10 y^2  \left(g_{1}^{(1, 2, 2)}\right)^2 \left(y^2 - 1\right) + \left(h_{1}^{(1, 2, 2)}\right)^2 \left(10 y^4 -  y^2 -12 \right)}{\left(-b^2\right)^4}, && \\
    & (c_u)^{(1, 2, 2)}_{0,0,2} =  -\frac{225 y^2  \kappa^2 m_2^2 }{2^{23} (y^2 - 1) } \,\, \frac{ 79 \left(g_{1}^{(1, 2, 2)}\right)^2 \left( y^2 - 1\right) + \left(h_{1}^{(1, 2, 2)}\right)^2 \left(79 y^2 - 115 \right)}{\left(-b^2\right)^4}, && \\
    & (c_b)^{(1, 2, 2)}_{1,0,1} =   -\frac{135 y   \kappa^2 m_2^2 }{2^{22} (y^2 - 1) } \,\, \frac{ 7 \left(g_{1}^{(1, 2, 2)}\right)^2 \left( y^2 - 1\right) + \left(h_{1}^{(1, 2, 2)}\right)^2 \left(7 y^2 - 3 \right)}{\left(-b^2\right)^4}, && \\
    & (c_b)^{(1, 2, 2)}_{0,1,1} = -\frac{135  \kappa^2  m_2^2 }{2^{20} (y^2 - 1)} \,\, \frac{\left(g_{1}^{(1, 2, 2)}\right)  \left(h_{1}^{(1, 2, 2)}\right) }{\left(-b^2\right)^4}, && \\
    & (c_n)^{(1, 2, 2)}_{1,0,1} = 6 (c_b)^{(1, 2, 2)}_{0,1,1}, && \\
    & (c_n)^{(1, 2, 2)}_{0,1,1} = - \frac{6}{ (y^2 - 1)} (c_b)^{(1, 2, 2)}_{1,0,1}. &&
\end{flalign}
\end{itemize}

\begin{itemize}[leftmargin = 0 pt]
    \item \scalebox{1.1}{$\left(1, 3, 2\right):$}
    \begin{flalign}
    & (c_u)^{(1, 3, 2)}_{2,0,0} = \frac{3 \kappa^2 m_1^2 m_2^2}{2^{22}} \frac{1}{(-b^2)^4}\nonumber \\ & \hspace{1.7cm}\left(\left(g_{1}^{(1, 3, 2)}\right)^2 \left(1485 y^4- 296 y^2 + 243 \right) + 75 \left(g_{1}^{(0, 2, 2)}\right)^2 \left(21 y^4- 14 y^2  + 9 \right) \right.  && \nonumber \\
    & \left. \hspace{2.7cm} + \, 5 (y^2 - 1) \left( \left(h_{1}^{(1, 3, 2)}\right)^2 \left(201 y^2 - 7 \right) +  105 \left(h_{1}^{(0, 2, 2)}\right)^2 \left(3 y^2 + 1 \right) \right)\right), && \\
    & (c_u)^{(1, 3, 2)}_{1,1,0} = \frac{189  \kappa^2  y m_1^2 m_2^2  }{2^{18} (y^2 - 1)}  \left(\frac{\left(g_{1}^{(1, 3, 2)}\right)  \left(h_{1}^{(1, 3, 2)}\right) }{\left(-b^2\right)^4}\right), &&   \\
    & (c_u)^{(1, 3, 2)}_{0,2,0} = \frac{3 \kappa^2 m_1^2 m_2^2}{2^{22} (y^2 - 1)} \frac{1}{(-b^2)^4} \nonumber \\
    & \hspace{1.5 cm}\left(\left(g_{1}^{(1, 3, 2)}\right)^2 \left(1485 y^4- 1094 y^2 + 537 \right) + 75 \left(g_{1}^{(0, 2, 2)}\right)^2 \left(21 y^4- 14 y^2 + 9 \right) \right. &&\nonumber \\
    & \left. \hspace{2.5cm} + \, 5 (y^2 - 1) \left( \left(h_{1}^{(1, 3, 2)}\right)^2 \left(201 y^2 + 35 \right) + 105 \left(h_{1}^{(0, 2, 2)}\right)^2 \left(3 y^2 + 1 \right) \right)\right), && \\
    & (c_u)^{(1, 3, 2)}_{0,0,2} =   \frac{15 \kappa^2 m_1^2 m_2^2}{2^{22} (y^2 - 1)} \frac{1}{(-b^2)^4} \nonumber && \\ & \hspace{1.5 cm}\left(\left(g_{1}^{(1, 3, 2)}\right)^2 \left(171 y^4 + 8 y^2 + 69 \right) + 15 \left(g_{1}^{(0, 2, 2)}\right)^2 \left(21 y^4- 14 y^2 + 9 \right) \right. &&\nonumber \\
    & \left. \hspace{3cm} + 5 (y^2 - 1)\left(\left(h_{1}^{(1, 3, 2)}\right)^2 \left(15 y^2 + 7 \right) + 21 \left(h_{1}^{(0, 2, 2)}\right)^2 \left(3 y^2 + 1 \right)\right) \right), && \\
    & (c_b)^{(1, 3, 2)}_{1,0,1} =  -\frac{9 y   \kappa^2 m_1^2 m_2^2 }{2^{20} (y^2 - 1) } \,\, \frac{  \left(g_{1}^{(1, 3, 2)}\right)^2 \left(7 y^2 - 3 \right) + 7 \left(h_{1}^{(1, 3, 2)}\right)^2 \left( y^2 -1 \right)}{\left(-b^2\right)^4}, && \\
    & (c_b)^{(1, 3, 2)}_{0,1,1} = -\frac{9  \kappa^2  m_1^2 m_2^2  }{2^{18} (y^2 - 1)} \,\, \frac{\left(g_{1}^{(1, 2, 2)}\right)  \left(h_{1}^{(1, 3, 2)}\right) }{\left(-b^2\right)^4}, && \\
    & (c_n)^{(1, 3, 2)}_{1,0,1} =  6 (c_b)^{(1, 3, 2)}_{0,1,1}, && \\
    & (c_n)^{(1, 3, 2)}_{0,1,1} =  -\frac{6}{ (y^2 - 1)} (c_b)^{(1, 3, 2)}_{1,0,1}. &&
\end{flalign}
\end{itemize}
\endgroup


\bibliographystyle{utphys}
\bibliography{refs.bib}

\providecommand{\href}[2]{#2}\begingroup\raggedright\begin{thebibliography}{10}

\bibitem{Einstein:1938yz}
A.~Einstein, L.~Infeld, and B.~Hoffmann, ``{The Gravitational equations and the problem of motion},'' \href{http://dx.doi.org/10.2307/1968714}{{\em Annals Math.} {\bfseries 39} (1938) 65--100}.

\bibitem{Ohta:1973je}
T.~Ohta, H.~Okamura, T.~Kimura, and K.~Hiida, ``{Physically acceptable solution of einstein's equation for many-body system},'' \href{http://dx.doi.org/10.1143/PTP.50.492}{{\em Prog. Theor. Phys.} {\bfseries 50} (1973) 492--514}.

\bibitem{Goldberger:2004jt}
W.~D. Goldberger and I.~Z. Rothstein, ``{An Effective field theory of gravity for extended objects},'' \href{http://dx.doi.org/10.1103/PhysRevD.73.104029}{{\em Phys. Rev. D} {\bfseries 73} (2006) 104029}, \href{http://arxiv.org/abs/hep-th/0409156}{{\ttfamily arXiv:hep-th/0409156}}.

\bibitem{Bertotti:1956pxu}
B.~Bertotti, ``{On gravitational motion},'' \href{http://dx.doi.org/10.1007/bf02746175}{{\em Nuovo Cim.} {\bfseries 4} no.~4, (1956) 898--906}.

\bibitem{Kerr:1959zlt}
R.~P. Kerr, ``{The Lorentz-covariant approximation method in general relativity I},'' \href{http://dx.doi.org/10.1007/bf02732767}{{\em Nuovo Cim.} {\bfseries 13} no.~3, (1959) 469--491}.

\bibitem{Bertotti:1960wuq}
B.~Bertotti and J.~Plebanski, ``{Theory of gravitational perturbations in the fast motion approximation},'' \href{http://dx.doi.org/10.1016/0003-4916(60)90132-9}{{\em Annals Phys.} {\bfseries 11} no.~2, (1960) 169--200}.

\bibitem{Westpfahl:1979gu}
K.~Westpfahl and M.~Goller, ``{Gravitational Scattering of Two Relativistic Particles in Post-Linear Approximation},'' \href{http://dx.doi.org/10.1007/BF02817047}{{\em Lett. Nuovo Cim.} {\bfseries 26} (1979) 573--576}.

\bibitem{Portilla:1980uz}
M.~Portilla, ``{Scattering of Two Gravitating Particles: Classical Approach},'' \href{http://dx.doi.org/10.1088/0305-4470/13/12/017}{{\em J. Phys. A} {\bfseries 13} (1980) 3677--3683}.

\bibitem{Bel:1981be}
L.~Bel, T.~Damour, N.~Deruelle, J.~Ibanez, and J.~Martin, ``{Poincar{\'e}-invariant gravitational field and equations of motion of two pointlike objects: The postlinear approximation of general relativity},'' \href{http://dx.doi.org/10.1007/BF00756073}{{\em Gen. Rel. Grav.} {\bfseries 13} (1981) 963--1004}.

\bibitem{LIGOScientific:2016aoc}
{\bfseries LIGO Scientific, Virgo} Collaboration, B.~P. Abbott {\em et~al.}, ``{Observation of Gravitational Waves from a Binary Black Hole Merger},'' \href{http://dx.doi.org/10.1103/PhysRevLett.116.061102}{{\em Phys. Rev. Lett.} {\bfseries 116} no.~6, (2016) 061102}, \href{http://arxiv.org/abs/1602.03837}{{\ttfamily arXiv:1602.03837 [gr-qc]}}.

\bibitem{LIGOScientific:2017vwq}
{\bfseries LIGO Scientific, Virgo} Collaboration, B.~P. Abbott {\em et~al.}, ``{GW170817: Observation of Gravitational Waves from a Binary Neutron Star Inspiral},'' \href{http://dx.doi.org/10.1103/PhysRevLett.119.161101}{{\em Phys. Rev. Lett.} {\bfseries 119} no.~16, (2017) 161101}, \href{http://arxiv.org/abs/1710.05832}{{\ttfamily arXiv:1710.05832 [gr-qc]}}.

\bibitem{Cheung:2018wkq}
C.~Cheung, I.~Z. Rothstein, and M.~P. Solon, ``{From Scattering Amplitudes to Classical Potentials in the Post-Minkowskian Expansion},'' \href{http://dx.doi.org/10.1103/PhysRevLett.121.251101}{{\em Phys. Rev. Lett.} {\bfseries 121} no.~25, (2018) 251101}, \href{http://arxiv.org/abs/1808.02489}{{\ttfamily arXiv:1808.02489 [hep-th]}}.

\bibitem{Bern:2019nnu}
Z.~Bern, C.~Cheung, R.~Roiban, C.-H. Shen, M.~P. Solon, and M.~Zeng, ``{Scattering Amplitudes and the Conservative Hamiltonian for Binary Systems at Third Post-Minkowskian Order},'' \href{http://dx.doi.org/10.1103/PhysRevLett.122.201603}{{\em Phys. Rev. Lett.} {\bfseries 122} no.~20, (2019) 201603}, \href{http://arxiv.org/abs/1901.04424}{{\ttfamily arXiv:1901.04424 [hep-th]}}.

\bibitem{Bern:2021yeh}
Z.~Bern, J.~Parra-Martinez, R.~Roiban, M.~S. Ruf, C.-H. Shen, M.~P. Solon, and M.~Zeng, ``{Scattering Amplitudes, the Tail Effect, and Conservative Binary Dynamics at O(G4)},'' \href{http://dx.doi.org/10.1103/PhysRevLett.128.161103}{{\em Phys. Rev. Lett.} {\bfseries 128} no.~16, (2022) 161103}, \href{http://arxiv.org/abs/2112.10750}{{\ttfamily arXiv:2112.10750 [hep-th]}}.

\bibitem{Dlapa:2021npj}
C.~Dlapa, G.~K{\"a}lin, Z.~Liu, and R.~A. Porto, ``{Dynamics of binary systems to fourth Post-Minkowskian order from the effective field theory approach},'' \href{http://dx.doi.org/10.1016/j.physletb.2022.137203}{{\em Phys. Lett. B} {\bfseries 831} (2022) 137203}, \href{http://arxiv.org/abs/2106.08276}{{\ttfamily arXiv:2106.08276 [hep-th]}}.

\bibitem{Dlapa:2021vgp}
C.~Dlapa, G.~K{\"a}lin, Z.~Liu, and R.~A. Porto, ``{Conservative Dynamics of Binary Systems at Fourth Post-Minkowskian Order in the Large-Eccentricity Expansion},'' \href{http://dx.doi.org/10.1103/PhysRevLett.128.161104}{{\em Phys. Rev. Lett.} {\bfseries 128} no.~16, (2022) 161104}, \href{http://arxiv.org/abs/2112.11296}{{\ttfamily arXiv:2112.11296 [hep-th]}}.

\bibitem{Driesse:2024xad}
M.~Driesse, G.~U. Jakobsen, G.~Mogull, J.~Plefka, B.~Sauer, and J.~Usovitsch, ``{Conservative Black Hole Scattering at Fifth Post-Minkowskian and First Self-Force Order},'' \href{http://dx.doi.org/10.1103/PhysRevLett.132.241402}{{\em Phys. Rev. Lett.} {\bfseries 132} no.~24, (2024) 241402}, \href{http://arxiv.org/abs/2403.07781}{{\ttfamily arXiv:2403.07781 [hep-th]}}.

\bibitem{Driesse:2024feo}
M.~Driesse, G.~U. Jakobsen, A.~Klemm, G.~Mogull, C.~Nega, J.~Plefka, B.~Sauer, and J.~Usovitsch, ``{Emergence of Calabi{\textendash}Yau manifolds in high-precision black-hole scattering},'' \href{http://dx.doi.org/10.1038/s41586-025-08984-2}{{\em Nature} {\bfseries 641} no.~8063, (2025) 603--607}, \href{http://arxiv.org/abs/2411.11846}{{\ttfamily arXiv:2411.11846 [hep-th]}}.

\bibitem{Reitze:2019iox}
D.~Reitze {\em et~al.}, ``{Cosmic Explorer: The U.S. Contribution to Gravitational-Wave Astronomy beyond LIGO},'' {\em Bull. Am. Astron. Soc.} {\bfseries 51} no.~7, (2019) 035, \href{http://arxiv.org/abs/1907.04833}{{\ttfamily arXiv:1907.04833 [astro-ph.IM]}}.

\bibitem{Punturo:2010zz}
M.~Punturo {\em et~al.}, ``{The Einstein Telescope: A third-generation gravitational wave observatory},'' \href{http://dx.doi.org/10.1088/0264-9381/27/19/194002}{{\em Class. Quant. Grav.} {\bfseries 27} (2010) 194002}.

\bibitem{LISA:2017pwj}
{\bfseries LISA} Collaboration, P.~Amaro-Seoane {\em et~al.}, ``{Laser Interferometer Space Antenna},'' \href{http://arxiv.org/abs/1702.00786}{{\ttfamily arXiv:1702.00786 [astro-ph.IM]}}.

\bibitem{ET:2025xjr}
{\bfseries ET} Collaboration, A.~Abac {\em et~al.}, ``{The Science of the Einstein Telescope},'' \href{http://arxiv.org/abs/2503.12263}{{\ttfamily arXiv:2503.12263 [gr-qc]}}.

\bibitem{Aoude:2023fdm}
R.~Aoude and A.~Ochirov, ``{Gravitational partial-wave absorption from scattering amplitudes},'' \href{http://arxiv.org/abs/2307.07504}{{\ttfamily arXiv:2307.07504 [hep-th]}}.

\bibitem{Jones:2023ugm}
C.~R.~T. Jones and M.~S. Ruf, ``{Absorptive effects and classical black hole scattering},'' \href{http://dx.doi.org/10.1007/JHEP03(2024)015}{{\em JHEP} {\bfseries 03} (2024) 015}, \href{http://arxiv.org/abs/2310.00069}{{\ttfamily arXiv:2310.00069 [hep-th]}}.

\bibitem{Chen:2023qzo}
Y.-J. Chen, T.~Hsieh, Y.-T. Huang, and J.-W. Kim, ``{On-shell approach to (spinning) gravitational absorption processes},'' \href{http://arxiv.org/abs/2312.04513}{{\ttfamily arXiv:2312.04513 [hep-th]}}.

\bibitem{Aoude:2024jxd}
R.~Aoude, A.~Cristofoli, A.~Elkhidir, and M.~Sergola, ``{Inelastic coupled-channel eikonal scattering},'' \href{http://dx.doi.org/10.1007/JHEP05(2025)136}{{\em JHEP} {\bfseries 05} (2025) 136}, \href{http://arxiv.org/abs/2411.02294}{{\ttfamily arXiv:2411.02294 [hep-th]}}.

\bibitem{Bautista:2024emt}
Y.~F. Bautista, Y.-T. Huang, and J.-W. Kim, ``{Absorptive effects in black hole scattering},'' \href{http://dx.doi.org/10.1103/PhysRevD.111.044043}{{\em Phys. Rev. D} {\bfseries 111} no.~4, (2025) 044043}, \href{http://arxiv.org/abs/2411.03382}{{\ttfamily arXiv:2411.03382 [hep-th]}}.

\bibitem{Goldberger:2005cd}
W.~D. Goldberger and I.~Z. Rothstein, ``{Dissipative effects in the worldline approach to black hole dynamics},'' \href{http://dx.doi.org/10.1103/PhysRevD.73.104030}{{\em Phys. Rev. D} {\bfseries 73} (2006) 104030}, \href{http://arxiv.org/abs/hep-th/0511133}{{\ttfamily arXiv:hep-th/0511133}}.

\bibitem{Porto:2007qi}
R.~A. Porto, ``{Absorption effects due to spin in the worldline approach to black hole dynamics},'' \href{http://dx.doi.org/10.1103/PhysRevD.77.064026}{{\em Phys. Rev. D} {\bfseries 77} (2008) 064026}, \href{http://arxiv.org/abs/0710.5150}{{\ttfamily arXiv:0710.5150 [hep-th]}}.

\bibitem{Goldberger:2020wbx}
W.~D. Goldberger and I.~Z. Rothstein, ``{Horizon radiation reaction forces},'' \href{http://dx.doi.org/10.1007/JHEP10(2020)026}{{\em JHEP} {\bfseries 10} (2020) 026}, \href{http://arxiv.org/abs/2007.00731}{{\ttfamily arXiv:2007.00731 [hep-th]}}.

\bibitem{Goldberger:2020fot}
W.~D. Goldberger, J.~Li, and I.~Z. Rothstein, ``{Non-conservative effects on spinning black holes from world-line effective field theory},'' \href{http://dx.doi.org/10.1007/JHEP06(2021)053}{{\em JHEP} {\bfseries 06} (2021) 053}, \href{http://arxiv.org/abs/2012.14869}{{\ttfamily arXiv:2012.14869 [hep-th]}}.

\bibitem{Holstein:2008sx}
B.~R. Holstein and A.~Ross, ``{Spin Effects in Long Range Gravitational Scattering},'' \href{http://arxiv.org/abs/0802.0716}{{\ttfamily arXiv:0802.0716 [hep-ph]}}.

\bibitem{Holstein:2008sw}
B.~R. Holstein and A.~Ross, ``{Spin Effects in Long Range Electromagnetic Scattering},'' \href{http://arxiv.org/abs/0802.0715}{{\ttfamily arXiv:0802.0715 [hep-ph]}}.

\bibitem{Vaidya:2014kza}
V.~Vaidya, ``{Gravitational spin Hamiltonians from the S matrix},'' \href{http://dx.doi.org/10.1103/PhysRevD.91.024017}{{\em Phys. Rev. D} {\bfseries 91} no.~2, (2015) 024017}, \href{http://arxiv.org/abs/1410.5348}{{\ttfamily arXiv:1410.5348 [hep-th]}}.

\bibitem{LIGOScientific:2016wkq}
{\bfseries LIGO Scientific, Virgo} Collaboration, T.~D. Abbott {\em et~al.}, ``{Improved analysis of GW150914 using a fully spin-precessing waveform Model},'' \href{http://dx.doi.org/10.1103/PhysRevX.6.041014}{{\em Phys. Rev. X} {\bfseries 6} no.~4, (2016) 041014}, \href{http://arxiv.org/abs/1606.01210}{{\ttfamily arXiv:1606.01210 [gr-qc]}}.

\bibitem{Maybee:2019jus}
B.~Maybee, D.~O'Connell, and J.~Vines, ``{Observables and amplitudes for spinning particles and black holes},'' \href{http://dx.doi.org/10.1007/JHEP12(2019)156}{{\em JHEP} {\bfseries 12} (2019) 156}, \href{http://arxiv.org/abs/1906.09260}{{\ttfamily arXiv:1906.09260 [hep-th]}}.

\bibitem{FebresCordero:2022jts}
F.~Febres~Cordero, M.~Kraus, G.~Lin, M.~S. Ruf, and M.~Zeng, ``{Conservative Binary Dynamics with a Spinning Black Hole at O(G3) from Scattering Amplitudes},'' \href{http://dx.doi.org/10.1103/PhysRevLett.130.021601}{{\em Phys. Rev. Lett.} {\bfseries 130} no.~2, (2023) 021601}, \href{http://arxiv.org/abs/2205.07357}{{\ttfamily arXiv:2205.07357 [hep-th]}}.

\bibitem{Akpinar:2025bkt}
D.~Akpinar, F.~Febres~Cordero, M.~Kraus, A.~Smirnov, and M.~Zeng, ``{First Look at Quartic-in-Spin Binary Dynamics at Third Post-Minkowskian Order},'' \href{http://dx.doi.org/10.1103/c2dh-tj4v}{{\em Phys. Rev. Lett.} {\bfseries 135} no.~4, (2025) 041602}, \href{http://arxiv.org/abs/2502.08961}{{\ttfamily arXiv:2502.08961 [hep-th]}}.

\bibitem{Akpinar:2025byi}
D.~Akpinar, ``{Scattering Gravitons off General Spinning Compact Objects to $\mathcal{O}(G^2 S^4)$},'' \href{http://arxiv.org/abs/2511.10280}{{\ttfamily arXiv:2511.10280 [hep-th]}}.

\bibitem{Bern:1994zx}
Z.~Bern, L.~J. Dixon, D.~C. Dunbar, and D.~A. Kosower, ``{One loop n point gauge theory amplitudes, unitarity and collinear limits},'' \href{http://dx.doi.org/10.1016/0550-3213(94)90179-1}{{\em Nucl. Phys. B} {\bfseries 425} (1994) 217--260}, \href{http://arxiv.org/abs/hep-ph/9403226}{{\ttfamily arXiv:hep-ph/9403226}}.

\bibitem{Bern:1994cg}
Z.~Bern, L.~J. Dixon, D.~C. Dunbar, and D.~A. Kosower, ``{Fusing gauge theory tree amplitudes into loop amplitudes},'' \href{http://dx.doi.org/10.1016/0550-3213(94)00488-Z}{{\em Nucl. Phys. B} {\bfseries 435} (1995) 59--101}, \href{http://arxiv.org/abs/hep-ph/9409265}{{\ttfamily arXiv:hep-ph/9409265}}.

\bibitem{Bern:1995db}
Z.~Bern and A.~G. Morgan, ``{Massive loop amplitudes from unitarity},'' \href{http://dx.doi.org/10.1016/0550-3213(96)00078-8}{{\em Nucl. Phys. B} {\bfseries 467} (1996) 479--509}, \href{http://arxiv.org/abs/hep-ph/9511336}{{\ttfamily arXiv:hep-ph/9511336}}.

\bibitem{Kawai:1985xq}
H.~Kawai, D.~C. Lewellen, and S.~H.~H. Tye, ``{A Relation Between Tree Amplitudes of Closed and Open Strings},'' \href{http://dx.doi.org/10.1016/0550-3213(86)90362-7}{{\em Nucl. Phys. B} {\bfseries 269} (1986) 1--23}.

\bibitem{Bern:2008qj}
Z.~Bern, J.~J.~M. Carrasco, and H.~Johansson, ``{New Relations for Gauge-Theory Amplitudes},'' \href{http://dx.doi.org/10.1103/PhysRevD.78.085011}{{\em Phys. Rev. D} {\bfseries 78} (2008) 085011}, \href{http://arxiv.org/abs/0805.3993}{{\ttfamily arXiv:0805.3993 [hep-ph]}}.

\bibitem{Bern:2010ue}
Z.~Bern, J.~J.~M. Carrasco, and H.~Johansson, ``{Perturbative Quantum Gravity as a Double Copy of Gauge Theory},'' \href{http://dx.doi.org/10.1103/PhysRevLett.105.061602}{{\em Phys. Rev. Lett.} {\bfseries 105} (2010) 061602}, \href{http://arxiv.org/abs/1004.0476}{{\ttfamily arXiv:1004.0476 [hep-th]}}.

\bibitem{Vines:2017hyw}
J.~Vines, ``{Scattering of two spinning black holes in post-Minkowskian gravity, to all orders in spin, and effective-one-body mappings},'' \href{http://dx.doi.org/10.1088/1361-6382/aaa3a8}{{\em Class. Quant. Grav.} {\bfseries 35} no.~8, (2018) 084002}, \href{http://arxiv.org/abs/1709.06016}{{\ttfamily arXiv:1709.06016 [gr-qc]}}.

\bibitem{Guevara:2018wpp}
A.~Guevara, A.~Ochirov, and J.~Vines, ``{Scattering of Spinning Black Holes from Exponentiated Soft Factors},'' \href{http://dx.doi.org/10.1007/JHEP09(2019)056}{{\em JHEP} {\bfseries 09} (2019) 056}, \href{http://arxiv.org/abs/1812.06895}{{\ttfamily arXiv:1812.06895 [hep-th]}}.

\bibitem{Chung:2018kqs}
M.-Z. Chung, Y.-T. Huang, J.-W. Kim, and S.~Lee, ``{The simplest massive S-matrix: from minimal coupling to Black Holes},'' \href{http://dx.doi.org/10.1007/JHEP04(2019)156}{{\em JHEP} {\bfseries 04} (2019) 156}, \href{http://arxiv.org/abs/1812.08752}{{\ttfamily arXiv:1812.08752 [hep-th]}}.

\bibitem{Guevara:2019fsj}
A.~Guevara, A.~Ochirov, and J.~Vines, ``{Black-hole scattering with general spin directions from minimal-coupling amplitudes},'' \href{http://dx.doi.org/10.1103/PhysRevD.100.104024}{{\em Phys. Rev. D} {\bfseries 100} no.~10, (2019) 104024}, \href{http://arxiv.org/abs/1906.10071}{{\ttfamily arXiv:1906.10071 [hep-th]}}.

\bibitem{Chung:2019duq}
M.-Z. Chung, Y.-T. Huang, and J.-W. Kim, ``{Classical potential for general spinning bodies},'' \href{http://dx.doi.org/10.1007/JHEP09(2020)074}{{\em JHEP} {\bfseries 09} (2020) 074}, \href{http://arxiv.org/abs/1908.08463}{{\ttfamily arXiv:1908.08463 [hep-th]}}.

\bibitem{Chung:2020rrz}
M.-Z. Chung, Y.-t. Huang, J.-W. Kim, and S.~Lee, ``{Complete Hamiltonian for spinning binary systems at first post-Minkowskian order},'' \href{http://dx.doi.org/10.1007/JHEP05(2020)105}{{\em JHEP} {\bfseries 05} (2020) 105}, \href{http://arxiv.org/abs/2003.06600}{{\ttfamily arXiv:2003.06600 [hep-th]}}.

\bibitem{Bern:2020buy}
Z.~Bern, A.~Luna, R.~Roiban, C.-H. Shen, and M.~Zeng, ``{Spinning black hole binary dynamics, scattering amplitudes, and effective field theory},'' \href{http://dx.doi.org/10.1103/PhysRevD.104.065014}{{\em Phys. Rev. D} {\bfseries 104} no.~6, (2021) 065014}, \href{http://arxiv.org/abs/2005.03071}{{\ttfamily arXiv:2005.03071 [hep-th]}}.

\bibitem{Kosmopoulos:2021zoq}
D.~Kosmopoulos and A.~Luna, ``{Quadratic-in-spin Hamiltonian at $ \mathcal{O} $(G$^{2}$) from scattering amplitudes},'' \href{http://dx.doi.org/10.1007/JHEP07(2021)037}{{\em JHEP} {\bfseries 07} (2021) 037}, \href{http://arxiv.org/abs/2102.10137}{{\ttfamily arXiv:2102.10137 [hep-th]}}.

\bibitem{Chen:2021kxt}
W.-M. Chen, M.-Z. Chung, Y.-t. Huang, and J.-W. Kim, ``{The 2PM Hamiltonian for binary Kerr to quartic in spin},'' \href{http://dx.doi.org/10.1007/JHEP08(2022)148}{{\em JHEP} {\bfseries 08} (2022) 148}, \href{http://arxiv.org/abs/2111.13639}{{\ttfamily arXiv:2111.13639 [hep-th]}}.

\bibitem{Bern:2022kto}
Z.~Bern, D.~Kosmopoulos, A.~Luna, R.~Roiban, and F.~Teng, ``{Binary Dynamics through the Fifth Power of Spin at O(G2)},'' \href{http://dx.doi.org/10.1103/PhysRevLett.130.201402}{{\em Phys. Rev. Lett.} {\bfseries 130} no.~20, (2023) 201402}, \href{http://arxiv.org/abs/2203.06202}{{\ttfamily arXiv:2203.06202 [hep-th]}}.

\bibitem{Bern:2023ity}
Z.~Bern, D.~Kosmopoulos, A.~Luna, R.~Roiban, T.~Scheopner, F.~Teng, and J.~Vines, ``{Quantum field theory, worldline theory, and spin magnitude change in orbital evolution},'' \href{http://dx.doi.org/10.1103/PhysRevD.109.045011}{{\em Phys. Rev. D} {\bfseries 109} no.~4, (2024) 045011}, \href{http://arxiv.org/abs/2308.14176}{{\ttfamily arXiv:2308.14176 [hep-th]}}.

\bibitem{Menezes:2022tcs}
G.~Menezes and M.~Sergola, ``{NLO deflections for spinning particles and Kerr black holes},'' \href{http://dx.doi.org/10.1007/JHEP10(2022)105}{{\em JHEP} {\bfseries 10} (2022) 105}, \href{http://arxiv.org/abs/2205.11701}{{\ttfamily arXiv:2205.11701 [hep-th]}}.

\bibitem{Aoude:2022thd}
R.~Aoude, K.~Haddad, and A.~Helset, ``{Classical Gravitational Spinning-Spinless Scattering at O(G2S{\ensuremath{\infty}})},'' \href{http://dx.doi.org/10.1103/PhysRevLett.129.141102}{{\em Phys. Rev. Lett.} {\bfseries 129} no.~14, (2022) 141102}, \href{http://arxiv.org/abs/2205.02809}{{\ttfamily arXiv:2205.02809 [hep-th]}}.

\bibitem{Aoude:2023vdk}
R.~Aoude, K.~Haddad, and A.~Helset, ``{Classical gravitational scattering amplitude at O(G2S1{\ensuremath{\infty}}S2{\ensuremath{\infty}})},'' \href{http://dx.doi.org/10.1103/PhysRevD.108.024050}{{\em Phys. Rev. D} {\bfseries 108} no.~2, (2023) 024050}, \href{http://arxiv.org/abs/2304.13740}{{\ttfamily arXiv:2304.13740 [hep-th]}}.

\bibitem{Gatica:2024mur}
J.~P. Gatica, ``{One-Loop Observables to Higher Order in Spin},'' \href{http://arxiv.org/abs/2412.02034}{{\ttfamily arXiv:2412.02034 [hep-th]}}.

\bibitem{Gatica:2023iws}
J.~P. Gatica, ``{The Eikonal Phase and Spinning Observables},'' \href{http://arxiv.org/abs/2312.04680}{{\ttfamily arXiv:2312.04680 [hep-th]}}.

\bibitem{Liu:2021zxr}
Z.~Liu, R.~A. Porto, and Z.~Yang, ``{Spin Effects in the Effective Field Theory Approach to Post-Minkowskian Conservative Dynamics},'' \href{http://dx.doi.org/10.1007/JHEP06(2021)012}{{\em JHEP} {\bfseries 06} (2021) 012}, \href{http://arxiv.org/abs/2102.10059}{{\ttfamily arXiv:2102.10059 [hep-th]}}.

\bibitem{Jakobsen:2021lvp}
G.~U. Jakobsen, G.~Mogull, J.~Plefka, and J.~Steinhoff, ``{Gravitational Bremsstrahlung and Hidden Supersymmetry of Spinning Bodies},'' \href{http://dx.doi.org/10.1103/PhysRevLett.128.011101}{{\em Phys. Rev. Lett.} {\bfseries 128} no.~1, (2022) 011101}, \href{http://arxiv.org/abs/2106.10256}{{\ttfamily arXiv:2106.10256 [hep-th]}}.

\bibitem{Jakobsen:2021zvh}
G.~U. Jakobsen, G.~Mogull, J.~Plefka, and J.~Steinhoff, ``{SUSY in the sky with gravitons},'' \href{http://dx.doi.org/10.1007/JHEP01(2022)027}{{\em JHEP} {\bfseries 01} (2022) 027}, \href{http://arxiv.org/abs/2109.04465}{{\ttfamily arXiv:2109.04465 [hep-th]}}.

\bibitem{Jakobsen:2022fcj}
G.~U. Jakobsen and G.~Mogull, ``{Conservative and Radiative Dynamics of Spinning Bodies at Third Post-Minkowskian Order Using Worldline Quantum Field Theory},'' \href{http://dx.doi.org/10.1103/PhysRevLett.128.141102}{{\em Phys. Rev. Lett.} {\bfseries 128} no.~14, (2022) 141102}, \href{http://arxiv.org/abs/2201.07778}{{\ttfamily arXiv:2201.07778 [hep-th]}}.

\bibitem{Jakobsen:2022zsx}
G.~U. Jakobsen and G.~Mogull, ``{Linear response, Hamiltonian, and radiative spinning two-body dynamics},'' \href{http://dx.doi.org/10.1103/PhysRevD.107.044033}{{\em Phys. Rev. D} {\bfseries 107} no.~4, (2023) 044033}, \href{http://arxiv.org/abs/2210.06451}{{\ttfamily arXiv:2210.06451 [hep-th]}}.

\bibitem{Jakobsen:2023ndj}
G.~U. Jakobsen, G.~Mogull, J.~Plefka, B.~Sauer, and Y.~Xu, ``{Conservative Scattering of Spinning Black Holes at Fourth Post-Minkowskian Order},'' \href{http://dx.doi.org/10.1103/PhysRevLett.131.151401}{{\em Phys. Rev. Lett.} {\bfseries 131} no.~15, (2023) 151401}, \href{http://arxiv.org/abs/2306.01714}{{\ttfamily arXiv:2306.01714 [hep-th]}}.

\bibitem{Chen:2024mmm}
G.~Chen and T.~Wang, ``{Dynamics of spinning binary at 2PM},'' \href{http://dx.doi.org/10.1007/JHEP12(2024)213}{{\em JHEP} {\bfseries 12} (2025) 213}, \href{http://arxiv.org/abs/2406.09086}{{\ttfamily arXiv:2406.09086 [hep-th]}}.

\bibitem{Bohnenblust:2024hkw}
L.~Bohnenblust, L.~Cangemi, H.~Johansson, and P.~Pichini, ``{Binary Kerr black-hole scattering at 2PM from quantum higher-spin Compton},'' \href{http://dx.doi.org/10.1007/JHEP07(2025)261}{{\em JHEP} {\bfseries 07} (2025) 261}, \href{http://arxiv.org/abs/2410.23271}{{\ttfamily arXiv:2410.23271 [hep-th]}}.

\bibitem{Haddad:2024ebn}
K.~Haddad, G.~U. Jakobsen, G.~Mogull, and J.~Plefka, ``{Spinning bodies in general relativity from bosonic worldline oscillators},'' \href{http://dx.doi.org/10.1007/JHEP02(2025)019}{{\em JHEP} {\bfseries 02} (2025) 019}, \href{http://arxiv.org/abs/2411.08176}{{\ttfamily arXiv:2411.08176 [hep-th]}}.

\bibitem{Ben-Shahar:2023djm}
M.~Ben-Shahar, ``{Scattering of spinning compact objects from a worldline EFT},'' \href{http://dx.doi.org/10.1007/JHEP03(2024)108}{{\em JHEP} {\bfseries 03} (2024) 108}, \href{http://arxiv.org/abs/2311.01430}{{\ttfamily arXiv:2311.01430 [hep-th]}}.

\bibitem{Kosower:2018adc}
D.~A. Kosower, B.~Maybee, and D.~O'Connell, ``{Amplitudes, Observables, and Classical Scattering},'' \href{http://dx.doi.org/10.1007/JHEP02(2019)137}{{\em JHEP} {\bfseries 02} (2019) 137}, \href{http://arxiv.org/abs/1811.10950}{{\ttfamily arXiv:1811.10950 [hep-th]}}.

\bibitem{Brito:2015oca}
R.~Brito, V.~Cardoso, and P.~Pani, ``{Superradiance}: {New Frontiers in Black Hole Physics},'' \href{http://dx.doi.org/10.1007/978-3-319-19000-6}{{\em Lect. Notes Phys.} {\bfseries 906} (2015) pp.1--237}, \href{http://arxiv.org/abs/1501.06570}{{\ttfamily arXiv:1501.06570 [gr-qc]}}.

\bibitem{Endlich:2016jgc}
S.~Endlich and R.~Penco, ``{A Modern Approach to Superradiance},'' \href{http://dx.doi.org/10.1007/JHEP05(2017)052}{{\em JHEP} {\bfseries 05} (2017) 052}, \href{http://arxiv.org/abs/1609.06723}{{\ttfamily arXiv:1609.06723 [hep-th]}}.

\bibitem{Page:1976df}
D.~N. Page, ``{Particle Emission Rates from a Black Hole: Massless Particles from an Uncharged, Nonrotating Hole},'' \href{http://dx.doi.org/10.1103/PhysRevD.13.198}{{\em Phys. Rev. D} {\bfseries 13} (1976) 198--206}.

\bibitem{Luna:2023uwd}
A.~Luna, N.~Moynihan, D.~O'Connell, and A.~Ross, ``{Observables from the Spinning Eikonal},'' \href{http://arxiv.org/abs/2312.09960}{{\ttfamily arXiv:2312.09960 [hep-th]}}.

\bibitem{Cristofoli:2021jas}
A.~Cristofoli, R.~Gonzo, N.~Moynihan, D.~O'Connell, A.~Ross, M.~Sergola, and C.~D. White, ``{The Uncertainty Principle and Classical Amplitudes},'' \href{http://arxiv.org/abs/2112.07556}{{\ttfamily arXiv:2112.07556 [hep-th]}}.

\bibitem{Aoude:2021oqj}
R.~Aoude and A.~Ochirov, ``{Classical observables from coherent-spin amplitudes},'' \href{http://dx.doi.org/10.1007/JHEP10(2021)008}{{\em JHEP} {\bfseries 10} (2021) 008}, \href{http://arxiv.org/abs/2108.01649}{{\ttfamily arXiv:2108.01649 [hep-th]}}.

\bibitem{Eden:1966dnq}
R.~J. Eden, P.~V. Landshoff, D.~I. Olive, and J.~C. Polkinghorne, {\em {The analytic S-matrix}}.
\newblock Cambridge Univ. Press, Cambridge, 1966.

\bibitem{Beneke:1997zp}
M.~Beneke and V.~A. Smirnov, ``{Asymptotic expansion of Feynman integrals near threshold},'' \href{http://dx.doi.org/10.1016/S0550-3213(98)00138-2}{{\em Nucl. Phys. B} {\bfseries 522} (1998) 321--344}, \href{http://arxiv.org/abs/hep-ph/9711391}{{\ttfamily arXiv:hep-ph/9711391}}.

\bibitem{Cheung:2020sdj}
C.~Cheung and M.~P. Solon, ``{Tidal Effects in the Post-Minkowskian Expansion},'' \href{http://dx.doi.org/10.1103/PhysRevLett.125.191601}{{\em Phys. Rev. Lett.} {\bfseries 125} no.~19, (2020) 191601}, \href{http://arxiv.org/abs/2006.06665}{{\ttfamily arXiv:2006.06665 [hep-th]}}.

\bibitem{Bern:2020uwk}
Z.~Bern, J.~Parra-Martinez, R.~Roiban, E.~Sawyer, and C.-H. Shen, ``{Leading Nonlinear Tidal Effects and Scattering Amplitudes},'' \href{http://dx.doi.org/10.1007/JHEP05(2021)188}{{\em JHEP} {\bfseries 05} (2021) 188}, \href{http://arxiv.org/abs/2010.08559}{{\ttfamily arXiv:2010.08559 [hep-th]}}.

\bibitem{Proca:1936fbw}
A.~Proca, ``{Sur la theorie ondulatoire des electrons positifs et negatifs},'' \href{http://dx.doi.org/10.1051/jphysrad:0193600708034700}{{\em J. Phys. Radium} {\bfseries 7} (1936) 347--353}.

\bibitem{Fierz:1939ix}
M.~Fierz and W.~Pauli, ``{On relativistic wave equations for particles of arbitrary spin in an electromagnetic field},'' \href{http://dx.doi.org/10.1098/rspa.1939.0140}{{\em Proc. Roy. Soc. Lond. A} {\bfseries 173} (1939) 211--232}.

\bibitem{Singh:1974qz}
L.~P.~S. Singh and C.~R. Hagen, ``{Lagrangian formulation for arbitrary spin. 1. The boson case},'' \href{http://dx.doi.org/10.1103/PhysRevD.9.898}{{\em Phys. Rev. D} {\bfseries 9} (1974) 898--909}.

\bibitem{Akpinar:2024meg}
D.~Akpinar, F.~Febres~Cordero, M.~Kraus, M.~S. Ruf, and M.~Zeng, ``{Spinning black hole scattering at $ \mathcal{O} $(G$^{3}$S$^{2}$): Casimir terms, radial action and hidden symmetry},'' \href{http://dx.doi.org/10.1007/JHEP03(2025)126}{{\em JHEP} {\bfseries 03} (2025) 126}, \href{http://arxiv.org/abs/2407.19005}{{\ttfamily arXiv:2407.19005 [hep-th]}}.

\bibitem{Alaverdian:2024spu}
M.~Alaverdian, Z.~Bern, D.~Kosmopoulos, A.~Luna, R.~Roiban, T.~Scheopner, and F.~Teng, ``{Conservative Spin-Magnitude Change in Orbital Evolution in General Relativity},'' \href{http://dx.doi.org/10.1103/PhysRevLett.134.101602}{{\em Phys. Rev. Lett.} {\bfseries 134} no.~10, (2025) 101602}, \href{http://arxiv.org/abs/2407.10928}{{\ttfamily arXiv:2407.10928 [hep-th]}}.

\bibitem{Alaverdian:2025jtw}
M.~Alaverdian, Z.~Bern, D.~Kosmopoulos, A.~Luna, R.~Roiban, T.~Scheopner, and F.~Teng, ``{Observables and Unconstrained Spin Tensor Dynamics in General Relativity from Scattering Amplitudes},'' \href{http://arxiv.org/abs/2503.03739}{{\ttfamily arXiv:2503.03739 [hep-th]}}.

\bibitem{Weinberg:1995mt}
S.~Weinberg, \href{http://dx.doi.org/10.1017/CBO9781139644167}{{\em {The Quantum theory of fields. Vol. 1: Foundations}}}.
\newblock Cambridge University Press, 6, 2005.

\bibitem{Cangemi:2023ysz}
L.~Cangemi, M.~Chiodaroli, H.~Johansson, A.~Ochirov, P.~Pichini, and E.~Skvortsov, ``{From higher-spin gauge interactions to Compton amplitudes for root-Kerr},'' \href{http://dx.doi.org/10.1007/JHEP09(2024)196}{{\em JHEP} {\bfseries 09} (2024) 196}, \href{http://arxiv.org/abs/2311.14668}{{\ttfamily arXiv:2311.14668 [hep-th]}}.

\bibitem{Cangemi:2022abk}
L.~Cangemi and P.~Pichini, ``{Classical limit of higher-spin string amplitudes},'' \href{http://dx.doi.org/10.1007/JHEP06(2023)167}{{\em JHEP} {\bfseries 06} (2023) 167}, \href{http://arxiv.org/abs/2207.03947}{{\ttfamily arXiv:2207.03947 [hep-th]}}.

\bibitem{Chandrasekhar:1975nkd}
S.~Chandrasekhar, ``{On the equations governing the perturbations of the Schwarzschild black hole},'' \href{http://dx.doi.org/10.1098/rspa.1975.0066}{{\em Proc. Roy. Soc. Lond. A} {\bfseries 343} no.~1634, (1975) 289--298}.

\bibitem{Teukolsky:1973ha}
S.~A. Teukolsky, ``{Perturbations of a rotating black hole. 1. Fundamental equations for gravitational electromagnetic and neutrino field perturbations},'' \href{http://dx.doi.org/10.1086/152444}{{\em Astrophys. J.} {\bfseries 185} (1973) 635--647}.

\bibitem{Arkani-Hamed:2017jhn}
N.~Arkani-Hamed, T.-C. Huang, and Y.-t. Huang, ``{Scattering amplitudes for all masses and spins},'' \href{http://dx.doi.org/10.1007/JHEP11(2021)070}{{\em JHEP} {\bfseries 11} (2021) 070}, \href{http://arxiv.org/abs/1709.04891}{{\ttfamily arXiv:1709.04891 [hep-th]}}.

\end{thebibliography}\endgroup

\end{document}